Universitat Politècnica de Catalunya

Departament de Física i Enginyeria Nuclear

Doctorat en Física Aplicada i Simulació en Ciències

# SECOND HARMONIC GENERATION IN SUB-DIFFRACTIVE TWO-DIMENSIONAL PHOTONIC CRYSTALS

Thesis presented to obtain the qualification of Doctor from the Universitat Politècnica de Catalunya

Cristian Nistor

Supervisors: Kestutis Staliunas
Jose Trull
Crina Cojocaru

Terrassa 2010

# Contents

# Introduction

The work presented in this thesis started 4 years ago when I came to the Universitat Politècnica de Catalunya (UPC) Barcelona, where I joined the group of "Nonlinear Dynamics, Nonlinear Optics and Lasers" (DONLL). One of the emerging topics of research in this group at that time was the study of nondiffractive propagation of light. This topic started from the idea that diffraction, a fundamental property of waves in homogeneous materials, can be manipulated (reduced or even suppressed) in spatially modulated materials.

At the time when I started my PhD a very interesting idea was suggested on the application of sub-diffraction in three wave mixing. The degenerated parametric generation was studied and it was shown that the flattening of the dispersion curve can allow efficient parametric amplification of narrow beams [Sta07b]. The efficiency of the parametric amplification was improved by two different effects: 1) due to the self-collimation the beam does not spread diffractively because the wave vectors of the Bloch modes lying on the flat segment of the spatial dispersion curve have equal longitudinal components and thus do not dephase mutually in propagation and 2) the phase matching holds simultaneously for all wave vectors belonging to the flat segment since the phase matching refers to the longitudinal components of the wave vectors of parametrically interacting waves $k_{1,II} + k_{2,II} = k_0$ and this means that the diffusive broadening of the beam can also be eliminated.

From the geometrical point of view, the wave diffraction can be described as Fourier decomposition of light beam into plane waves which, in propagation, acquire phase shifts depending on their propagation direction. The dephasing of the plane wave components results in the diffractive broadening of the beam, as showed in Fig. 1(d). In the case of photonic crystals (PC), which are structures with a periodic distribution of the dielectric constant whose period is comparable to the wavelength of light [Yab87, Joh87], light propagation is governed by its dispersion surfaces [Kos99]. Different examples of dispersion surfaces are represented in Fig. 1(a), 1(b) and 1(c), where the direction of propagation of light in the PC is normal to the dispersion surface. The light propagation in PC can follow one of the three cases: normal diffraction (Fig. 1(a)), anti-diffraction (Fig. 1(b)) and zero-diffraction (situation also called self-collimation or non-diffractive propagation, illustrated in Fig. 1(c)). The beam evolution in space (calculated by FDTD simulation of the light propagation in a homogeneous material) corresponding to the case of normal diffraction is represented in Fig. 1(d), where the diffractive broadening of the beam is visible. The light propagation in the case of zero-diffraction in a PC is represented in Fig. 1(e), where the beam propagates well-collimated.

Other related studies of the group include different aspects of non-diffractive light propagation: short pulse propagation under sub-diffractive conditions [Loi07, Loi07b, Sta06], diffraction manipulation in Fabry-Perot resonators filled by two-dimensional (2D) and three-dimensional (3D) photonic crystals [Pec09, Ili08, Sta07], spatially localized modes (cavity solitons) in Kerr-nonlinear resonators [Sta08b, Ego07], self-collimation in gain-loss modulated materials [Sta09b], diffraction management in Bose-Einstein condensates (BECs) [Sta09, Sta08, Sta06b] and self-collimation of sonic and ultrasonic beams [Sol09, Esp07, Per07].

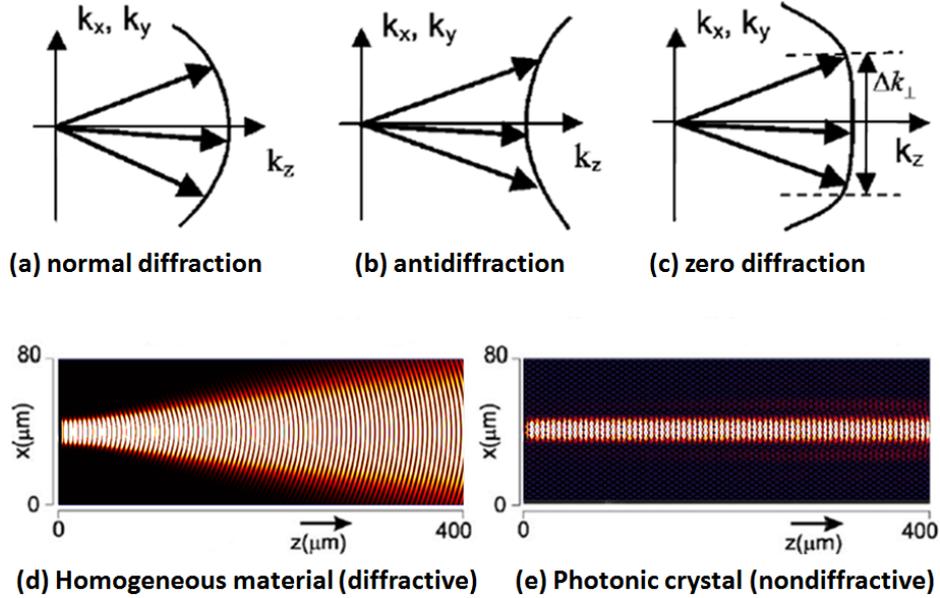

**Fig. 1** Diffraction surfaces in the case of normal diffraction (a), antidiffraction (b) and zero-diffraction (c) and numerical simulation for light propagation in the case of normal diffraction (d) and zero-diffraction (e)

The idea promoted in [Sta07b] essentially defined the direction of my PhD work. The natural way of scientific research is to bring an idea towards its realization in concrete systems, to experiments and applications. My PhD work was focused to study the effects of self-collimation regimes over the parametric nonlinear interactions, with emphasis on second harmonic (SH) generation of very narrow beams, with diameter of the order of wavelength.

The main goal of this study was to design a realistic 2D PC structure that is able to efficiently generate SH using extremely narrow beams. The most relevant feature of this structure is the fact that both input and generated beams propagate in self-collimation regimes within the structure. However, at the same time, two other important conditions have to be fulfilled: the phase matching and a sufficiently large nonlinear coupling between the two interacting modes.

The *first objective* of my work was to study the light propagation in a 2D PC and to find a realistic structure where two beams with frequencies $\omega$ and $2\omega$ can propagate simultaneously without diffraction, in phase matching and with a good spatial overlap of the modes of the fields. The *second objective* was the numerical simulation of the nonlinear interaction between the two beams inside the PC structure and the optimization of the sample to obtain the maximum efficiency of the process.

Initially, we have considered an ideal 2D PC, infinite in the third direction and made of a material without chromatic dispersion. We have made a systematical study of different configurations of the PC where we have used the radius of the air holes and the lattice geometry as parameters. As a result, we found a structure which simultaneously ensures the phase matching and self-collimation for both frequencies. The nonlinear simulations using the finite-difference time-domain (FDTD) method show that in this case a higher efficiency of the nonlinear process with respect to the plane waves in homogenous materials situation can be obtained. The results were published in [Nis08] and are described with all details in Chapter 2 of the thesis.

In our first studies we have neglected the dispersion of material and the finite extent of the crystal in the third dimension. However, these two parameters should be taken into account in a realistic sample where the material dispersion is always present and the finite size of the PC in the third dimension may cause losses. We have studied a 2D PC etched on top of a planar waveguide, which affects the dispersion and losses of the structure. After a systematic search using different geometries of the 2D PC lattice, different materials (with different refractive indices) and combining different waveguiding modes of the planar structure it appeared that achieving phase matching and self-collimation simultaneously for both waves under realistic conditions is a quite difficult issue. Sometimes the self-collimation conditions for both frequencies are fulfilled but the phase matching is absent, and sometimes the phase matching condition is satisfied but the self-collimation is no more present for both frequencies. Some of these "unsuccessful" schemes were summarized in [Nis09] and are described in Chapter 3 of the thesis. As a result of this detailed study, we have finally found one promising configuration, where the different planar modes came into play. The dispersion surfaces calculated in this case show that both FW and SH wave propagate without diffraction at the phase matched frequencies. This fact was confirmed by the FDTD simulations. As the structure has a finite size in the third dimension, a very important factor for the efficiency of the nonlinear process is represented by the out-of-plane losses. For this configuration, the losses vanish for both FW and SH wave. The characterization of this scheme was published in [Nis10b] and the detailed calculations are presented in Chapter 3.

The extremely rich dispersive properties of the photonic crystals lead to highly unusual refractive properties, significant enhancement of the light matter interaction due to low group velocities and novel possibilities for realizing phase-matching (PM) in nonlinear optical applications. By a careful choice of parameters (modulation amplitude and period), one can compensate the dispersion of the material, as well as the induced dispersion in planar structures. In this way we combine two ideas: 1) the dispersion of material and of planar waveguide can be compensated, and the PM can be reached in principle in PCs for plane waves propagating along a particular direction due to distortion of dispersion curves belonging to different bands; 2) if the PM is achieved in one particular direction it can be extended in angular space, due to the distortion (the flattening) of the diffraction curves. In addition, the PM can also be obtained in a broad spectral range given by the similar slopes of the dispersion curves of the two waves. As a result we propose a geometry for efficient parametric wave mixing, and in particular for the SH generation of narrow beams, that was published in [Nis10] and it is described with all details in Chapter 4 of the thesis.

The aim of the calculations described in Chapter 3 and Chapter 4 is to design a realistic configuration and eventually to realize the SH generation of narrow beams in the laboratory of an interested group. At least one laboratory manifested interest in our theoretical predictions of the realizable schemes and invested time and money for building the structures and organizing the experiments. To the moment of presenting the thesis the first samples were already fabricated but the experiments didn't start yet. The real samples are described in Chapter 3 and we expect that the experiments confirm well our predictions, which would be the ideal recognition of our work.

The first chapter is an introductory one, containing a short overview of the theoretical background of the problems treated along the thesis. The overview starts with a general presentation of photonic crystals (definitions, characteristics, fabrication, and applications). We also describe the main numerical methods used for the study of light propagation in PCs, treating with more details the plane wave expansion (PWE) and the finite-difference time-domain (FDTD) methods, the methods that were used extensively in the studies presented in the thesis. We make a short description of the influence that material dispersion and diffraction have over the light propagation and nonlinear processes in dielectric media. Furthermore, it is shown how photonic crystals can be used to overcome the negative effects that they have over the efficiency of the nonlinear processes. The last part of the introductory chapter contains a brief introduction about the narrow beams nonlinear interaction in photonic crystals with emphasis on parametric processes (in particular, on second harmonic generation), which is the direct subject of this thesis.

The last chapter of the thesis summarizes the results presented in this thesis outlining the actual state of the project and a possible continuation of this topic.

Chapter 1

# 1. Photonic Crystals: Analytical and Numerical Tools for Linear and Nonlinear Materials

With the "discovery" of the photon and with the subsequent invention of the Ruby laser by T.H. Mainman in 1960, optics turned into photonics. The lasers offered coherent radiation and short pulses, something that was unavailable in classical optics. This led to a revolution in many fields such as nonlinear optics, telecommunications, atomic and molecular spectroscopy and biology. The future potential of optics and photonics is enormous since the progress in all of the above areas is strongly related to the availability of novel or improved materials and structures.

The continuous progress during the last decade in the fabrication of photonic nanostructures results in a rich variety of one- , two- and three-dimensional periodic structures made of materials with different dielectric constant. These structures show new and fascinating linear-, nonlinear- and quantum-optical properties leading to unprecedented control over light propagation and light-matter interaction.

In this chapter we make a general presentation of the principal characteristics of photonic crystals and some of their applications. We describe also the main computational techniques for the study of the light propagation in periodic dielectric structures with emphasis on the finite-difference time-domain and plane wave expansion methods that were used extensively to obtain the results presented in the thesis. In the last part of the chapter we make a brief introduction about parametric nonlinear processes with emphasis on the second harmonic generation, which is the process that we consider along the thesis, together with a description of non-diffractive propagation of light and narrow beams interaction in photonic crystals.

## 1.1 Photonic crystals: general overview

Wave propagation in periodic media was first studied by Lord Rayleigh in 1887 [Ray887], who showed that one-dimensional (1D) periodic media have a narrow frequency band gap prohibiting light propagation. This band gap is angle-dependent, due to the different periodicities experienced by light propagating at different incidence angles. This produces a reflected color that varies with the angle. The term "photonic crystals" appeared 100 years later in the work of Yablonovich [Yab87] and S. John [Joh87] and it was associated with two- and three- dimensional periodic structures. However, these structures have also a longer history, since in 1914 C.G. Darwin [Dar14] derived the "dynamical theory of X-ray diffraction" that includes the effect of X-ray standing waves, finite mini-gaps, etc. In 1993 E. Yablonovich raised the question that if ordinary crystals have band gaps for X-ray waves, which are also electromagnetic waves, does it mean that every ordinary crystal is a "photonic crystal"? [Yab93] Finally, it appeared that the very small index contrasts for X-rays lead to dynamical diffraction that is essentially one-dimensional, even if the crystals themselves are three-dimensional [Yab07]. Therefore, a conventional definition of a *photonic crystal* has two requirements: high index contrast and a 2D or 3D periodicity.

In nature, color producing structures such as multilayer reflectors and photonic crystals are found in a wide variety of living organisms [Par98] from marine polychaete worms [Par01], birds [Vig06] (as shown in Fig. 1.1(a) and Fig. 1.1(b)) to fruits [Lee91] and beetles [Wel07]. Also, there are several butterfly species that are known to have color-producing structure [Kin05], as illustrated in Fig. 1.2(a) and Fig. 1.2(b) for two exemplars of Karner blue butterfly from the Lycaenid family. The electron microscopic investigation of the photonic crystal structure found on the wing scales of two Lycaenid butterflies (Fig. 1.2(c) and Fig. 1.2(d)) shows that one of the possible functions of the photonic crystal structure is the body thermal regulation, reducing significantly the penetration of light with certain wavelengths [Bir03].

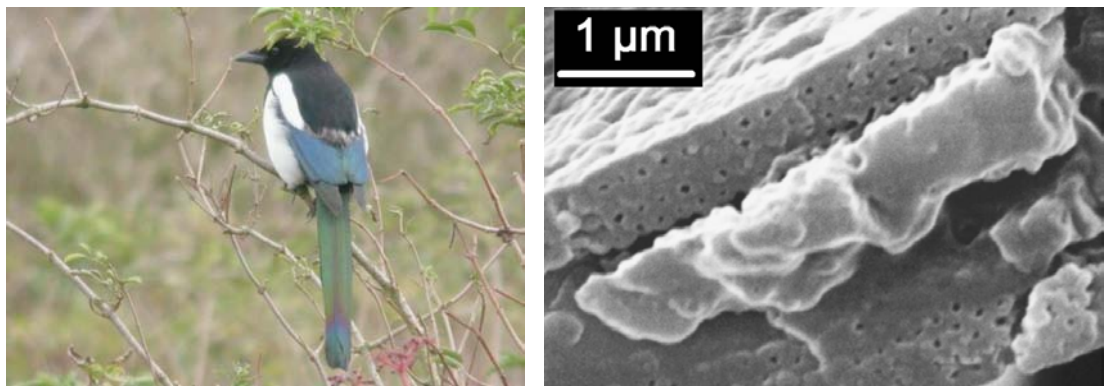

**Fig. 1.1** (a) The black-billed magpie *Pica pica* has yellow-green iridescent feather forming a long tail; (b) The cross-section of barbules of a yellow-green tail feature show that the cortex surrounding the central core is a thin film of keratine and melanine containing cylindrical air holes regularly distributed [Vig06]

Photonic crystals appear in nature not only in living creatures. Opal's color play (Fig. 1.3(a)) creates a very special attraction and fascination since long time ago. Only in the 1960s scientists could analyze opals with electron microscopy and discovered that a microscopic structure of silica spheres (illustrated in Fig. 1.3(b)) cause this behavior.

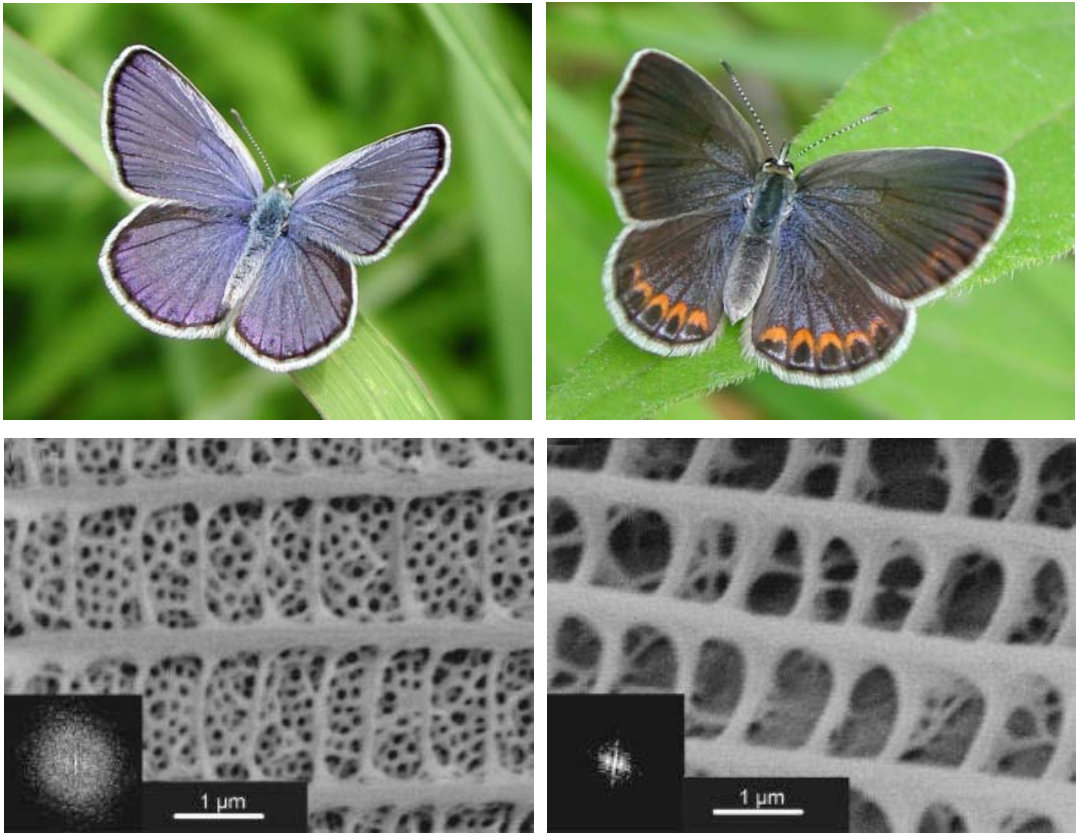

**Fig. 1.2** (a) and (b)Two exemplars of butterflies from Lycaenid family; (c) and (d) SEM images showing the fine structure found on the wing scales for a blue butterfly and for a brown one, respectively, from the Lycaenid family [Bir03]

After Yablonovitch [Yab87] and John [Joh87] joined the tools of classical electromagnetism and solid-state physics and introduced the concepts of omnidirectional photonic band gaps in two and three dimensions, this generalization led to many subsequent developments in the fabrication, theory and application of PCs, from integrated optics to negative refraction and fibers that guide light in air.

The regularly repeating internal regions of high and low dielectric constant affect the propagation of electromagnetic waves in a similar way as the periodic potential affects the electron motion in a semiconductor crystal creating allowed and forbidden electronic energy bands. In the PC, photons propagate in a medium with periodic dielectric constant $\varepsilon(\vec{r}+\vec{a}) = \varepsilon(\vec{r})$, where $a$ is the lattice constant. The periodic variation can be one- , two- or three-dimensional. This periodicity imposes a limitation on the propagating modes which, according to the Bloch's theorem (that states that the eigenfunction of a wave in a periodic potential can be written as the product of a plane wave envelope function and a periodic function) can be written as $\vec{E}_{\vec{k}}(\vec{r}) = \vec{u}_{\vec{k}}(\vec{r})e^{i(\vec{k}\cdot\vec{r}-\omega_k t)}$ where $\vec{u}_{\vec{k}}(\vec{r}) = \vec{u}_{\vec{k}}(\vec{r}+\vec{a})$. The wave vector $\vec{k}$ is called the Bloch vector and it is related to the frequency $\omega_k$ through the dispersion relation. Those frequencies for which the wave vector $\vec{k}$ has real values represent the propagating modes and the corresponding photons are allowed to propagate through the crystal. However, for the frequency range for which $\vec{k}$ becomes imaginary the photons cannot propagate and they will be totally reflected, giving rise to regions called photonic band gaps.

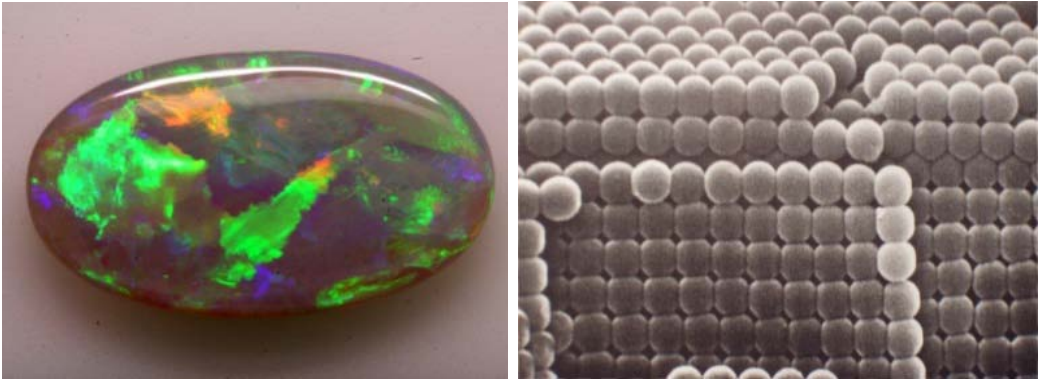

**Fig. 1.3** (a) Image of an opal and (b) SEM image of an opal showing the silica spheres structure

If some defects are introduced in the crystal (analogue to the electronic dopants) it leads to the appearance of localized electromagnetic states: linear waveguides and point-like cavities.

### 1.1.1 Light propagation in one-dimensional periodic media

A 1D periodic medium can be made by alternating layers of materials with different refractive indices $n_1$ and $n_2$ generally of different thickness $d_1$ and $d_2$, where $a = d_1 + d_2$ is the lattice constant. The classic approach for analyzing this kind of structure considers a plane wave propagating through the periodic medium and calculates the multiple reflections that take place at each interface. The relation between the fields at each side of the interface can be written in matrix form. Through matrix multiplication over the different layers of the crystal, the total reflection and transmission can be calculated. This method is known as the transfer matrix method. It can be shown that the structure acts as a perfect mirror for a certain wavelength range near $\lambda = 2m(d_1 n_1 + d_2 n_2)$, where $m$ is an integer number. This condition is known as the Bragg condition and the structure (represented in Fig. 1.4) is known as Bragg mirror [Yeh88]. Another way to analyze the properties of Bragg mirrors is based on the dispersion relation or, the band diagram. We focus on this approach since it can be used to describe also the more general cases of 2D and 3D periodic media.

1D periodic structures are homogeneous in the plane $(x, y)$ perpendicular to the direction of periodicity, which will be considered to be the $z$-direction. One important attribute of Bloch states is that the solution with the wave vector $k_z$ is identical to the solutions with wave vectors $k_z + mG$, where $G = 2\pi/a$ is the reciprocal lattice constant. Therefore, it is only needed to consider $k_z$ in the range $-\pi/a < k_z \leq \pi/a$. This region is called the first Brillouin zone and can be further reduced by symmetry considerations to $0 < k_z \leq \pi/a$, which is called the irreducible Brillouin zone.

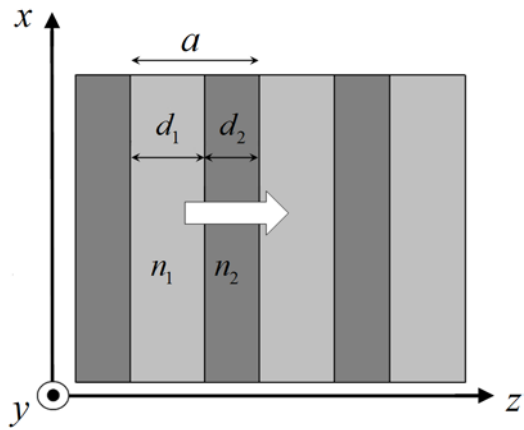

**Fig. 1.4** One-dimensional PC made by alternating layers of materials with different refractive indices $n_1$ and $n_2$ and different thicknesses $d_1$ and $d_2$; the lattice constant is $a = d_1 + d_2$

Fig. 1.5 shows the dispersion relation $\omega_n(k_z)$ for a homogeneous material (Fig. 1.5(a)) and for a 1D periodic structure made by two alternating layers with different refractive indices (Fig. 1.5(b)). In the case of the homogeneous material a "formal" periodicity $a$ has been assigned and therefore the dispersion line is folded back when it reaches the edges of the irreducible Brillouin zone. In the 1D periodic structure the dispersion line is split at the band edges, $k_z = 0$ and $k_z = \pi/a$ giving rise to frequency intervals in which no mode can propagate through the structure, regardless of the value of $k_z$. This frequency range, the photonic band gap (PBG), appears because the modes at the band edges are coupled one with another due to the presence of the periodic structure resulting in a frequency splitting. The PBG is generally larger when the index contrast between the alternating layers of the periodic structure is higher [Joa95].

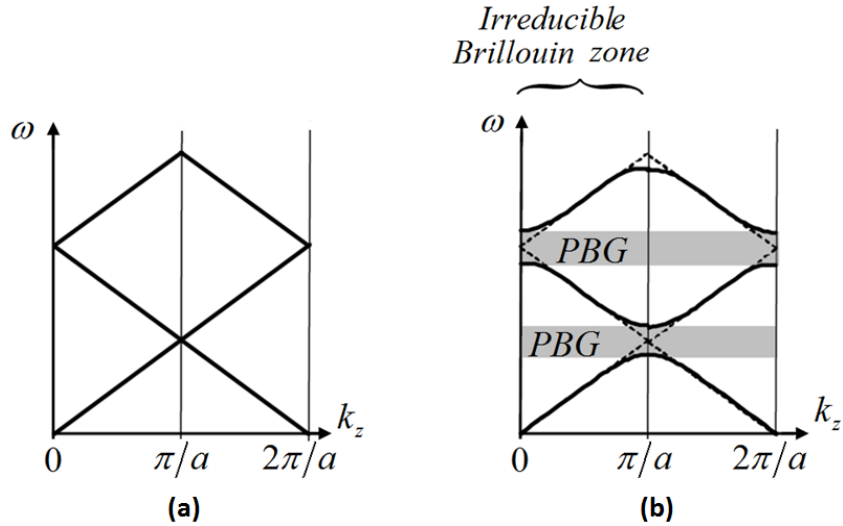

**Fig. 1.5** Dispersion relation for a homogeneous material (a) and 1D periodic structure (b) formed by two alternating layers with different refractive indices; in the periodic structure the dispersion line is split at the band edges giving rise to ranges of frequencies, called photonic band gaps (PBG), in which no mode can propagate through the structure, regardless of $k_z$

## 1.1.2 Light propagation in two-dimensional photonic crystals

A 2D PC is a periodic dielectric medium along two of its axis and homogeneous along the third axis. The plane of periodicity is determined by the primitive lattice vectors that form the basic cell. The primitive lattice vectors are defined as the smallest vectors pointing from one lattice point to another. Therefore, any lattice point can be obtained by an integer linear combination of the lattice vectors. Fig. 1.6 shows a cross section in the plane of periodicity for two of the most common lattices used in photonic crystals, the triangular lattice (or rhombic lattice, represented in Fig. 1.6(a)) and the square lattice (Fig. 1.6(b)) of air holes. The basic cell corresponding to each lattice is represented by dashed line. The lattice vectors $a_1$ and $a_2$, in both lattices have the same modulus $a$.

The homogeneous direction is along the $z$-axis while the plane of periodicity is the $xy$-plane. If the light is propagating in the plane of periodicity, the wave vector component in the $z$-direction is zero: $k_z = 0$. Similarly to 1D periodic media, the analysis of the $k$ values is also reduced to the irreducible Brillouin zone. The first Brillouin zones for the triangular and square lattices are shown in Fig. 1.6(c) and Fig. 1.6(d), respectively. The Brillouin zone has a hexagonal shape for the triangular lattice and a square shape for the square case.

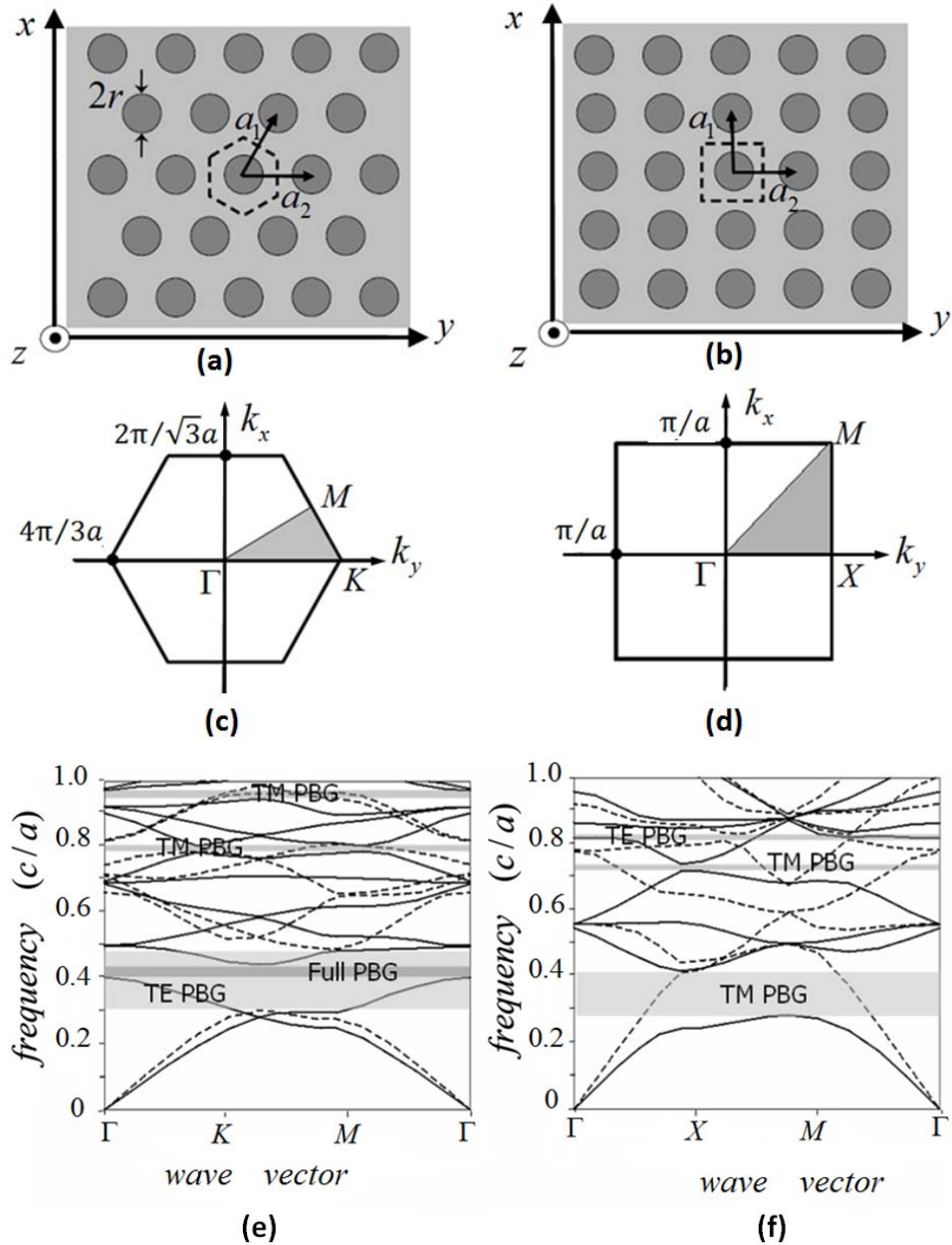

**Fig. 1.6** Triangular lattice (a) and square lattice (b) and their corresponding hexagonal (c) and squared (d) shaped Brillouin zones. The band diagram is calculated for a triangular lattice of air holes with $r = 0.45a$ etched in dielectric material with $n = 3.45$ (e), and for a square lattice of dielectric rods with $n = 3.45$ and $r = 0.2a$ in air (f). The solid lines show the TM polarized propagating modes and the dashed lines show TE polarized propagating modes

Although in both of the above cases the shape of the Brillouin zone in the reciprocal space coincides with the shape of the basic cell in the real space (for the triangular lattice the Brillouin zone is rotated with 90 deg. with respect to the basic cell), generally it is not the case. The irreducible Brillouin zone corresponds to the highlighted regions and it is usually defined by the $\Gamma$, $M$ and $X$ points in the square lattice and by the $\Gamma$, $M$ and $K$ points in the triangular lattice. For 2D PCs, if we consider different propagation directions in the plane, we should analyse a surface dispersion relation in the space $(\omega, k_x, k_y)$. The band diagram, the plot of frequency as a function of the wave vector, is typically represented along the edge of the irreducible Brillouin zone, because along this contour the maximum and minimum frequencies for each band are obtained. The mirror symmetry of the structure allows to classify the modes by separating them into two uncoupled polarizations: the transverse-electric modes (*TE*-polarization), that have the electric field confined in the plane of periodicity and the magnetic field normal to the plane ($E_x$, $E_y$ and $H_z$) and transverse–magnetic modes (*TM*-polarization) that have the magnetic field confined in the plane of periodicity and the electric field normal to the plane ($H_x$, $H_y$ and $E_z$). This definition is not rigorously correct for PCs, but it is generally accepted as adopted from the TM and TE polarizations definition in the planar structures.

The band diagram for a triangular lattice of air holes is represented in Fig. 1.6(e). The holes are etched in dielectric material and the solid lines correspond to the propagating modes with *TM* polarization while the dashed lines correspond to the propagating modes with *TE* polarization. A PBG appears for *TE* polarization between the first and second band and other narrower PBGs appear at higher frequencies. Fig. 1.6(f) shows the band diagram for the square lattice of dielectric rods in air and in this the PBGs are mostly given for the *TM* polarization. It can be shown that *TM* PBGs are usually predominant in the case of dielectric rods lattices while *TE* PBGs are predominant in air holes lattices [Joa95].

In Fig. 1.6(e) there is an overlap between the PBG for the *TM* polarization with the PBG for *TE* polarization. The frequency range where the periodic structure has a PBG for both *TE* and *TM* polarizations is called full photonic band gap, where no modes of radiation can propagate through the structure, regardless of its polarization. As in the 1D periodic media, the PBG is broader when the index contrast of the periodic structure increases. On the other hand, the central frequency of the PBG and the position of the bands can be controlled by varying the radius of the cylinders (or holes) and the geometry of the lattice.

### 1.1.3 Planar photonic crystals

Full control of light propagation can be achieved in 3D periodic media. However, due to fabrication difficulties of 3D PC at optical wavelengths the planar photonic crystals, also known as photonic crystals slabs, were proposed [Wit02, Ili04]. This kind of structures are fabricated by combining the current processing technologies developed by the microelectronics industry [Kra96, Joh99, Chu00] together with the growth of layers by well established epitaxial methods.

Planar photonic crystals combine 2D periodic structures with a slab waveguide in the vertical direction that confines light in the third dimension. Some examples of planar PC are represented in Fig. 1.7. The vertical symmetry of the structure, the slab thickness and the index contrast between the core slab and the claddings play an important role in determining the properties of planar photonic crystals.

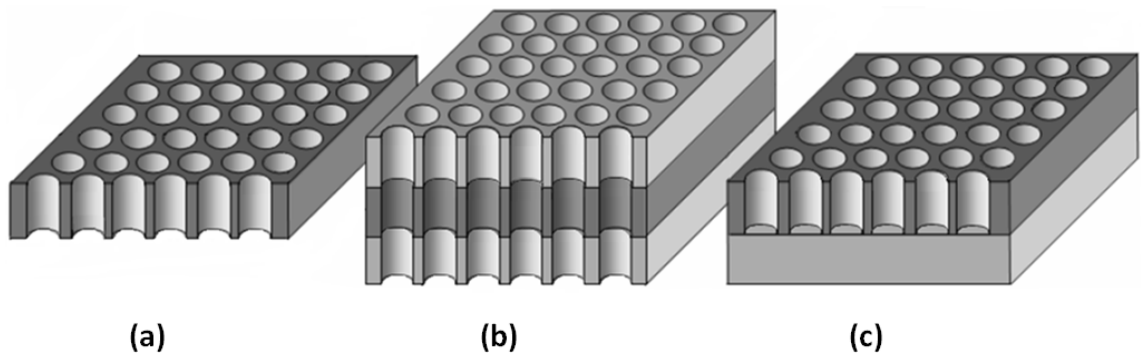

**Fig. 1.7** Schematic representation of planar PCs formed with homogeneous materials as claddings above and below the core slab ((a) and (c)) or with claddings having the same periodicity as the core slab (b) but with a lower refractive index. The planar structures can have a symmetric structure (b), or asymmetric(c).

For 2D PC, the electromagnetic fields can be decoupled into two transversely polarized modes, *TE* and *TM* modes. However, the polarization in planar photonic crystals cannot be so clearly defined because the structure becomes inhomogeneous in the vertical direction. Therefore, the symmetry or asymmetry of the structure in the vertical direction has a large influence on the polarization [Joh99, Qiu02].

In planar photonic crystals, due to the presence of a horizontal symmetry plane bisecting the slab, the guided modes can be classified according to whether they are even or odd with respect to reflections through this plane. For the fundamental TE and TM modes, the even modes are similar to TE modes while odd modes correspond to TM modes.

On the other hand, in the case of asymmetric structures in the vertical direction, as illustrated in Fig. 1.7(c) where the lower and upper claddings have different refractive indices, the guided modes can no longer be classified as even or odd modes and they couple to each other due to symmetry breaking. However, it is still possible to distinguish them as *TE* -like and *TM* -like when the guided modes are tightly confined in the slab core [Qiu02].

The band diagrams calculated for 2D PC cannot be applied directly to the photonic crystal slabs, because they apply only to a structure that is infinitely extended in the third dimension. Moreover, the 2D bands correspond only to states that have no wave vector component in the vertical direction (perpendicular to the plane of periodicity). The finite thickness of the slab requires the inclusion of the vertical wave vector components that produce a continuum of states that represents perhaps the most important feature of the projected band diagram, the element that distinguishes slabs from ordinary 2D photonic crystals. This continuum of states is called *light cone* and consists of radiation modes that are extended infinitely in the region outside the slab. Guided modes, which are states localized to the plane of the slab, can only exist in the regions of the band diagram that are below the light cone [Joh99].

The projected band structure for a PC slab is plotted in Fig. 1.8. Any state that lies below the light cone in the band diagram cannot couple with modes in the bulk background. Therefore, the discrete bands below the light cone are guided: the states are infinitely extended within the plane of the slab, but decay exponentially into the background region. This confinement is analogous to total internal reflection and it appears because the guided modes see a higher effective index in the slab than in the background. When a guided mode reaches the edge of the light cone, it becomes a radiative state: it extends, although with low amplitude, infinitely far into the background and cannot be used to permanently confine light within the slab.

There are many studies about the 2D PC waveguides (2D PC-WGs) and about the losses that appear in this kind of structures. There are two main approaches to study the problem of losses for modes which are inside the light cone. One of them consists in frequency domain methods which calculate the complex propagation constant for the waveguide modes, the imaginary part giving then the loss directly [Lal02, Had02, And03, Sau03]. An alternative approach is the use of the time domain methods where the total input and output powers are found and from these a loss per unit length can be calculated. In this case, the FDTD method is used [Cry02].

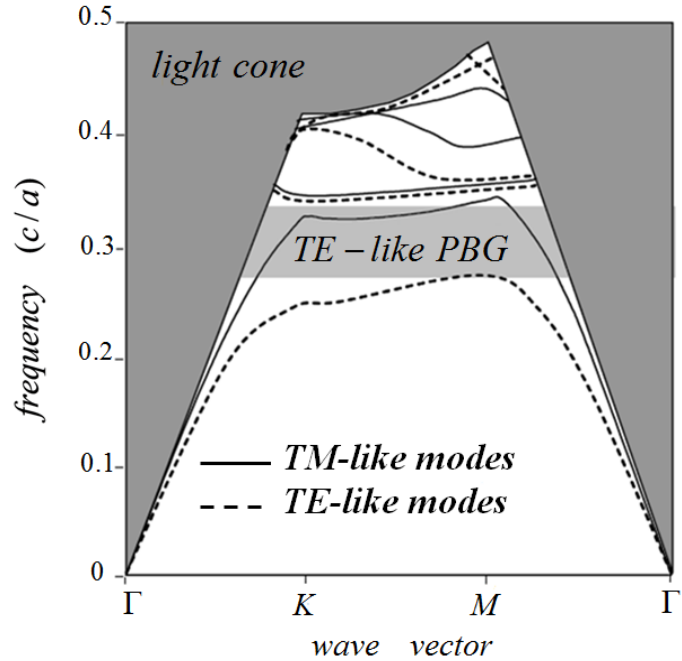

**Fig. 1.8** Band diagram for a planar PC forms by a triangular lattice of air holes with $r = 0.3a$ etched in a medium with $n = 3.45$; the core slab thickness is $2a$.

As in the case of 2D PC, planar PCs have photonic band gaps, but of a different type. A "band gap" in this case is a range of frequencies in which no guided modes exist. It is not a true band gap because there are still radiation modes at those frequencies. The presence of the radiation modes in the gap has the consequence that resonant cavity modes will eventually decay into the background.

Photonic crystals slabs are an important system in the practical application of PCs, and the band structure formalism provides a powerful tool in their analysis. The presence of the light cone means that a complete band gap is impossible and, moreover, the presence of the gap in the guided modes was used in applications [Fan97]. However, perhaps the most important advantage of the PC slabs is that they allow exciting results from 2D PCs to be implemented more easily on optical and infrared length scales.

### 1.1.4 Three-dimensional photonic crystals

Full control of light propagation can only be achieved in 3D periodic media. There are a wide variety of possible geometries for realizing 3D periodic media and, similarly to 2D PCs, the optical properties are usually analyzed from the band diagram calculated in the irreducible Brillouin zone.

The first experimental demonstration of photonic bandgaps was carried out in the microwave regime using a 3D structure [Yab89]. The structure, called sometimes "Yablonovite", consists of a dielectric medium drilled along three of the axes of a diamond lattice, as illustrated schematically in Fig. 1.9(a) and by a SEM image in Fig. 1.9(b), and it was the first geometry analyzed to achieve a full 3D PBG [Ho90]. However, the fabrication of this structure for optical wavelengths is difficult and different configurations were investigated.

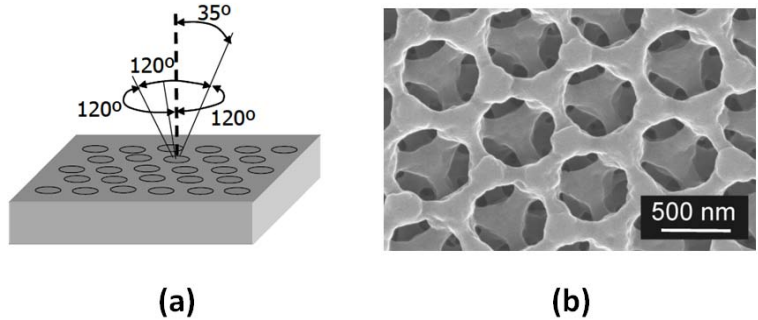

**Fig. 1.9** Schematic representation (a) and SEM image of one Yablonovite sample

Other geometries used for the fabrication of 3D PCs at optical wavelengths are the "woodpile" (illustrated in Fig. 1.10(a)), the opal (Fig. 1.10(b)) and inverse-opal configurations (Fig. 1.10(c)). The applications of this kind of structures cover a very large area, from low threshold lasers through control of spontaneous emission and compact routing via low-loss optical waveguides in 3D to high–quality factor resonators by introducing a defect [Oga04, Nod07, Ish09, Rad09, Ima06, and Rin08].

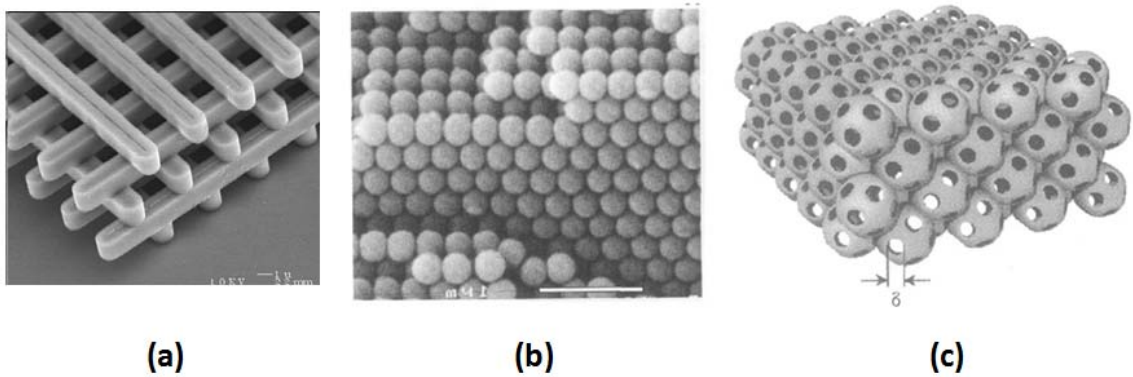

**Fig. 1.10** Representation of the woodpile (a), the opal (b) and inverse-opal (c) configurations

### 1.1.5 Fabrication of two-dimensional photonic crystals

The PCs fabrication technologies have a great pace of advancement and in the last years became very mature. In the following paragraphs the basic techniques used in the process of fabrication are shortly described.

First, we consider the case of the 2D PC structures. All of the fabrication steps are implemented on a top of a substrate. Therefore, the substrate choice appears as an essential factor for the performance of the fabricated structure. In the case of PCs, there are three common possibilities for making the substrate: Silicon (Si), Gallium Arsenide (GaAs) and Silicon-on-insulator (SOI).

Silicon processing is the most common and mature compared to the others, due to the fact that it was behind the micro-chip revolution and it has evolved in a more practical, cost efficient and well developed technology. On the other hand, the SOI substrates are used for their higher quality mechanical characteristics, but they are less common compared to the first two choices. For the fabrication of nonlinear PCs there are usually used GaAs substrates due to their more appropriate optical properties (such as the high dielectric constant) even the processing techniques are more difficult and less mature compared to the Silicon technology.

For covering the wafer with new layers of material, a deposition process is required. The two main deposition processes are Physical Vapor Deposition (PVD) and Chemical Vapor Deposition (CVD) [Fra04].

In the PVD, the material that we want to deposit on top of the wafer is evaporated and then it climbs up and sticks to the surface of the wafer that is attached to the upper part of the system. Instead, CVD is a totally chemical process where the material that is deposited on the wafer is transformed into gas form and sent through the wafers in a closed system. The deposition is realized as a result of the chemical reactions occurring between the wafers and the gas, resulting very highly uniform deposited layers. Low pressure chemical vapor deposition (LPCVD), schematically illustrated in Fig. 1.11, is a method that provides even better uniformity of the deposition process.

Pattern transfer is the next step in the fabrication and it consists from a lithography process followed by an etching process. The lithography transfers a pattern image onto the wafer which can be covered by a mask, e.g. PMMA, or not, as in the case of maskless litography. The two most widely used lithography techniques are electron-beam (e-beam) lithography and standard photolithography [McC00].

In the e-beam lithography, as schematically represented in Fig. 1.12, the pattern image is transferred using an electron beam which is sent onto the wafer in a specific configuration which represents the shape of the proposed structure. The main advantage of this method is that it permits to avoid the diffraction limit of light and make features on the nanometer scale, down to few nanometers. The standard photolithography is a much faster method but the main disadvantage in this case is a poor resolution compared to the e-beam method.

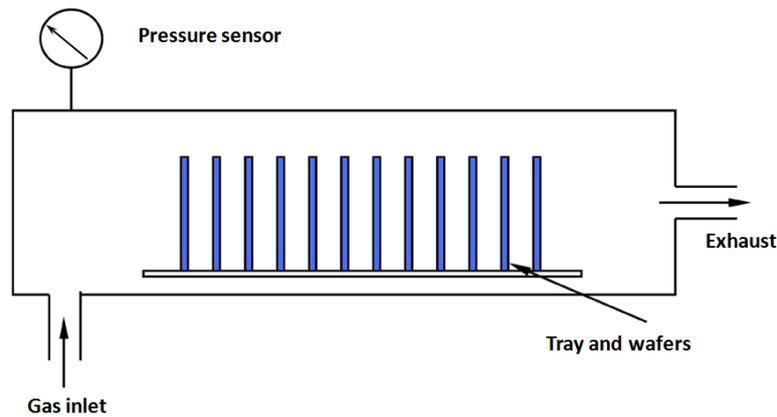

**Fig. 1.11** Schematic representation of a typical low pressure CVD installation

After the image of the structure is transferred onto the substrate, in the next step the structure is actually created in the etching process. Most widely used methods for etching are Reactive Ion Etching and Wet (Chemical) Etching [Fra04].

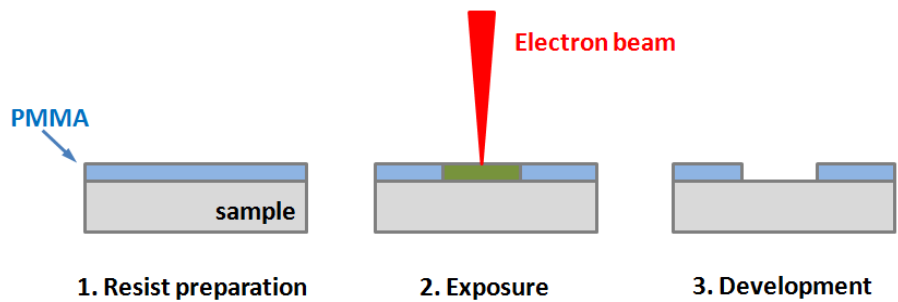

**Fig. 1.12** Schematic representation of the e-beam litography process where the sample is coated by a PMMA solution and after exposing selected areas to a beam of high-energy electrons the sample is immersed in a developer solution top selectively remove resist from the exposed areas

In the Reactive Ion Etching (RIE), chemically reactive plasma is generated and high energy ions from the plasma attack the wafer surface and react with it, removing material deposited on wafers, as schematically shown in Fig. 1.13. Some of the main advantages of this method are the perpendicular sidewalls and a high selectivity, in terms of deciding which materials should be etched. Meanwhile, the Wet Etching is a chemical process where the wafer is put into a suitable chemical solvent which etches away the required sections of the wafer ensuring also a high selectivity. However, the main disadvantage of this technique is the poor process control which makes that some unwanted extra parts of the substrate are extracted too.

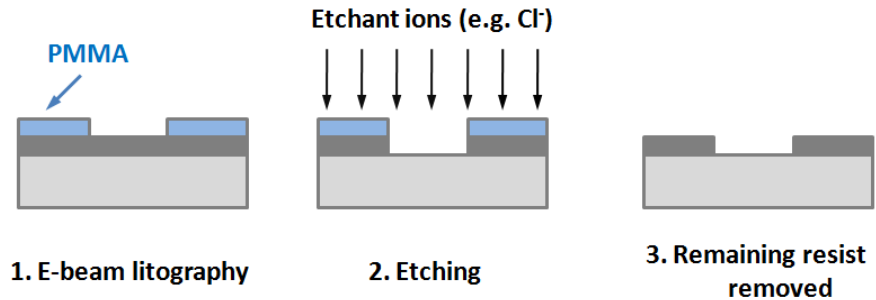

**Fig. 1.13** Schematic representation of the RIE process where, after the pattern transfer using the e-beam lithography, for example, the ions are accelerated in an electric field to produce a vertical etch after which the remaining resist is removed

Other techniques used for fabrication of 3D PCs are colloidal self-assembly [Ste08], two-photon or ink direct writing [Shi09] and multibeam interference lithography [Hyu06]. However, despite the remarkable progress in the fabrication of 2D PCs, there remain significant challenges for the fabrication of 3D crystal structures with the required nanometer scale precision.

We made an overview of several fabrication techniques, the number of which is increasing strongly during the last years. More complete overviews can be found in [Weh08] and [Fre10], for example.

## 1.2 Computational techniques

Since the PCs are generally complex, high index-contrast, 2D or 3D systems, numerical computations makes a crucial part of the theoretical analysis and also is relevant for the design of the experimental structures. Such computational methods typically fall into three categories: *time-domain "numerical experiments"* that model the time-evolution of the fields with arbitrary starting conditions in a discretized system (e.g. finite-difference time-domain method); *definite-frequency transfer method* where the scattering matrices are computed in some basis to extract the transmission or reflection through the structure, and *frequency-domain methods* to directly extract the Bloch fields and frequencies. The first two categories intuitively correspond to directly measurable quantities such as transmission (although they can also be used to compute e.g. eigenvalues), whereas the frequency-domain method is more abstract, yielding to band diagrams that provide a guide to interpretation of measurements as well as starting-point for device design and semi-analytical methods.

The band diagrams presented in this thesis are obtained using the plane wave expansion method, while the simulations for light propagation in linear and nonlinear materials presented in this work are obtained using the finite-difference time-domain technique. These two computational techniques that were used extensively for obtaining the results presented in the thesis are described in detail in this section along with some other methods used in the study of PCs.

### 1.2.1 Maxwell's equations in periodic media

The study of wave propagation in 3D periodic media was pioneered by Felix Bloch in 1928, extending an 1883 theorem in one dimension by G. Floquet. Bloch proves that waves in such a 3D periodic medium can propagate without scattering and their behavior is characterized by a periodic envelope function multiplied by a plane wave. Although Bloch studied quantum mechanics, leading to the surprising result that electrons in a conductor scatter only from imperfections and not from the periodic ions, the same technique can be applied to electromagnetism by casting Maxwell's equations as an eigenproblem in analogy with Schrödinger's equation. This allows us to take advantage of some well established results from quantum mechanics, such as orthogonality of modes, the variational theorem and perturbation theory.

All of classical electromagnetism, including the propagation of light in PCs, is governed by the four macroscopic Maxwell's equations. For the moment we are interested in the eigenmodes of the radiation field, as the interaction between field and matter will be discussed in later chapters and, therefore, we assume here that free charges and the electric current are absent. In this case the Maxwell's equations in the most general form are given by:

$$\nabla \cdot \vec{D}(\vec{r},t) = 0 \tag{1.1}$$

$$\nabla \cdot \vec{B}(\vec{r},t) = 0 \tag{1.2}$$

$$\nabla \times \vec{E}(\vec{r},t) = -\frac{\partial}{\partial t}\vec{B}(\vec{r},t) \tag{1.3}$$

$$\nabla \times \vec{H}(\vec{r},t) = \frac{\partial}{\partial t}\vec{D}(\vec{r},t) \tag{1.4}$$

where $\vec{E}$ and $\vec{H}$ are the electric and magnetic fields, respectively, and $\vec{D}$ and $\vec{B}$ are the electric displacement and magnetic induction fields.

To obtain the wave equations derived from Maxwell's equations there are used the constitutive relations that relate $\vec{D}$ to $\vec{E}$ and $\vec{B}$ to $\vec{H}$. It is assumed that the magnetic permeability is equal to that in free space $\mu_0$ and therefore:

$$\vec{B}(\vec{r},t) = \mu_0 \vec{H}(\vec{r},t) \qquad (1.5)$$

It is assumed for the moment that the dielectric constant is real, isotropic and perfectly periodic with respect to the spatial coordinate $\vec{r}$ and it does not depend on frequency. The dielectric constant of free space is noted by $\varepsilon_0$ and the relative dielectric constant of the periodic medium by $\varepsilon(\vec{r})$. The electric displacement is thus given by:

$$\vec{D}(\vec{r},t) = \varepsilon_0 \varepsilon(\vec{r}) \vec{E}(\vec{r},t) \qquad (1.6)$$

Direct substitution of equations (1.5) and (1.6) into Maxwell's equations (1.1)-(1.4) gives:

$$\nabla \cdot \{\varepsilon(\vec{r})\vec{E}(\vec{r},t)\} = 0 \qquad (1.7)$$

$$\nabla \cdot \vec{H}(\vec{r},t) = 0 \qquad (1.8)$$

$$\nabla \times \vec{E}(\vec{r},t) = -\mu_0 \frac{\partial}{\partial t} \vec{H}(\vec{r},t) \qquad (1.9)$$

$$\nabla \times \vec{H}(\vec{r},t) = \varepsilon_0 \varepsilon(\vec{r}) \frac{\partial}{\partial t} \vec{E}(\vec{r},t) \qquad (1.10)$$

After the elimination of $\vec{E}(\vec{r},t)$ or $\vec{H}(\vec{r},t)$ in (1.9) and (1.10), there are obtained the following wave equations:

$$\frac{1}{\varepsilon(\vec{r})}\nabla\times\{\nabla\times\vec{E}(\vec{r},t)\}=-\frac{1}{c^2}\frac{\partial^2}{\partial t^2}\vec{E}(\vec{r},t) \quad (1.11)$$

$$\nabla\times\{\frac{1}{\varepsilon(\vec{r})}\nabla\times\vec{H}(\vec{r},t)\}=-\frac{1}{c^2}\frac{\partial^2}{\partial t^2}\vec{H}(\vec{r},t) \quad (1.12)$$

where $c$ stands for the light velocity in free space:

$$c=\frac{1}{\sqrt{\varepsilon_0\mu_0}} \quad (1.13)$$

The periodicity of $\varepsilon(\vec{r})$ implies:

$$\varepsilon(\vec{r}+\vec{a_i})=\varepsilon(\vec{r}) \quad (i=1,2,3) \quad (1.14)$$

where $\{\vec{a_i}:i=1,2,3\}$ are the elementary lattice vectors of the PC. Because of this spatial periodicity $\varepsilon(\vec{r})$ and its inverse $\varepsilon^{-1}(\vec{r})$ can be expended into Fourier series. With this purpose there are introduced the elementary reciprocal lattice vectors $\{\vec{b_i}:i=1,2,3\}$ and the reciprocal lattice vectors $\{\vec{G}\}$:

$$\vec{a_i}\cdot\vec{b_j}=2\pi\delta_{i,j} \quad (1.15)$$

$$\vec{G}=l_1\vec{b_1}+l_2\vec{b_2}+l_3\vec{b_3} \quad (1.16)$$

where $\{\vec{l_i}:i=1,2,3\}$ are arbitrary integers and $\delta_{i,j}$ is Kronecker's delta. $\varepsilon^{-1}(\vec{r})$ is expressed as:

$$\frac{1}{\varepsilon(\vec{r})}=\sum_{\vec{G}} K(\vec{G})\exp(i\vec{G}\cdot\vec{r}) \quad (1.17)$$

For a particular distribution $\varepsilon(r)$ one can solve the wave equation (1.12) and find the modes $\vec{H}(\vec{r})$ with the corresponding frequencies, subject to the transversality condition. Once $\vec{H}(\vec{r})$ is known, the electric field $\vec{E}(\vec{r})$ can be obtained using the equation (1.11).

Following this procedure ensures that also $\vec{E}$ satisfies the transversality condition $\nabla\cdot\vec{E}=0$, as the divergence of a curl is zero. Therefore, choosing to formulate the problem in terms of $\vec{H}(\vec{r})$ and not $\vec{E}(\vec{r})$ is just an issue of mathematical convenience, since in this way it is needed to impose one transversality condition, rather than two [Joa95].

### 1.2.2 Plane wave expansion method

In the analysis of a particular structure one first tries to understand how the extended field modes propagate through the structure. This means that for a particular dielectric structure one needs to find the allowed frequencies (eigenfrequencies) propagating inside the PC and the field distribution for these modes. There are several techniques capable to perform such band calculations (calculate the dispersion relations), and one of the most widely used is the plane-wave expansion (PWE) method.

In this method, from the wave equations (1.11) and (1.12), we look for solutions of these equations of the form:

$$\vec{E}(\vec{r},t) = \vec{E}(\vec{r})e^{-i\omega t} \qquad (1.18)$$

$$\vec{H}(\vec{r},t) = \vec{H}(\vec{r})e^{-i\omega t} \qquad (1.19)$$

where $\omega$ is the eigenfrequency, and $\vec{E}(\vec{r})$ and $\vec{H}(\vec{r})$ are the eigenfunctions of the wave equations. The eigenfunctions should satisfy the following eigenvalue equations:

$$\hat{\Theta}_E \vec{E}(\vec{r}) \equiv \frac{1}{\varepsilon(\vec{r})} \nabla \times \{\nabla \times \vec{E}(\vec{r})\} = \frac{\omega^2}{c^2} \vec{E}(\vec{r}) \qquad (1.20)$$

$$\hat{\Theta}_H \vec{H}(\vec{r}) \equiv \nabla \times \{\frac{1}{\varepsilon(\vec{r})} \nabla \times \vec{H}(\vec{r})\} = \frac{\omega^2}{c^2} \vec{H}(\vec{r}) \qquad (1.21)$$

where the two operators $\hat{\Theta}_E$ and $\hat{\Theta}_H$ are defined by the first equality in each of the above equations.

Since $\varepsilon$ is a periodic function of the spatial coordinates $\vec{r}$, one can apply Bloch's theorem in equations (1.20) and (1.21) as in the case of electronic wave equation in ordinary crystals with a periodic potential due to the regular array of atoms. $\vec{E}(\vec{r})$ and $\vec{H}(\vec{r})$ are then characterized by a wave vector $\vec{k}$ in the first Brillouin zone and a band index $n$ and expressed as:

$$\vec{E}(\vec{r}) = \vec{E}_{\vec{k}n}(\vec{r}) = \vec{u}_{\vec{k}n}(\vec{r})e^{i\vec{k}\cdot\vec{r}} \qquad (1.22)$$

$$\vec{H}(\vec{r}) = \vec{H}_{\vec{k}n}(\vec{r}) = \vec{v}_{\vec{k}n}(\vec{r})e^{i\vec{k}\cdot\vec{r}} \qquad (1.23)$$

where $\vec{u}_{\vec{k}n}$ and $\vec{v}_{\vec{k}n}$ are periodic vectorial functions that satisfy the following relations:

$$\vec{u}_{\vec{k}n}(\vec{r}+\vec{a}_i) = \vec{u}_{\vec{k}n}(\vec{r}) \text{ for } i=1,2,3 \qquad (1.24)$$

$$\vec{v}_{\vec{k}n}(\vec{r}+\vec{a}_i) = \vec{v}_{\vec{k}n}(\vec{r}) \text{ for } i=1,2,3 \qquad (1.25)$$

Due to the spatial periodicity of these functions, they can be expanded in Fourier series like $\varepsilon^{-1}(\vec{r})$ in equation (1.17). The Fourier expansion leads to the following form of the eigenfunctions:

$$\vec{E}_{\vec{k}n}(\vec{r}) = \sum_{\vec{G}} \vec{E}_{\vec{k}n}(\vec{G})\exp\{i(\vec{k}+\vec{G})\cdot\vec{r}\} \qquad (1.26)$$

$$\vec{H}_{\vec{k}n}(\vec{r}) = \sum_{\vec{G}} \vec{H}_{\vec{k}n}(\vec{G})\exp\{i(\vec{k}+\vec{G})\cdot\vec{r}\} \qquad (1.27)$$

Substituting equations (1.17), (1.26) and (1.27) into (1.20) and (1.21), the following eigenvalue equations for the expansion coefficients $\{\vec{E}_{\vec{k}n}(\vec{G})\}$ and $\{\vec{H}_{\vec{k}n}(\vec{G})\}$ are obtained:

$$-\sum_{\vec{G}'} K(\vec{G}-\vec{G}')(\vec{k}+\vec{G}')\times\{(\vec{k}+\vec{G}')\times\vec{E}_{\vec{k}n}(\vec{G}')\} = \frac{\omega_{\vec{k}n}^2}{c^2}\vec{E}_{\vec{k}n}(\vec{G}) \qquad (1.28)$$

$$-\sum_{\vec{G}'} K(\vec{G}-\vec{G}')(\vec{k}+\vec{G}')\times\{(\vec{k}+\vec{G}')\times\vec{H}_{\vec{k}n}(\vec{G}')\} = \frac{\omega_{\vec{k}n}^2}{c^2}\vec{H}_{\vec{k}n}(\vec{G}) \qquad (1.29)$$

where $\omega_{\vec{k}n}$ denotes the eigen-angular frequency of $\vec{E}_{\vec{k}n}(\vec{r})$ and $\vec{H}_{\vec{k}n}(\vec{r})$. By solving one of these two sets of equations numerically one can obtain the dispersion relation of the eigenmodes, or the photonic band structure.

The plane wave method can also be extended to calculate transmission spectra [Sak01] and modal characteristics [Yok02].

### 1.2.3 Finite-difference time-domain method

The finite-difference time-domain (FDTD) method is one of the most frequently used computational techniques for simulation of electromagnetic phenomena. Based on central-difference approximations in space and time of Maxwell's equations, it provides a direct solution of time-dependent electromagnetic fields in a region of the space.

When looking at the Maxwell's equations, it can be seen that the change in the $E$-field in time (the time derivative) is dependent on the change in the $H$-field across space (the curl). This results in the basic FDTD time-stepping relation that, at any point in space, the updated value of the $E$-field in time is dependent on the stored value of the E-field and the numerical curl of the local distribution of $H$-field in space [Yee66].

The $H$-field is time-stepped in similar manner. At any point in space, the updated value of the $H$-field is dependent on the stored value of the $H$-field and the numerical curl of the local distribution of the $E$-field in space. Iterating the $E$-field and $H$-field updates results in a marching-in-time process in which data analogue to the continuous electromagnetic waves considered propagate in a numerical grid stored in the computer memory, as described with more details in Appendix A.

This description holds true for 1D, 2D and 3D techniques. Yee proposed in [Yee66] the spatially staggering of the vector components for the $E$-field and $H$-field about rectangular unit cells of a Cartesian computational grid so that each $E$-field vector component is located midway between a pair of $H$-field vector components, and conversely. This scheme, now known as a Yee lattice and represented in Fig. 1.14, has proven to be very robust and remains at the core of many current FDTD software applications.

Furthermore, Yee proposed a leapfrog scheme for marching in time in which the $E$-field and $H$-field updates are staggered so that $E$-field updates are conducted midway during each time-step between successive $H$-field updates, and conversely. As an advantage, this time-stepping scheme avoids the need to solve simultaneous equations and yields to dissipation-free numerical wave propagation. As a disadvantage of this scheme, it establishes an upper bound on the time-step to ensure numerical stability [Taf05].

As every modeling technique, the FDTD method has strengths and weaknesses. One of the advantages of the FDTD technique is that it is versatile and intuitive and can be easily understood how to use it and what to expect from a given model. Also, it is a time-domain method, which means that when a broadband pulse (such a Gaussian pulse) is used as the source, then the response of the system over a wide range of frequencies can be obtained with a single simulation. The FDTD technique allows specifying the material at all points within the computational domain and in this way a wide variety of linear and nonlinear dielectric and magnetic materials can be naturally and easily modeled.

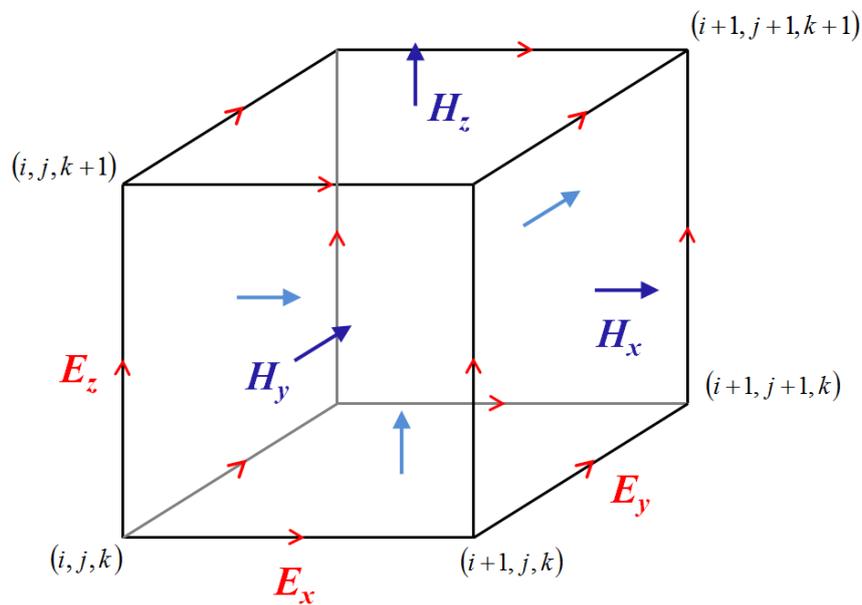

**Fig. 1.14** Representation of a standard Cartesian Yee cell used for FDTD: the electric field components form the edges of the cube and the magnetic field components form the normal to the faces of the cube.

The main disadvantage of the FDTD technique follows from the fact that the entire computational domain must be gridded and the grid spatial increment must be sufficiently fine to resolve both the smallest electromagnetic wavelength and the smallest geometrical feature in the model. This means that very large computational domains result in very large solutions in time and configurations with long thin features, like wires, are difficult to model with the FDTD method because of the excessive large computational space required. Also, the FDTD method calculates the electromagnetic fields directly in all points of the space domain, and for calculating the far field values some amount of post-processing is needed [Taf05]. Another disadvantage of the FDTD technique comes from the finite size of the computational domain. This requires introducing some artificial boundaries into the simulation space and there are a number of available highly effective absorbing boundary conditions (as the perfectly matched layer, for example) to simulate an infinite unbounded computational domain. However, the boundary conditions can only minimize the errors introduced by the truncation of the computational domain, as the weak absorption for fields propagating parallel to the boundaries or the nonzero reflections from the boundaries.

### 1.2.4 Korringa-Kohn-Rostoker (KKR) method

Another method used for the band calculation is the KKR method, introduce by Korringa [Kor47], Kohn and Rostocker [Koh54]. The characteristic feature of this method is the use of multiple scattering theory to solve the Schrödinger equation. In this way the problem is split into two parts: first, the scattering problem of a single potential in free space is solved and second, the multiple scattering is taken into account by demanding that the incident wave to each scattering centre should be the sum of the outgoing waves from all other scattering centers.

The technique was successively adapted [Oht80, Mod87, Mod01] for the vectorial resolution of the Maxwell equation and later was implemented for numerical calculations of light propagation in PCs by Stefanou [Ste92, Ste00].

In this model the 3D crystal is treated as a superposition of planes with a 2D periodic distribution of the objects of spherical symmetry. The planes are considered perpendicular to a particular crystallographic direction. A spherical multipolar expansion is used to express the incident and the scattered fields by this plane of spheres. The field outside the sphere is considered as a superposition of the incident and scattered fields for the same sphere. Next, the fields inside and outside the sphere are expanded in a similar way and by applying the contour condition on the surface of the sphere the scattered field in terms of the incident field was found.

Then, one can find the coefficients of the multipolar expansion of the scattered fields starting from a linear equation system of the incident field through a transition matrix and structural Green functions that depends on the geometry of the lattice. Next, imposing the Bloch condition in the direction perpendicular to the planes (along the crystallographic axis), one gets to an eigenvalue equation from which the photonic bands of the structure can be obtained.

The layer KKR formulation permits to obtain the total field on each side of a 3D structure of finite thickness formed by a stack of planes with the same 2D symmetry using the transfer matrix to describe the interaction between planes.

This method allows the computation of finite structures because it doesn't require periodicity in the direction of propagation. It permits also to model systems with imperfections as well as calculation of absorption by introducing an imaginary part in the dielectric constant. Lately, the technique was used with good results in modeling the real opals [Dor07, Dor08, and Loz09]. However, a main disadvantage of this method is that it doesn't permit to simulate nonlinear systems.

### 1.2.5  Order-N spectral method

For traditional schemes of photonic band calculations, as the plane wave method and the KKR method, the complexity and calculation time scales like $N^3$, where N is proportional to the size of the system. The reason is that usually a basis expansion is employed and the photonic structure is represented by an eigenvalue problem by diagonalizing a $N \times N$ matrix. The diagonalization of such a matrix needs $O(N^3)$ operations [Cha95].

Such standard $N^3$ methods serve well for ideal, perfect periodic systems with not complicated unit cells. However, for the structures presenting a defect or a certain degree of disorder it is necessary to study larger domains, which leads to an increasing of the necessary computational resources. Therefore, new methods that scale in a more favorably way with the system size were developed. It has been demonstrated that photonic structures calculations can be treated with a computational effort that scales linearly with the size of the system [Cha95]. These methods are called order-N methods.

The order N-spectral method employs discretization of the Maxwell equations in both the spatial and time domain and the integration of the Maxwell equations is made in the time domain. The spectral intensities are obtained by a Laplace transform and, if the initial field intensities are random numbers, the spectral intensities correspond to the density of states (DOS). The local density of modes and normal mode amplitudes can also be obtained [Cha98]. A slightly modified version of the order N method that permits to use this technique for calculating photonic bands for idealized metals and other dispersive materials has been proposed [Arr99].

In the category of algorithms called "finite–difference time-domain" methods, the order-N spectral method become one of the most efficient theoretical techniques for analyzing photonic crystals with a very large area of applications [Oht09, Li09, and Chu08].

## 1.3 Fundamentals of parametric nonlinear optical processes

Nonlinear optics studies the effects and phenomena that occur as a consequence of the interaction of intense coherent light with matter. Between the different processes observed it can be mentioned the alteration of the optical properties of the material, or the frequency conversion that occurs when the response of a material system to an applied optical field depends in a nonlinear way on the strength of the optical field (for example, the second-harmonic generation occurs as a result of the part of the atomic response that scales quadratically with the strength of the applied field).

In order to describe more precisely what means the optical nonlinearity, it is considered how the dipole moment per unit of volume, or polarization $P(t)$, of a material depends on the strength $E(t)$ of the applied optical field. The optical response can be described by expressing the polarization $P(t)$ as a power series in the field strength $E(t)$:

$$P(t) = \varepsilon_0 \left[ \chi^{(1)} E(t) + \chi^{(2)} E^2(t) + \chi^{(3)} E^3(t) + ... \right] \quad (1.30)$$

where $\varepsilon_0$ is the permittivity of free space, $\chi^{(1)}$ is the linear susceptibility and $\chi^{(2)}$ and $\chi^{(3)}$ are the second- and third-order nonlinear susceptibilities, respectively.

The physical processes that occur as a result of non-zero second-order polarization $P^{(2)}(t) = \varepsilon_0 \chi^{(2)} E^2(t)$ tend to be distinct from those that occur as a result of nonzero third order nonlinear polarization $P^{(3)}(t) = \varepsilon_0 \chi^{(3)} E^3(t)$. Second-order nonlinear optical interactions can occur only in noncentrosymmetrical crystals, i.e. only in crystals that do not show inversion symmetry, while the third-order nonlinear optical interactions can occur for both centrosymmetric and noncentrosymmetric media.

The nonlinear processes can be classified in parametric and nonparametric processes. Typical parametric processes include frequency conversions (sum- and

difference- frequency generation, optical parametric oscillation, etc.) and are described by a real susceptibility.

### 1.3.1 The wave equation for nonlinear optical media

Consider the curl Maxwell equations:

$$\nabla \times \vec{E} = -\frac{\partial \vec{B}}{\partial t} \tag{1.31}$$

$$\nabla \times \vec{H} = \frac{\partial \vec{D}}{\partial t} \tag{1.32}$$

and assume that the material is nonmagnetic ($\vec{B} = \mu_0 \vec{H}$) but it can be nonlinear in the sense that the fields $\vec{D}$ and $\vec{E}$ are related by:

$$\vec{D} = \varepsilon_0 \vec{E} + \vec{P} \tag{1.33}$$

where the polarization vector $\vec{P}$ depends nonlinearly on the local value of the electric field strength $\vec{E}$ as shown in (1.30).

Taking the curl of the equation (1.31) and changing the order of space and time derivatives on the right-hand side of the resulting equation and use eq. (1.32) to replace $\nabla \times \vec{B}$ by $\mu_0 (\partial \vec{D}/\partial t)$ one obtains:

$$\nabla \times \nabla \times \vec{E} + \mu_0 \frac{\partial^2}{\partial t^2} \vec{D} = 0 \tag{1.34}$$

Using eq. (1.33) to eliminate $\vec{D}$ from this equation one obtains:

$$\nabla \times \nabla \times \vec{E} + \frac{1}{c^2}\frac{\partial^2}{\partial t^2}\vec{E} = -\frac{1}{\varepsilon_0 c^2}\frac{\partial^2 \vec{P}}{\partial t^2} \qquad (1.35)$$

where on the right-hand side of this equation $\mu_0$ is replaced by $1/\varepsilon_0 c^2$.

Using an identity from the vector calculus:

$$\nabla \times \nabla \times \vec{E} = \nabla(\nabla \cdot \vec{E}) - \nabla^2 \vec{E} \qquad (1.36)$$

and assuming that the contribution of $\nabla(\nabla \cdot \vec{E})$ in equation (1.36) is negligible because when $\vec{E}$ is a transverse, infinite plane wave the term $\nabla \cdot \vec{E}$ vanishes, the wave equation takes the form:

$$\nabla^2 \vec{E} - \frac{1}{c^2}\frac{\partial^2}{\partial t^2}\vec{E} = \frac{1}{\varepsilon_0 c^2}\frac{\partial^2 \vec{P}}{\partial t^2} \qquad (1.37)$$

where $P$ can be written as :

$$P(t) = P_L(t) + P_{NL}(t) = \varepsilon_0 \chi^{(1)} E(t) + \varepsilon_0 \left[\chi^{(2)} E^2(t) + \chi^{(3)} E^3(t) + ...\right]$$
$$(1.38)$$

By separating the linear and nonlinear contributions one can write:

$$\nabla^2 \vec{E} - \frac{n^2}{c^2}\frac{\partial^2}{\partial t^2}\vec{E} = \frac{1}{\varepsilon_0 c^2}\frac{\partial^2 \vec{P}_{NL}}{\partial t^2} \text{ with } n^2 = 1 + \chi^{(2)} = \varepsilon_r \qquad (1.39)$$

where $\varepsilon_r$ is the relative dielectric constant and, for example, for the quadratic nonlinearity case we have $P_{NL} = \varepsilon_0 \chi^{(2)} E^2(t)$. Writing the equation in this way it can be observed that the $\vec{P}_{NL}$ acts as a source for electromagnetic radiation.

### 1.3.2 Coupled wave equations for sum frequency generation

Next, we consider sum-frequency generation in a lossless nonlinear optical medium. We assume the configuration shown in Fig. 1.15 where the incident waves enter onto the nonlinear medium at normal incidence.

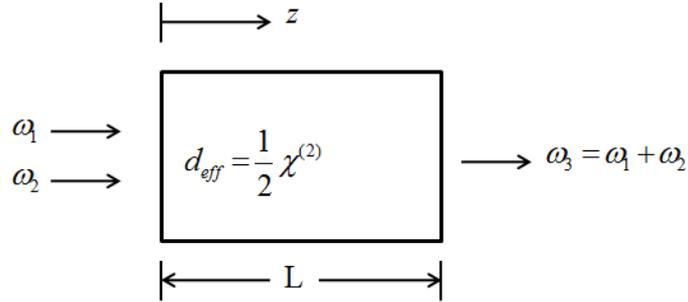

**Fig. 1.15** Schematic representation of the sum-frequency generation process

In the absence of the nonlinear source term the solution to the wave equation (1.37) for a plane wave at frequency $\omega_3$ propagating in the $+z$ direction is:

$$E(z,t) = A_3 e^{i(k_3 z - \omega_3 t)} + c.c. \qquad (1.40)$$

where $k_3 = \dfrac{n_3 \omega_3}{c}$ and $n_3$ is the linear refractive index for the $\omega_3$ frequency and the amplitude of the wave $A_3$ is constant.

When introducing the nonlinear term, $A_3$ will become a slowly varying function of $z$. Representing the nonlinear source term appearing in eq. (1.37) as:

$$P_3(z,t) = P_3 e^{-i\omega_3 t} + c.c. \qquad (1.41)$$

where $P_3 = 4\varepsilon_0 d_{eff} E_1 E_2$, with $d_{eff} = \dfrac{1}{2}\chi^{(2)}$. Considering the applied fields ($i = 1,\ 2$) are:

$$E_i(z,t) = E_i e^{-i\omega_i t} + c.c. \text{ with } E_i = A_i e^{ik_i z} \qquad (1.42)$$

the nonlinear polarization can be written as:

$$P_3 = 4\varepsilon_0 d_{eff} A_1 A_2 e^{i(k_1+k_2)z} \tag{1.43}$$

After the substitution of eqs. (1.40), (1.41) and (1.42) into the wave equation (1.37) and, since the fields depend only on the longitudinal coordinate $z$ when replacing $\nabla^2$ by $d^2/dz^2$, one obtains:

$$\frac{d^2 A_3}{dz^2} + 2ik_3 \frac{dA_3}{dz} = \frac{-4d_{eff}\omega_3^2}{c^2} A_1 A_2 e^{i(k_1+k_2-k_3)z} \tag{1.44}$$

The first term of this equation is usually neglected, as being much smaller than the second. This approximation is known as the slowly varying amplitude approximation and the fractional change in $A_3$ on a distance of the order of an optical wavelength should be much smaller than unity. When this approximation is made, eq. (1.44) becomes:

$$\frac{dA_3}{dz} = \frac{2id_{eff}\omega_3^2}{k_3 c^2} A_1 A_2 e^{i\Delta k z} \tag{1.45}$$

where $\Delta k = k_1 + k_2 - k_3$ is called the wave vector (or momentum) mismatch.

Eq. (1.45) is known as a coupled-amplitude equation, because it shows how the amplitude of the $\omega_3$ wave varies as a consequence of its coupling to the $\omega_1$ and $\omega_2$ waves. In general, the spatial variation of the $\omega_1$ and $\omega_2$ waves must also be taken into consideration, and it can be obtained analogous equations for $\omega_1$ and $\omega_2$ fields:

$$\frac{dA_1}{dz} = \frac{2id_{eff}\omega_1^2}{k_1 c^2} A_3 A_2^* e^{-i\Delta k z} \tag{1.46}$$

$$\frac{dA_2}{dz} = \frac{2id_{eff}\omega_2^2}{k_2 c^2} A_3 A_1^* e^{-i\Delta k z} \tag{1.47}$$

The solution of this system of coupled equations (1.45), (1.46) and (1.47) permits to calculate the evolution of the fields $E_1, E_2$ and $E_3$ propagating inside the nonlinear crystal.

### 1.3.3 Phase matching condition

As long as the conversion of the input fields into the sum-frequency field is not too large, the amplitudes $A_1$ and $A_2$ of the input fields can be considered as constants. In the case when $\Delta k = 0$ the equation (1.45) can be readily integrated to obtain $A_3(z)$. The amplitude $A_3$ of the sum-frequency wave increases linearly with $z$, so its intensity increases quadratically with $z$. The relation $\Delta k = 0$ is known as the condition of perfect phase matching. When this condition is fulfilled, the generated wave maintains a fixed phase relation with respect to the nonlinear polarization and it is able to extract energy most efficiently from the incident waves. From a microscopic point of view, when the phase matching condition is fulfilled the individual atomic dipoles that constitute the material system are properly phased with respect to the wave $A_3$, so that the field emitted by each dipole adds coherently in the forward direction.

When the phase matching (PM) condition is not satisfied, the amplitude of the sum-frequency $\omega_3$ field after the propagation in nonlinear medium is given by integrating eq. (1.45) from $z = 0$ to $z = L$, yielding to:

$$A_3 = \frac{2id_{eff}\omega_3^2 A_1 A_2}{k_3 c^2} \int e^{i\Delta k z} dz = \frac{2id_{eff}\omega_3^2 A_1 A_2}{k_3 c^2}\left(\frac{e^{i\Delta k L} - 1}{i\Delta k}\right) \tag{1.48}$$

The intensities of the waves are given by $I_i = \frac{1}{2} n_i \varepsilon_0 c |A_i|^2$ (with $i = 1,2,3$) and it is obtained:

$$I_3 = \frac{8 d_{eff}^2 \omega_3^2 I_1 I_2}{n_1 n_2 n_3 \varepsilon_0 c^3} sinc^2\left(\frac{\Delta k L}{2}\right) \tag{1.49}$$

where $inc(x) = \frac{sin(x)}{x}$.

The effect of the wave vector mismatch is included in the factor $sinc^2\left(\frac{\Delta k L}{2}\right)$ and the intensity of the field at a particular distance is plotted in Fig. 1.16 as a function of the mismatch. It can be noted that the efficiency of the three-wave mixing process strongly decreases as $\Delta k L/2$ increases, with the appearance of some oscillations with zero values at $L = \frac{2m\pi}{\Delta k}$. The reason of these oscillations is that if $L$ is greater than approximately $\pi/\Delta k$, the output wave can get out of phase from that of its driving polarization, and the power can start flowing back from $\omega_3$ wave to $\omega_1$ and $\omega_2$ waves. For this reason the coherent length of the interaction is defined as:

$$L_{coh} = \pi/\Delta k \qquad (1.50)$$

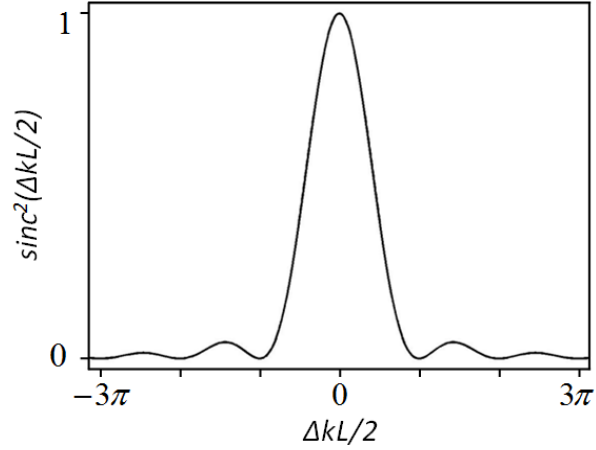

**Fig. 1.16** Effects of wave vector mismatch on the efficiency of sum-frequency generation

From eq. (1.49) it can be seen that a substantial decrease of the efficiency of the sum-frequency generation process is expected when the condition of phase matching $\Delta k \cdot L \leq \pi$ is not satisfied. The phase-matching condition $\Delta k = 0$ is often difficult to achieve because most of materials show an effect known as normal dispersion: the refractive index is an increasing function of frequency. As a result, the condition for perfect phase matching with collinear beams:

$$\frac{n_1 \omega_1}{c} + \frac{n_2 \omega_2}{c} = \frac{n_3 \omega_3}{c} \qquad (1.51)$$

where $\omega_1 + \omega_2 = \omega_3$, cannot be achieved. For the case of second-harmonic generation, with $\omega_1 = \omega_2$ and $\omega_3 = 2\omega_1$, these conditions require that

$$n(\omega_1) = n(2\omega_1) \qquad (1.52)$$

which is not possible when $n(\omega)$ increases monotonically with $\omega$.

The most common technique for obtaining the phase matching condition is the use of birefringence (the dependence of the refractive index on the direction of polarization) of materials [Boy08]. Careful control of propagation direction for the interacting frequencies allows the PM in such materials (angle tuning technique). However, in this case the PM is obtained only for particular directions and polarizations of the waves. The adjustment of the temperature in some birefringent crystals also helps to reach the PM. This method is called noncritical PM and is relatively insensitive to a slight misalignment of the beams.

However, not all crystals show birefringence, as in the case of the crystals belonging to the cubic crystal system which are optically isotropic and thus cannot be phase-matched through the birefringence method, such as for example GaAs which has a large nonlinear coefficient but a small birefringence. In this case, another technique known as quasi-phase-matching (QPM) could be used. QPM method uses a periodic modulation of the sign of the nonlinear susceptibility to compensate the refractive index dispersion. The presence of the modulation with period $\Lambda$ introduces a wave-vector $G = \frac{2\pi}{\Lambda}$ which compensates the phase mismatch. The condition for PM in the case of QPM becomes: $(k_1 + k_2 - k_3) - G = \Delta k - G = 0$. Therefore, in order to obtain maximum efficiency, the PM requires the period $\Lambda = \frac{2\pi}{\Delta k} = 2L_{coh}$. During the last years there has been an increasing interest in the development of the QPM materials for second harmonic generation, optical parametric oscillator and pulse compression [Reh08, Vod04, Sch02, Sal00]. Compared to the perfect phase-matched case, QPM leads to lower conversion (as illustrated in Fig. 1.17). Also, a large phase mismatch could require QPM materials with impractically small poling periods.

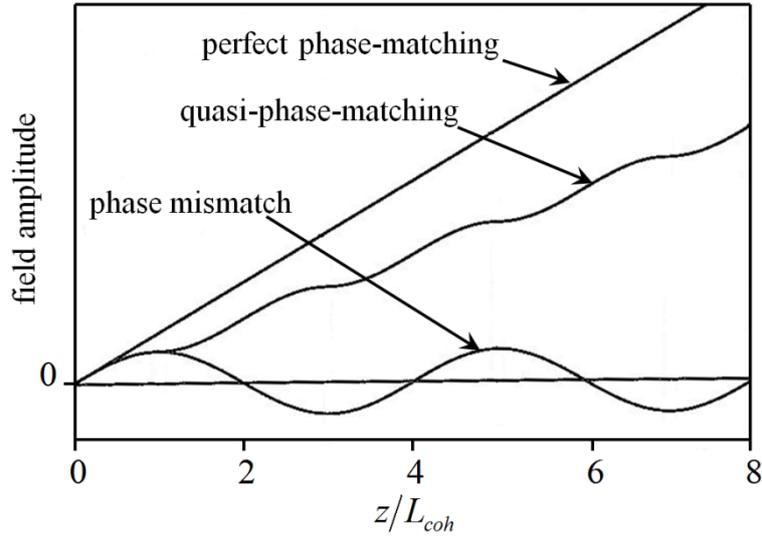

**Fig. 1.17** Comparison of spatial variation of the field amplitude in a nonlinear optical interaction for three different phase matching conditions: perfect phase-matching, quasi-phase-matching and a wave vector mismatch.

A more recently discussed possibility to obtain phase matching uses the tailoring of mode dispersion in nanoscale waveguides [DiF06, Ish09]. In particular, it was shown that collinear phase matching between TE and TM modes becomes possible for sub-wavelength dimensions of a slot waveguide [DiF06] or a high-index guiding slab [Ish09]. In such structures only the particular guided modes can be propagated. In the case of self-collimation in PCs, not only the narrow beams but arbitrary light patterns (as described with more details in Section 1.5) can be propagated and amplified. While the wave-guiding breaks the translational invariance of the position of narrow beams, the self collimation is transversally invariant.

The phase matching technique that we apply for obtaining the results presented in this thesis relies on the use of PCs. It is known that due to the periodic modulation of the refractive index in PCs, the dispersion curve can be strongly distorted in both frequency- and space- domain. The phase-matching in this case occurs not due to the compensation of $\Delta k$ by the corresponding periodicity of the material, but rather due to the distortion of the dispersion curve itself. In the next chapters more details about how phase matching can be obtained in PCs will be presented.

### 1.3.4 Second harmonic generation

In the case of second harmonic generation process, schematically represented in Fig. 1.18, there are only two frequencies involved: the fundamental frequency $\omega_1$ and the second harmonic frequency $\omega_2 = 2\omega_1$. In this case the coupled-amplitude equations for the two frequency components can be obtained analogously to those used for the case of sum-frequency generation, and it follows that:

$$\frac{dA_1}{dz} = \frac{2id_{eff}\omega_1^2}{k_1 c^2} A_2 A_1^* e^{-i\Delta kz} \quad (1.53)$$

$$\frac{dA_2}{dz} = \frac{id_{eff}\omega_2^2}{k_2 c^2} A_1^2 e^{i\Delta kz} \quad (1.54)$$

For the SH intensity it gives:

$$I_{2\omega}(L) = 2\frac{\omega^2 d_{eff}^2}{n_1^2 n_2 \varepsilon_0 c^3} I_\omega^2(0) L^2 sinc^2\left(\frac{\Delta kL}{2}\right) \quad (1.55)$$

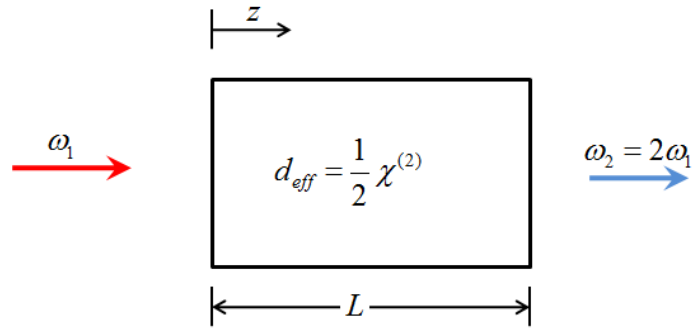

**Fig. 1.18** Schematic representation of the second-harmonic generation process

In order to describe the particularities of the parametric processes in photonic crystals, we will consider in this thesis the case of the second harmonic generation in 2D PCs in both bulk and planar waveguide configurations. The phase matching is obtained using the dispersion engineering offered by the PCs and, in order to increase the efficiency of the nonlinear process, simultaneous nondiffractive propagation regimes for both fundamental and second harmonic waves, a broad spectral range phase matching and, in the case of planar configurations, the vanishing of the out-of-plane losses for both waves are required.

## 1.4 Non-diffractive propagation of light

Diffraction is nothing else but the apparent bending of waves around small obstacles and the spreading out of waves past small openings. Diffraction occurs with all waves, including sound waves, water waves, matter waves (due to the wave-like behavior of small-momentum particles) and electromagnetic waves (such as visible light, x-rays and radio waves).

Therefore, the formalism of diffraction describes the way in which wave-beams of finite extent propagate in free space. Consider a complex wave-function $U(\vec{r},t)$ that satisfies the wave equation:

$$\nabla^2 U - \frac{1}{c^2}\frac{\partial^2 U}{\partial t^2} = 0 \qquad (1.56)$$

where $\nabla^2$ is the Laplace operator $\nabla^2 = \partial^2/\partial x^2 + \partial^2/\partial y^2 + \partial^2/\partial z^2$ in Cartesian coordinates. If one writes $U(\vec{r},t) = U(\vec{r})\exp(i\omega t)$, where $\omega$ is the frequency, the wave equation leads to a differential equation for the complex amplitude $U(\vec{r})$:

$$\nabla^2 U + k^2 U = 0 \qquad (1.57)$$

known as the Helmholtz equation, where $k = \omega/c$ is the wavenumber.

In the paraxial approximation the complex amplitude becomes:

$$U(\vec{r}) = A(\vec{r})\exp(-ikz) \qquad (1.58)$$

where the variation of the envelope $A(\vec{r})$ and its derivative with position $z$ must be slow within the distance of a wavelength $\lambda = 2\pi/k$, for which the wave approximately maintains its plane wave nature.

Therefore, substituting (1.58) into the Helmholtz equation (1.57) and neglecting $\partial^2 A/\partial^2 z$ in comparison with $k\,\partial A/\partial z$ or $k^2 A$ lead to a partial differential equation for the complex envelope $A(\vec{r})$:

$$\nabla_T^2 A - i2k \frac{\partial A}{\partial z} = 0 \tag{1.59}$$

where $\nabla_T^2 = \partial^2/\partial x^2 + \partial^2/\partial y^2$ is the transverse Laplacian operator. This equation is known as the paraxial Helmholtz equation (or the slowly varying envelope approximation of the Helmholtz equation). One of the most interesting and a useful solution of (1.59) is the Gaussian beam:

$$A(\vec{r}) = \frac{A_1}{q(z)} \exp\left[-ik \frac{\rho^2}{2q(z)}\right] \quad \text{with } q(z) = z + iz_0 \tag{1.60}$$

where $q(z)$ is called the q-parameter of the beam and $z_0$ is known as the Rayleigh length. In order to separate the amplitude and the phase of this complex envelope one can write the function $1/q(z) = 1/(z + iz_0)$ as:

$$\frac{1}{q(z)} = \frac{1}{R(z)} - i \frac{\lambda}{\pi w^2(z)} \tag{1.61}$$

where $w(z)$ and $R(z)$ are the beam width and the wavefront radius of curvature, respectively, represented in Fig. 1.19. Substituting (1.61) into (1.60) and using (1.58) it is obtained:

$$U(\vec{r}) = A_0 \frac{w_0}{w(z)} \exp\left[-\frac{\rho^2}{w^2(z)}\right] \exp\left[-ikz - ik \frac{\rho^2}{2R(z)} + i\zeta(z)\right] \tag{1.62}$$

in which $A_0 = A_1/iz_0$ has been defined for convenience and $\rho$ represents the distance from the beam axis (for point along the axis $\rho(0) = 0$). The phase term is $\zeta(z) = \tan^{-1} \frac{z}{z_0}$ and the radius of curvature of the wavefront is:

$$R(z) = z\left[1 + \left(\frac{z_0}{z}\right)^2\right] \quad (1.63)$$

The dependence of the beam width on $z$ is given by the relation:

$$w(z) = w_0\sqrt{1 + \left(\frac{z}{z_0}\right)^2} \quad (1.64)$$

where $w_0 = \sqrt{\frac{\lambda z_0}{\pi}}$ is the minimum width, at $z = 0$. Therefore, the Rayleigh length represents the distance after which the beam width is equal to $\sqrt{2}$ of its minimum value, and it is given by:

$$z_0 = \frac{\pi w_0^2}{\lambda} \quad (1.65)$$

As it can be seen from equation (1.62) the phase of a Gaussian beam is:

$$\varphi(\rho, z) = kz - \zeta(z) + \frac{k\rho^2}{2R(z)} \quad (1.66)$$

While the first term corresponds to the phase of a plane wave, the second term is given by the phase retardation $\zeta(z)$ that ranges from $-\pi/2$ at $z = -\infty$ to $\pi/2$ at $z = +\infty$, phenomenon known as Gouy effect [Sal07].

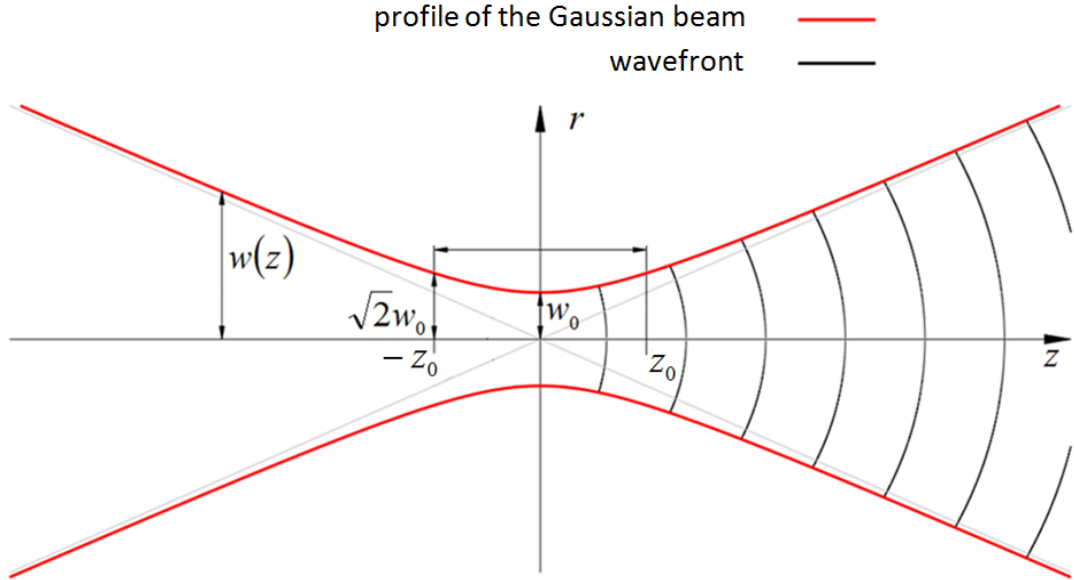

**Fig. 1.19** Gaussian beam width $w(z)$ as a function of the axial distance $z$, where $w_0$ is the beam waist, $z_R$ is the Rayleigh length

However, the third term of equation (1.66) shows that the phase of a Gaussian beam depends not only on $z$ but also varies with the distance from the beam axis for a constant $z$. This term is responsible for the wavefront bending, since the surface at constant phase satisfies the relation $k[z + \rho^2/2R(z)] - \zeta(z) = 2\pi q$ that can be approximated to the equation of a paraboloidal surface with radius of curvature $R(z)$ [Sal07].

The diffractive broadening of light beams is a fundamental phenomenon limiting the performance of many linear and nonlinear optical devices. A geometrical interpretation of diffraction is that light beams of arbitrary shapes can be Fourier decomposed into plane waves, which in propagation acquire phase shifts depending on their propagation angles. This dephasing of the plane wave components results in a diffractive broadening of the light beam. Fig. 1.20 (a) illustrates normal diffraction in propagation through homogeneous materials where the longitudinal component of the wave vector depends on the propagation angle $k_{II} \equiv k_z = \sqrt{|\vec{k}|^2 - |\vec{k}_\perp|^2}$ and $\vec{k}_\perp = (k_x, k_y)$. In general, the normal or positive diffraction means that the surfaces of constant frequency are concave in the wave vector domain $\vec{k} = (k_x, k_y, k_z)$.

It is known that diffraction can become negative for materials with refractive index modulated in a direction perpendicular to the propagation. The negative diffraction, as illustrated in Fig. 1.20 (b), geometrically means that the surfaces of constant frequency are convex in the wave vector domain. The negative diffraction was predicted for electromagnetic waves [Mor01, Gar07], for acoustic waves [Yan02, Tor04] and for matter waves [Zob99, Ost06] in propagation through modulated media. The change of the sign of diffraction is extremely interesting for propagation in nonlinear materials, as it results in a change between the nonlinear self-focusing and defocusing, and allowing, e.g. the bright solitons in defocusing media.

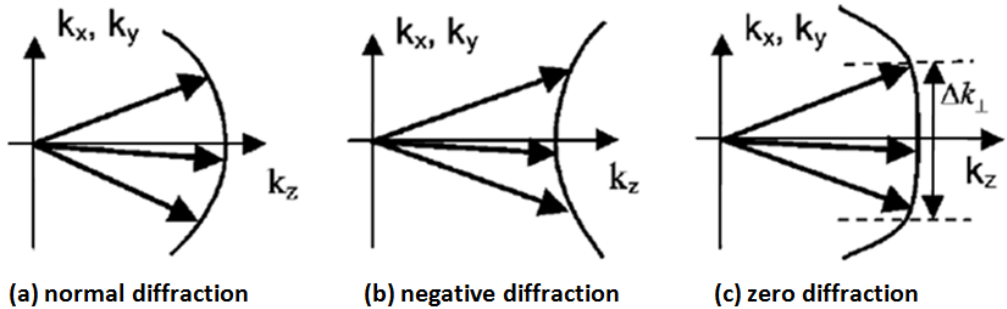

**Fig. 1.20** Representation of the surfaces of constant frequency in the case of normal diffraction (a), negative diffraction (b) and zero diffraction (c)

If the diffraction at the two edges of a propagation band for a periodic structure is of opposite signs, this means that it must vanish at some point within the band leading to the disappearance of diffraction, as illustrated in Fig. 1.20 (c). Zero diffraction physically means that light beams can propagate without diffractive broadening, as illustrated in Fig. 1.21(b) where the non-diffractive propagation of light in a 2D PC is represented. For comparison, in Fig. 1.21(a) diffractive light propagation in a homogeneous medium is represented. The non-diffractive propagation of light beams in 2D PC has been shown numerically [Kos99, Chi03], experimentally [Ili04, Pra04] and analytically [Sta06c].

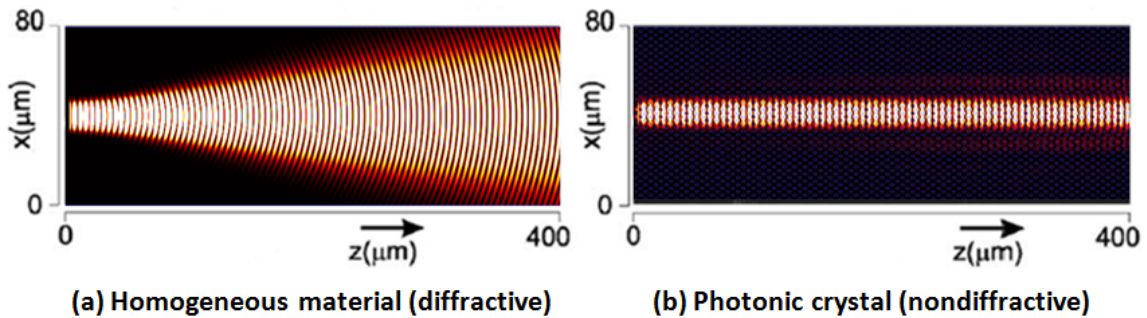

**Fig. 1.21** Representation of the diffractive broadening of the beam in homogeneous materials (a) and of the non-diffractive propagation that can be obtained in photonic crystals (b)

The light propagation in a PC is governed by its dispersion surface. A typical branch of a dispersion surface in reciprocal space (k-domain) for a PC is schematically shown by a thick curve in Fig. 1.22. The curvature of the dispersion surface turns from downward to upward and the incident light comes from the top of the figure. The direction of propagation is normal to the dispersion surface at the tie points because the energy velocity integrated over a unit cell is identical with the group velocity, which is given by $v_g = \nabla_k \omega(k)$, where $\omega(k)$ is the optical frequency at the wave vector $k$ [Kos99].

As mentioned above, the diffraction of electromagnetic waves in homogeneous media is traditionally characterized by the Rayleigh length. More recently, it has been shown that in PCs the diffraction of stationary light beams can be described by the same parameter as the Rayleigh length, but related to PC structure [Loi07]. For planar 2D PC with rhombic symmetry (which in particular cases can be reduced to the square and hexagonal types of symmetry) this parameter for light propagating in the direction of symmetry axis can be written as follows:

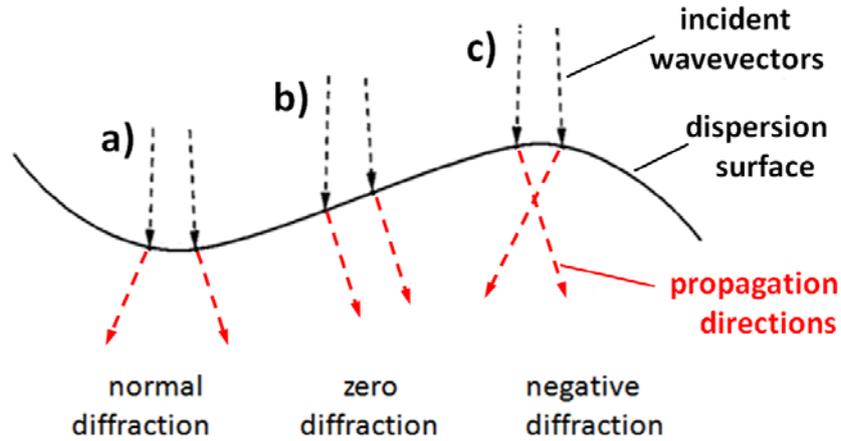

**Fig. 1.22** A schematic representation of the nondiffractive propagation (b) of light in PC together with the normal diffraction (a) and negative diffraction (c) cases; the arrows above the dispersion surface indicate incident wave vectors in reciprocal space and the arrows below the dispersion surface indicate the energy flow.

$$L^{(PC)}_{Rayleigh} = n_{eff} \pi w_\perp^2 / (d_2 \lambda_0) \qquad (1.67)$$

where $w_\perp$ is the half width of the input beam and $d_2$ is the leading order diffraction coefficient. The coefficient $d_2$ can be positive (which means normal diffraction), negative (which means negative diffraction) or equal to zero. The latter case corresponds to the sub-diffractive propagation of stationary light beams in PCs. In this case, the characteristic length $L_{Rayleigh}^{(PC)}$ becomes equal to infinity and the diffraction of the stationary light beams should be described by the higher order diffraction terms. In particular, under the sub-diffractive regimes along symmetry direction of PCs based on rhombic lattice (the square lattice being a particular case of the rhombic lattice) the next non-zero coefficient is that of the fourth order $d_4$. The diffraction length in this case (under sub-diffractive regime) can be written as follows:

$$L_{d4}^{(PC)} = \pi^2 (q_\perp^2 + q_{II}^2) w_\perp^4 / (4 d_4 \lambda_0) \tag{1.68}$$

where $q_{\perp(II)} = 2\pi / \Lambda_{\perp(II)}$ and $\Lambda_{\perp(II)}$ is the period of refractive index modulation perpendicular (along) to the direction of beam propagation.

It has been shown that the diffraction of pulsed light beams in PCs could be described also in a similar way [Sta06]. However, in this case the situation is more complicated since both diffraction and dispersion are involved. Dispersive broadening depends on the initial pulse duration and it is responsible for the symmetric broadening of the pulse in time. Diffraction, instead, is responsible for the symmetric broadening of beam in space. In addition, there appears a mixed dispersion-diffraction parameter that depends on the initial pulse duration and on the beam width at the input point and describes the spatial-temporal broadening and asymmetric distortion in time of the pulse. This dispersion-diffraction parameter is an exceptional characteristic of PCs, since in homogeneous materials diffraction broadening and dispersive spreading are decoupled one from another.

## 1.5 Narrow beams interaction in nonlinear photonic crystals

Diffraction of light beams is one of the factors which are limiting the efficiency of many parametric nonlinear processes, among others, the efficiency of second harmonic generation for narrow beams. There are two physical mechanisms that limit the gain. One is the diffractive broadening of the beam which results in a decrease of the top intensity and, eventually, of the efficiency. The other mechanism is the phase mismatch, since all spatial Fourier components of the generated beam can never be simultaneously matched with the components of the pump wave. The latter limitation leads to the filtering of the spatial Fourier components and the narrowing of the beam in the angular space, which eventually results in a diffusive broadening of the generated beam as represented in Fig. 1.23(a) where it is illustrated the situation in homogeneous materials with $\chi^{(2)}$ nonlinearity.

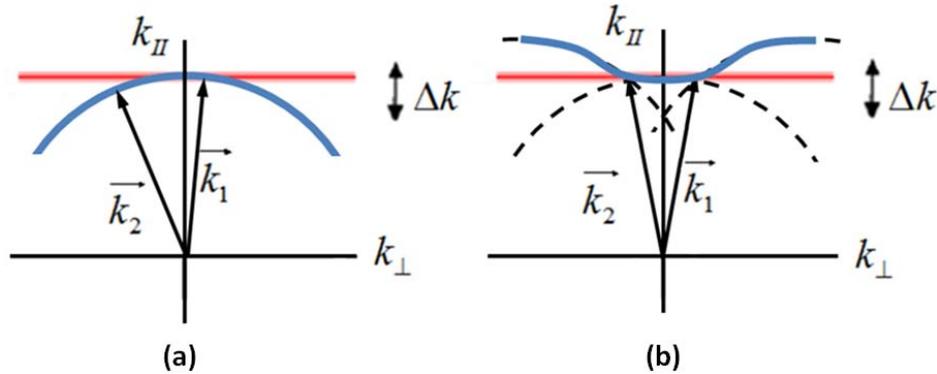

**Fig. 1.23** Spatial dispersion curves in case of normal diffraction (a) and subdiffraction (b); The different wave components with $\vec{k_1}$ and $\vec{k_2}$ dephase in propagation and are never simultaneously phase matched with the pump in case (a) but at self collimation they can be simultaneously phase-matched (b); the horizontal line denotes phase matching condition

In [Sta07b] it was shown that a 2D PC with uniformly distributed quadratic nonlinearity $\chi^{(2)}$ give rise to flat segments in the spatial dispersion curves leading to sub-diffractive propagation (self-collimation) regimes, as illustrated in Fig. 1.23(b). This results in several effects that can improve the performance of the parametric nonlinear process for narrow beams: 1) the beam does not spread due to self-collimation effect because in this case all wave vectors lying on a flat segment of the spatial dispersion curve have equal longitudinal components and thus do not dephase one with respect to another in propagation; 2) the phase matching condition holds simultaneously for all wave vectors belonging to the flat segment, since the phase matching refers to the longitudinal components of the wave vectors of the interacting waves $k_{1,II} + k_{2,II} = k_0$. This means that the diffusive broadening can also be eliminated.

Analytically, the light propagation in 2D PCs, as described in [Loi07c] can be treated by expanding the electromagnetic field into a set of spatial harmonic modes. After the substitution of these components into the Maxwell's equations and elimination of the magnetic components a set of coupled equations is obtained for the complex amplitudes of the harmonics of the electric field. These equations lead (by the diagonalization of the system) to the relationship between the longitudinal $k_{II}$ and transverse $k_\perp$ components of the wave vector for each Bloch mode (transverse dispersion relation). In the limit of small modulation amplitude of the refractive index, the dispersion curves are a set of circles. These circles are mutually shifted by the reciprocal lattice vectors of the PC lattice, as shown in Fig. 1.24. The modulation of the refractive index breaks the degeneracy at the crossing points and creates bandgaps in the spatial wave number domain. For particular values of the modulation amplitude the transverse dispersion curves develop straight segments, indicating the vanishing of diffraction. In Fig. 1.24(a) the circles representing the dispersion of the uncoupled harmonic modes in the first band are shown: the intersection of the circles results in a subdiffractive contour of a square shape. In Fig. 1.24(b) the circles corresponding to the fourth band are represented, and the subdiffractive contour given by their intersection has a triangular shape in this case. The appearance of nondiffractive propagation regimes in different bands is very useful especially in the case of frequency mixing processes (as second harmonic generation, for example), because it permits simultaneously nondiffractive propagation for more of the frequencies involved in the nonlinear process (e.g. for both FW and SH frequencies in the case of second harmonic generation).

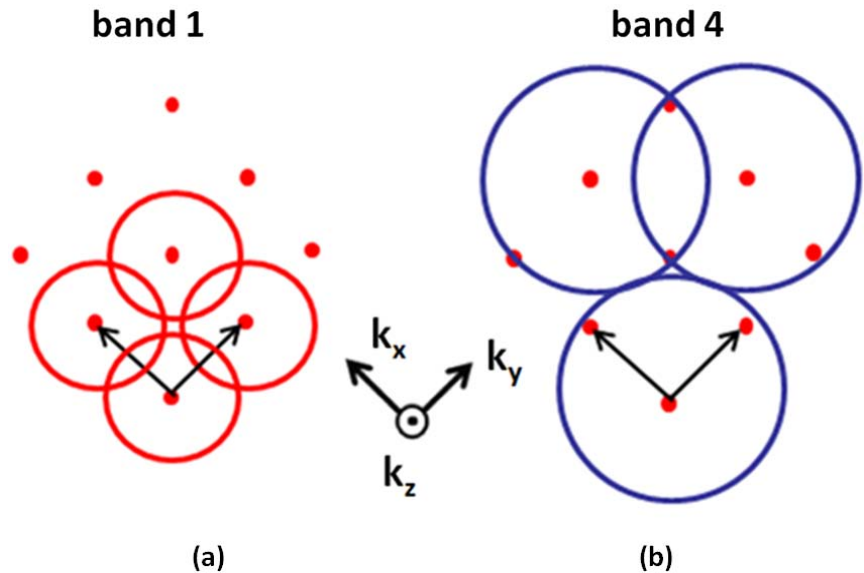

**Fig. 1.24** The set of circles corresponding to the dispersion curves (iso-frequency lines in the $k$-vector space) of the uncoupled harmonic modes in the first band (a) and for the fourth band (b); the subdiffractive area obtained has a square shape in the first band (a) and a triangular shape in the fourth band (b)

Moreover, in 2D PCs designed to ensure simultaneous non-diffractive propagation regimes for both FW and SH wave, a broad angular range phase matching can be obtained. In [Nis08] the case of SH generation in such structure is presented. The idea could be applied also to other parametric processes. This effect is illustrated in Fig. 1.25 showing that whereas only the collinear wave components can be phase matched in the case of normal diffraction structures (as illustrated in Fig. 1.25 (a)), all the components of the flattened diffraction surface can be phase matched at self-collimation (as shown in Fig. 1.25 (b)) and thus increasing the efficiency of the nonlinear process for narrow beams.

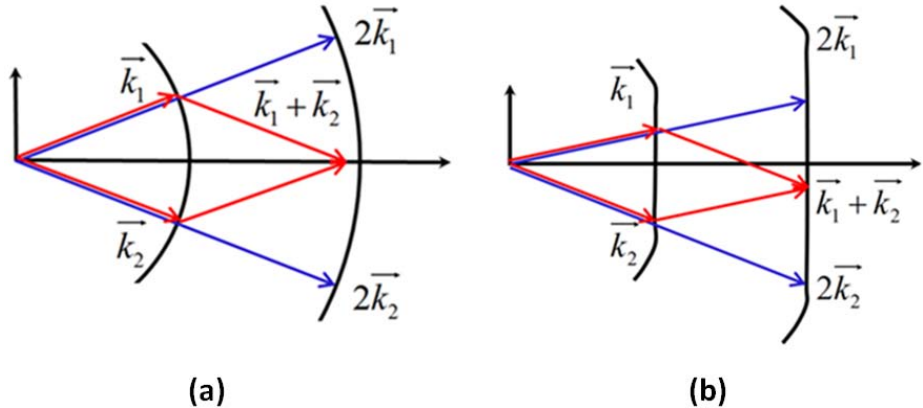

**Fig. 1.25** In the case of normal diffraction (a) only the collinear plane wave components can be phase matched and for flattened diffraction surfaces all components in a range given by the size of the flat segment can be phase matched (b)

The above described ideas on nondiffractive propagation developed in the field of nondiffractive optics were exported to other systems, too. One such system that has many analogies with nonlinear optical systems is the Bose-Einstein condensates, as the widely accepted description of dynamical behavior of Bose-Einstein condensates is the Gross-Pitaevskii equation, which is similar to nonlinear Schrödinger equation in nonlinear optics. Therefore, the "nondiffractive" results from optics can be extended to condensates that remain localized even in the absence of trap [Sta09, Sta08, Sta06b]. Another field of export is acoustics, more precisely the field of sonic crystals, as it was demonstrated theoretically and experimentally [Sol09, Esp07, Per07] that also the acoustics waves can propagate nondiffractivelly in periodic materials.

# 2. Second Harmonic Generation in Subdiffractive Two-Dimensional Photonic Crystals

Dispersive properties of periodic dielectric media result in a complex photonic band structure where the allowed propagation bands show strong anisotropy. The anisotropic propagation of Bloch waves inside PCs leads to a number of interesting phenomena such as super-prism or ultra-refractive phenomena [Kos98, Gra00], self-collimation [Kos99], photon focusing and internal diffraction [Etc96, Chi01].

In this chapter we address the problem of SH generation with narrow beams in 2D PCs tuned to self-collimation regime, made of *ideal dispersionless materials*. The self-collimation regime, as described in the previous chapter, leads to an enhancement of the nonlinear interaction given by the phase matching of the wave vectors lying on the flat segments in the spatial dispersion curves. In this chapter we show that simultaneous self-collimation regimes for both interacting waves lead to a highly degenerated process where all wave vectors lying on the flat segments can be phase matched (not only the collinear wave vectors, as in the case of homogeneous materials), creating conditions for a further increasing of the nonlinear interaction. Another very important factor for the efficiency of the nonlinear interaction is the nonlinear parametric coupling of the two modes. Therefore, we calculate the coupling coefficient and we compare it with the value corresponding to the coupling of plane waves.

Along this chapter we consider the case of the SH generation in a 2D PC made of a square lattice of air holes etched in an ideal dispersionless material. We make a systematical study using the radius of the air holes as parameter and considering different photonic bands for the two frequencies with the purpose to obtain the propagation regimes that we are looking for in order to increase the nonlinear interaction inside the crystal.

## 2.1 Self collimation

Unlike the spatial solitons, where the nonlinearity of homogeneous medium counteracts the natural spreading of the beam due to diffraction, the formation of the self-collimated beams in PCs is a purely linear phenomenon. The natural spreading of the beam is suppressed by the crystal anisotropy, in such a way that all wave vectors building the beam lead to a power flux in the same direction. This can be realized when the wave vectors lie on the flat regions of an isofrequency surface of a PC. Self-collimation occurs on the inflection points of an isofrequency surface, where the curvature of the surface tends to zero [Kos99]. Typically, a flat region spreads over very limited wave vector range centered at the wave vector ending at the inflection point. Therefore, exact self-collimation can occur only for very limited orientations of the beam with respect to the crystal lattice and for limited beam widths.

In our study we consider the SH generation, so radiation with frequencies $\omega$ and $2\omega$, will propagate within the structure. In order to achieve an efficient process, both interacting waves should propagate simultaneously at self collimation regimes. In order to reach this goal we should study first in detail the conditions for self-collimation in 2D PCs.

The general representation of the band diagram along a contour (as shown in Fig. 1.6), useful when one looks for PBGs, is not enough in our case. To find self-collimation regimes inside the crystal we need to calculate the dispersion relation over the whole area of the irreducible Brillouin zone.

For illustration, we consider the case of a 2D PC made by a square lattice of air holes in a dielectric material. We consider a configuration where the radius of the holes is $r = 0.3 \cdot a$ ($a$ is the lattice constant) and the material has the dielectric constant $\varepsilon = 7.1289$. The band structure for this PC is calculated using the PWE method (described in the previous chapter). The calculations were made on an area covering the first Brillouin zone and in Fig. 2.1(a) the first photonic band for TE polarization is shown. In Fig. 2.1(b) the isofrequency contours corresponding to this band are plotted and it can be observed that for particular frequencies, around the one represented with red line, flat segments appear in the isofrequency contours, meaning that for wave vectors lying on these flat segments (represented with blue arrows in Fig. 2.1(b)) the corresponding beam will propagate without diffraction.

From Fig. 2.1 one can see that it is difficult to obtain nondiffractive propagation for both FW and SH wave in the same band, since the self collimation appears usually close to the band edge. Therefore, in order to obtain simultaneously self-collimated propagation for both waves, it is needed to look for self-collimation regimes in different photonic bands.

With this purpose we plot in Fig. 2.2 the band diagram of the structure calculated for the points along the **Γ-X-M-Γ** contour which is limiting the irreducible Brillouin zone(as the maximum and minimum frequencies for each band appear along this path due to symmetry considerations). The red dashed line represents the frequency for which non-diffractive propagation regime appears in the first band (as illustrated in Fig. 2.1) and this frequency would represent the fundamental one. Blue dashed line represents the double of this frequency, which would correspond to the SH, and it can be seen that the SH frequency appears in the fourth band for the direction in which we obtained self-collimation, given by wave vectors from the interval **M-Γ**.

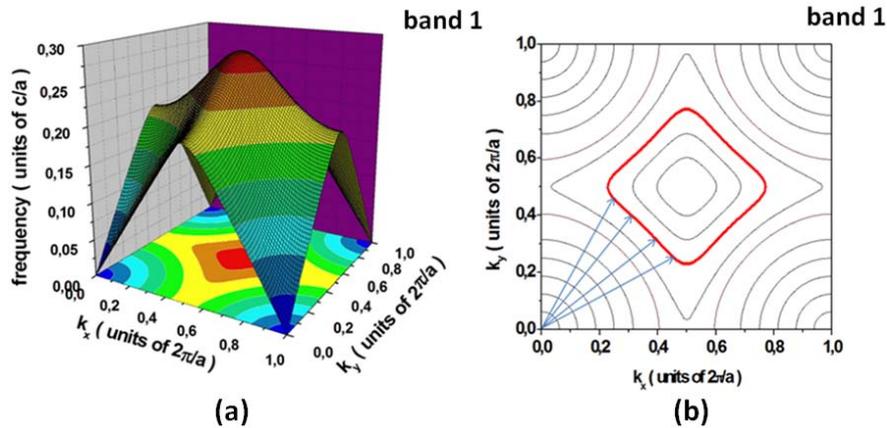

**Fig. 2.1** a) Representation of the first photonic band for TE polarization, for a 2D PC made by a square lattice of air holes with radius $r = 0.3 \cdot a$ (where $a$ is the lattice constant) in a material with the dielectric constant $\varepsilon = 7.1289$; b) the isofrequency contours where it can be observed the self-collimation regime for the frequencies around the one represented with red line.

In order to look if simultaneous self-collimation regimes for both FW (in the first band) and SH wave (in the fourth band) can be obtained along the same propagation direction (since we are interested in this thesis in the case of collinear SH generation) we plot in Fig. 2.3 the isofrequency contours for bands 2-10 with the idea to look if it is possible to combine frequencies from other bands, too.

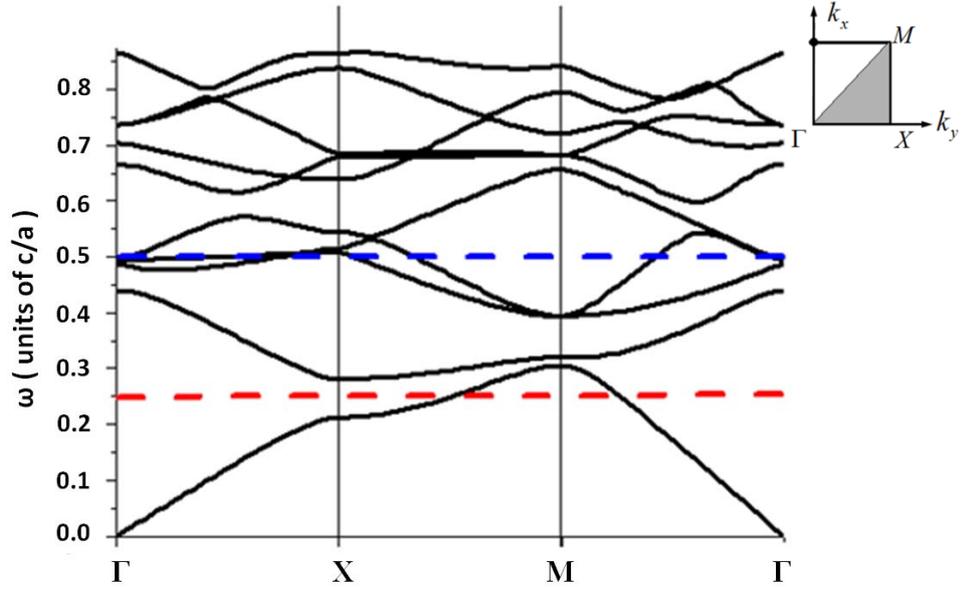

**Fig. 2.2** Band diagram for TE polarization for a PC made by a square lattice of air holes with radius $r = 0.3 \cdot a$ (where $a$ is the lattice constant) in a material with the dielectric constant $\varepsilon = 7.1289$

Whereas there are bands where it is difficult to obtain nondiffractive propagation, as in band 3 (Fig. 2.3(b)), there are also bands where self-collimation regimes can be obtained in the same propagation direction as in band 1. This is the case of band 4 (Fig. 2.3(c)) or band 6 (Fig. 2.3(e)). In band 2 (Fig. 2.3(a)), band 5 (Fig. 2.3(d)) or band 7 (Fig. 2.3(f)) the self collimation is possible, but in a different propagation direction. In band 8 (Fig. 2.3(g)), band 9 (Fig. 2.3(h)) or band 10 (Fig. 2.3(i)) it could be also obtained nondiffractive propagation regimes. However, the stronger degeneracy and folding of higher bands (bands 8-10) make that the same frequency appears in different bands and, moreover, inside the same band it appears with different wavevectors, not only with the one that is eventually phase matched, reducing in this way the coupling of light into the phase matched (and eventually self-collimated) modes. Anyway, using frequencies from higher bands also could bring an advantage, since these frequencies are larger with respect to those from the lower bands and therefore for using the same wavelength, the corresponding lattice period will be larger, which makes the structure easier to fabricate.

As illustrated in Fig. 2.3, for this configuration the self-collimation along the same direction for frequencies appearing in band 2, band 5 and band 7 can be obtained. Moreover, looking at the values of these frequencies it can be seen that the frequencies from band 7, where we obtain self-collimation regimes, are close to the double of those from band 2 for which light can propagate without diffraction. This means that, after a further tuning of the structure, one could try to obtain simultaneously self-collimation for both FW, in band 2, and SH wave, in band 7. However, in these bands the width of the flat segment indicating the vanishing of diffraction is rather small, especially in band 7, compared to the flat segment that could be obtained in band 4, for example. Moreover, in band 4 there is a wider range of frequencies for which the diffraction is strongly reduced and, as any beam has a spectral width (it doesn`t contain only one frequency, the self-collimated one), it is important that diffraction become strongly reduced for more spectral components of the beam.

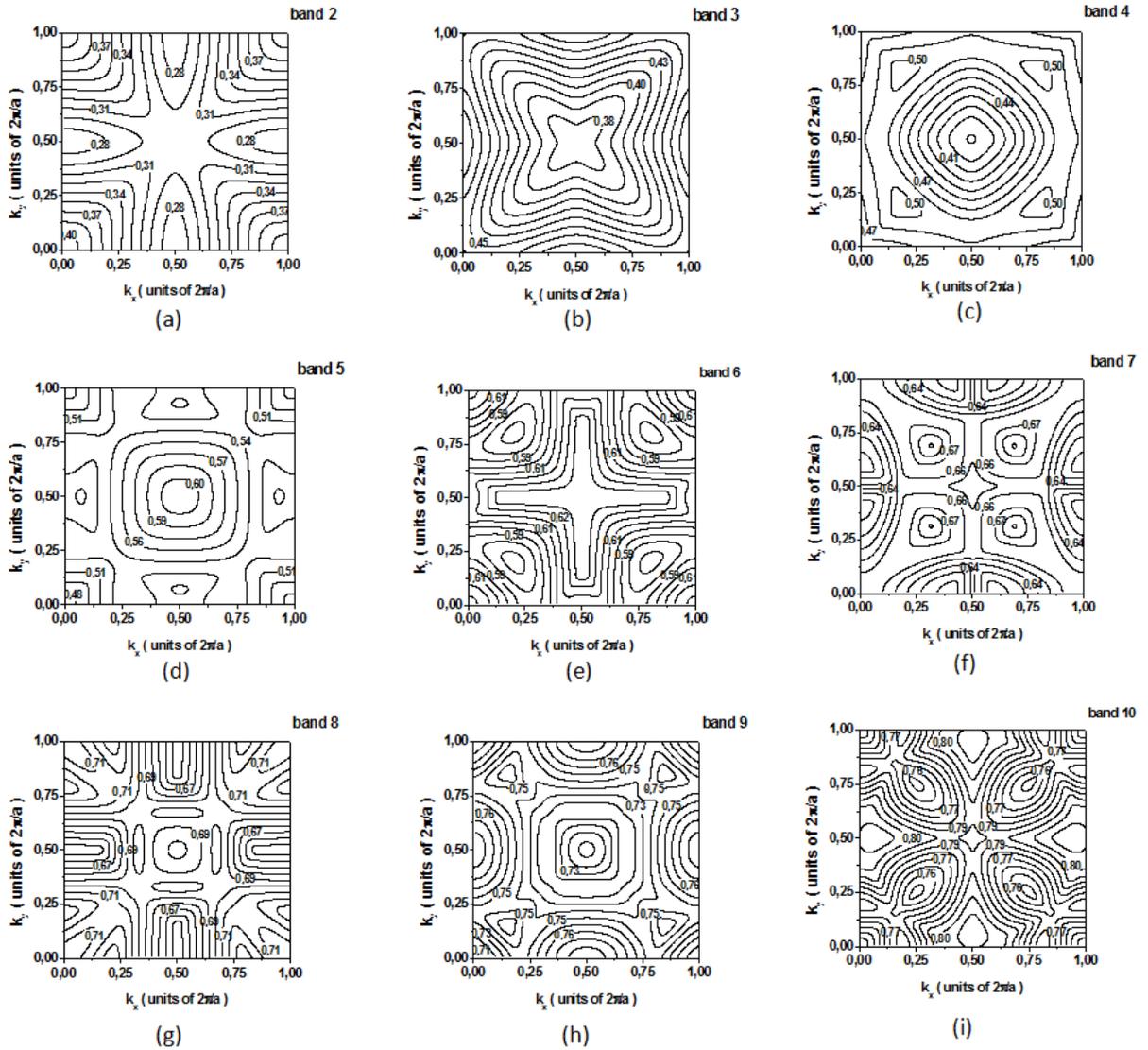

**Fig. 2.3** The isofrequency contours for bands 2-10 for TE polarization

As the folding and degeneracy of bands is weaker for lower bands, we are looking for configurations where the FW frequency appears in the first band, situation that should ensure a better coupling of light at a particular frequency and wave vector. In the band 1 the nondiffractive propagation is obtained along the direction of $k_x = k_y$. Self-collimation along the same direction is obtained also in band 4 and band 6. The case when the FW frequency is in band 1 and the SH one is in band 6 will be treated with more details in the Chapter 4 of the thesis. In this chapter we will consider the case when the FW frequency is in band 1 and the SH in band 4.

Regarding the frequencies for which the self-collimation regime appears in the fourth band, represented in Fig. 2.4(a), it can be noticed that these frequencies are close to the double of those for which we obtain self-collimation in the first band. Therefore, one could simultaneously obtain nondiffractive propagation conditions for both FW and SH frequencies, as confirmed by Fig. 2.4(b) which represents the isofrequency contours for the FW frequency $\omega_{FW} = 0.25$ in the first band for TE polarization (with red line) and for the SH frequency $\omega_{SH} = 0.5$ in the fourth band for TE polarization.

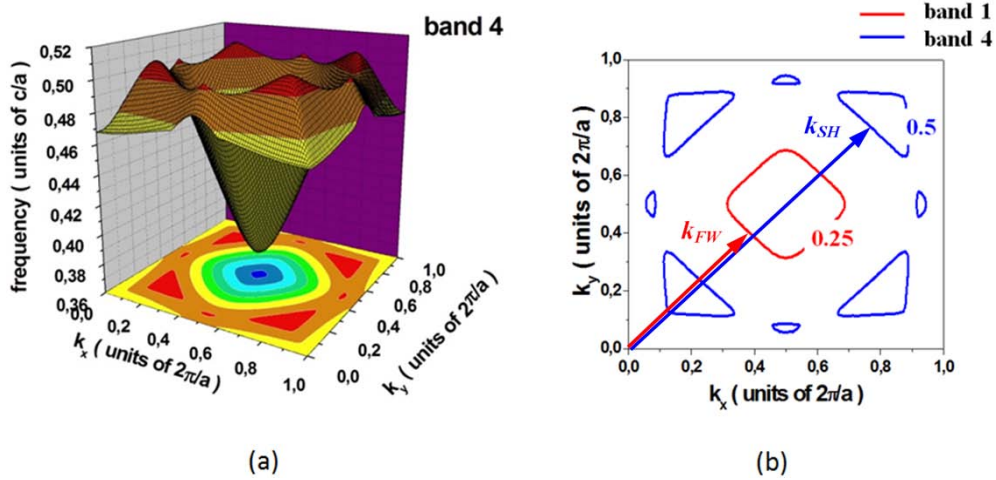

**Fig. 2.4** (a) Representation of the fourth band for TE polarization; (b) the isofrequency contours for the FW frequency (with red line) in the first band and for the SH frequency (with blue line) in the fourth band show simultaneously self-collimation regimes for both FW and SH frequencies

In order to confirm the simultaneous non-diffractive propagation at the FW and SH frequencies, we made 2D FDTD simulations for the propagation of light in the PC structure presented above using a commercial program, Crystal Wave. The propagation of radiation at the two frequencies was investigated separately in the linear case, using sources for both FW and SH waves. In Fig. 2.5 the nondiffractive linear propagation of narrow beams with a diameter of 1.5 μm (corresponding to Rayleigh distances of $z_0^{FW} = 2.66$ μm for FW and $z_0^{SH} = 5.32$ μm for the SH wave) at the FW (Fig. 2.5(a)) and SH frequencies (Fig. 2.5(b)) is illustrated, showing that in both cases the calculated frequencies for the flattening of the isofrequency contours from the harmonic expansion are compatible with self-collimation regimes. The diffractive broadening of the beams in homogeneous material is represented with the dashed lines and it indicates a strong decrease of diffraction for both waves. The FW source was placed outside the structure, while the SH source was placed inside the crystal in conformity with the real situation when the SH wave is generated inside the PC.

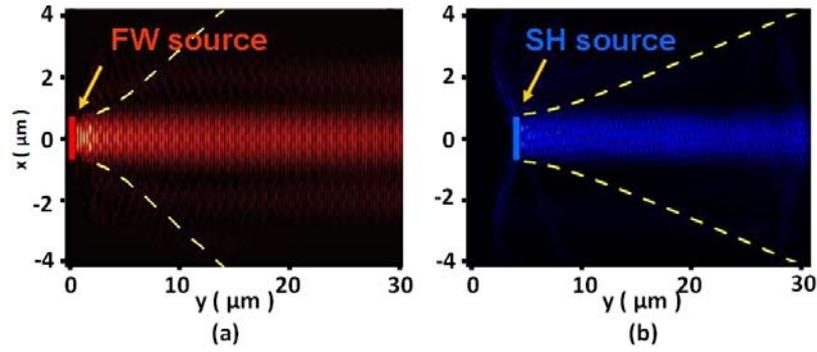

**Fig. 2.5** Linear propagation of FW (a) and SH wave (b) at zero diffraction points calculated by FDTD technique

## 2.2 Phase matching condition

Apart from the simultaneous self-collimation of both FW and SH wave, the phase matching of the two waves is another important factor for the efficiency of SH generation, as explained in the previous chapter. In order to evaluate the phase mismatch induced in the structure, we calculate the dispersion curves for the FW and SH wave. The two dispersion curves are calculated along the direction for which the non-diffractive propagation regime is obtained, $k_x = k_y$, shown by dashed black lines in Fig. 2.6(a) and Fig. 2.6(b). When we represent on the same plot the dispersion curve for the SH and the "double" of the dispersion curve for the FW in Fig. 2.6(c), it can be seen that the two dispersion intersect at a given point corresponding to the phase matching condition, since at this point $\omega_{SH} = 2\omega_{FW}$ and $k_{SH} = 2k_{FW}$. The isofrequency contour for the FW frequency (represented by red line in Fig. 2.6(d)) shows that at the FW phase-matched frequency the diffraction is strongly reduced. For this particular configuration the crossing point of the two curves appears in a region corresponding to the top of the propagation band for the SH wave. The isofrequency contour for the SH wave is represented by a small circle, as illustrated with blue line in Fig. 2.6(d). The fact that the isofrequency contour is represented by a small circle, apart from a strong diffraction, means also that there is a very small range of wave-vectors for which light can propagate at this frequency in this particular band.

In order to obtain simultaneous phase matching and non-diffractive propagation for both waves, it is needed to shift the crossing point of the two dispersion curves to an area characterized by flat segments for both FW and SH wave.

One way to shift the crossing point at different frequencies is to change the radius of the air holes. For example, in Fig. 2.7(a) and Fig. 2.7(c) the dispersion curves for structures with the radius $r = 0.1a$ and $r = 0.2a$, respectively, are represented. The position of the phase match point (the phase matched frequencies and the wave vectors) changes when the radius of the air holes is changed. From the isofrequency contours for the phase matched frequencies represented in Fig. 2.7(b) and Fig. 2.7(d) the diffraction for both FW and SH wave can be estimated. It can be noticed that for the both waves the diffraction is reduced compared to that obtained in homogeneous material.

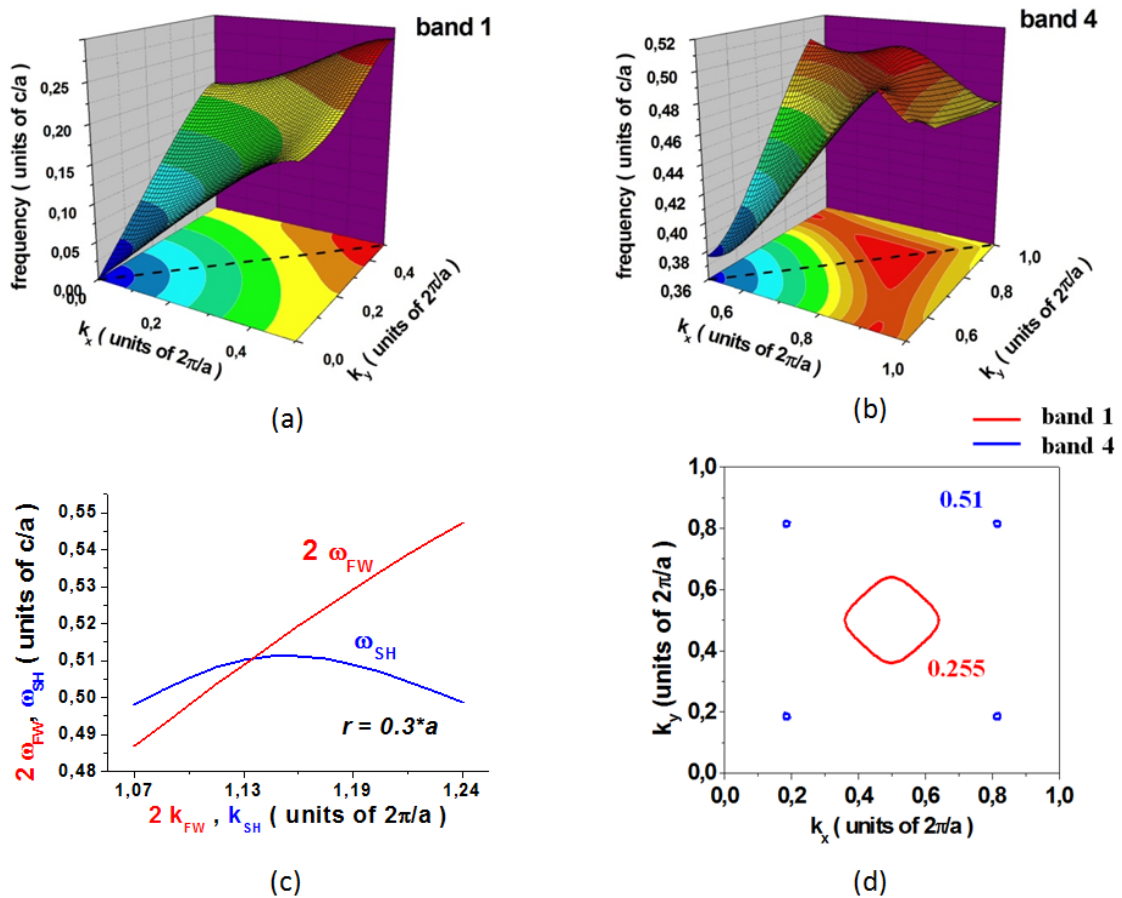

**Fig. 2.6** The dispersion curves are calculated along the direction of $k_x = k_y$ as illustrated in (a) for the FW and in (b) for the SH wave; the crossing point of the two dispersion curves (c) corresponds to the phase-matching condition; the isofrequency contours for the phase-matched frequencies plotted in (d) show that for the FW frequency (with red line) there is a subdiffractive propagation regime but for the SH wave the isofrequency contours are represented by a small circle, as the crossing point of the two curves is very close to the top of the propagation band for the SH wave

A more systematic way to solve this problem is to calculate by linear mode expansion the self collimation areas for both waves and the phase-matching curve for different values of the radius, as represented in Fig. 2.8. The exact zero diffraction point has a sense only for infinite propagation length. For a limited propagation length the reduction of diffraction below some limits is relevant. Therefore, we have calculated not an exact point in parametric space corresponding to zero diffraction, but rather regions corresponding to sufficiently small diffraction, in particular less than 10% with respect to the diffraction in homogeneous material. This is equivalent to the increase of the Rayleigh length of the corresponding beams by 10 times.

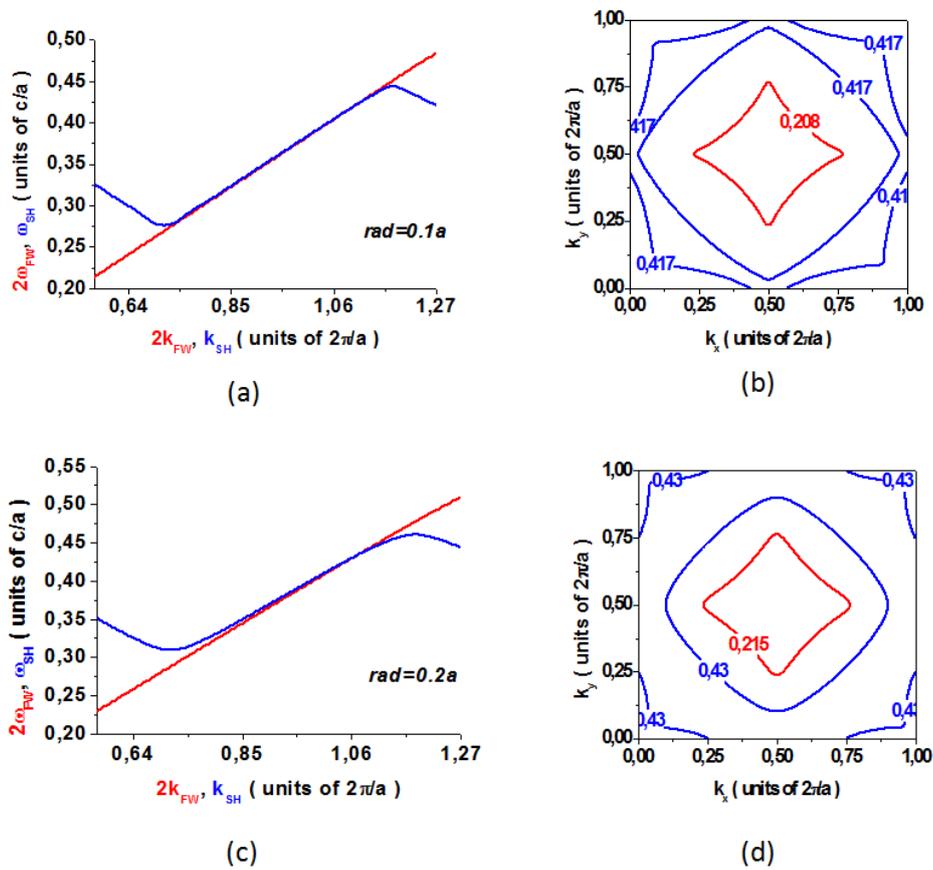

**Fig. 2.7** Dispersion curve for FW and the "double" of dispersion curve for SH wave for the structure when the radius has the value $r = 0.1a$ (a) and $r = 0.2a$ (b); the isofrequency contours for the frequencies corresponding to phase matching (c) and (d), respectively

The curves represented in Fig. 2.8 show that the phase matching condition (represented by the black line) is satisfied in the entire range of the radii considered, but for different values of the FW frequency. As already noticed from the previous calculations, for a radius around the value $r = 0.3 \cdot a$ the nondiffractive propagation conditions for both FW and SH waves are fulfilled simultaneously, but the phase matching is achieved for a different frequency. A very interesting region appears for a value of the radius around $r = 0.25 \cdot a$, because the small diffraction areas for the two waves are close one another and strong reduction of diffraction for both waves and simultaneous phase matching can be expected.

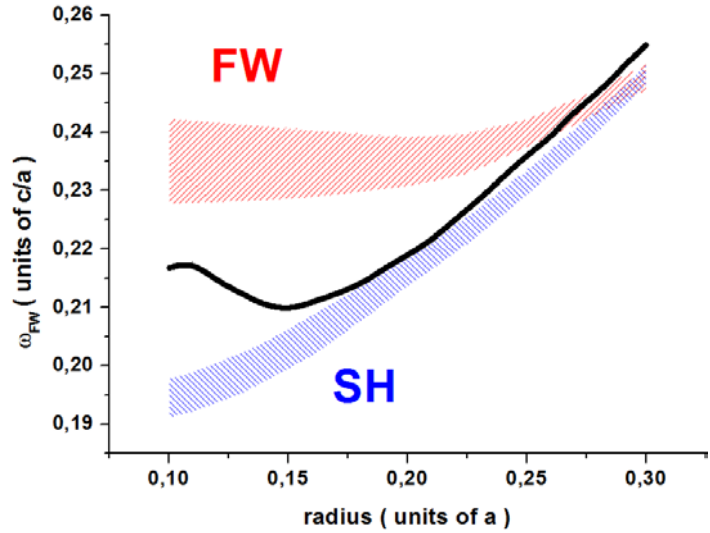

**Fig. 2.8** Representation of the surfaces of small diffraction (10% of transverse beam spreading in homogeneous media) in a plane (radius, frequency) for FW (red) and SH (blue) and the phase-matching curve (black)

The dispersion curves calculated for the structure with radius $r = 0.25 \cdot a$, plotted in Fig. 2.9(a) permit to obtain the phase matching. The isofrequency contours for the phase-matched frequencies represented in Fig. 2.9(b) illustrate that indeed, for this configuration the non-diffractive propagation for both FW and SH wave is obtained simultaneously with the phase matching condition. The crossing point of the dispersion curves (corresponding to the phase matching condition) for this configuration are at $\omega_{FW} = 0.235$ ($\omega_{FW} = 0.47$) and $k_{FW} = 0.545$ ($k_{SH} = 1.09$).

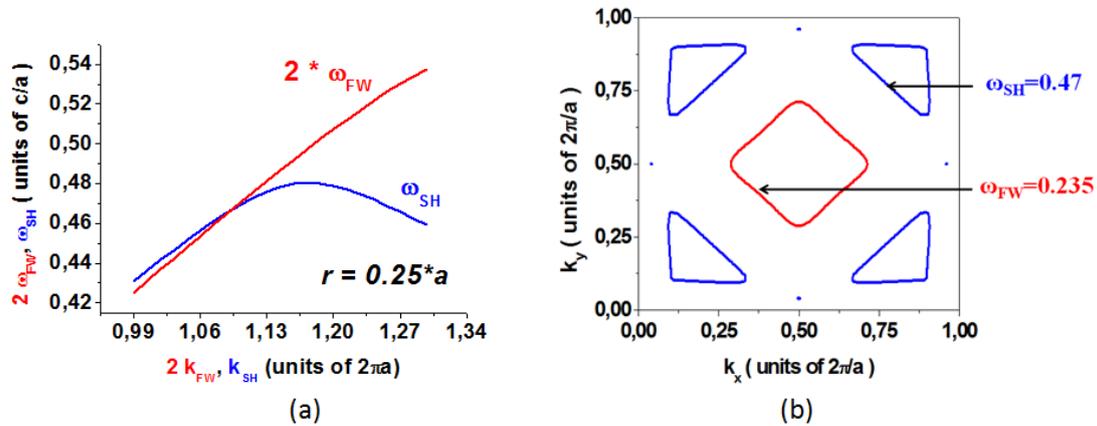

**Fig. 2.9** (a) Dispersion curves for the FW and SH wave; (b) the isofrequency contours for the phase-matched frequencies.

In this section we show that by properly choosing the parameters of a 2D PC (especially the air holes radius, which is an important parameter for the search in these conditions), one can obtain the situation in which the periodicity of the structure leads to a distortion of the dispersion surfaces in such way that it allows simultaneous self-collimation regimes for both FW and SH wave together with phase matching.

## 2.3 Nonlinear parametric coupling

The third condition that had to be fulfilled for an efficient harmonic generation is a sufficiently strong overlap between the two interacting waves. This condition is trivial in homogeneous case since the plane waves overlap by 100%. In case of PCs the propagation eigenmodes are Bloch waves. The transverse intensity distribution in different Bloch modes is different and the waves do not necessarily overlap in space. Without a spatial overlap of the two modes it is not possible to transfer energy between the different Bloch modes even when the phase matching is achieved. Therefore, the spatial overlap between the envelopes of the corresponding Bloch modes should be calculated. The strength of the nonlinear interaction is characterized by the Hamiltonian of interaction which, for parametric processes can be calculated from the relation:

$$H_{\text{int}} = \int d^3 r P_\omega^{(-)} E_\omega^{(+)} + H.c. \qquad (2.1)$$

where $P_\omega^{(-)}$ is the polarization and $E_\omega^{(\pm)}$ is the scalar part of the electric field operator oscillating as $e^{\mp i\omega t}$. For materials that have only one non-vanishing term in the nonlinear susceptibility tensor, the polarization can be written as:

$$P_\omega^{(-)} = \varepsilon_0 \chi E_{2\omega}^{(-)} E_\omega^{(+)} = \varepsilon_0 \chi \varepsilon_{2\omega} \varepsilon_\omega E_{2\omega} E_\omega^* e^{i(k_{2\omega}-k_\omega)} e^{-i\omega t} \qquad (2.2)$$

with $\chi$ the relevant nonlinear susceptibility tensor element. After the integration, at phase matching ($k_{2\omega} = 2k_\omega$) the expression for the Hamiltonian of interaction is obtained:

$$H_{int} \propto i\left(E_1^2 E_2^* + E_1^{*2} E_2\right) \qquad (2.3)$$

Here, $E_1$ and $E_2$ are the spatial envelopes of the Bloch modes of the FW and SH wave, respectively. Therefore, in order to evaluate the efficiency of the nonlinear coupling the cross-correlation between the functions $E_1^2$ and $E_2$ are calculated and normalized in such way that unity would correspond to the perfect matching of the modes (the interaction of plane waves gives unity under this normalization):

$$K = \frac{\left|\int_M \left(E_1^2 E_2^*\right) d\vec{r}\right|}{\left(\int_C \left|E_1^4\right| d\vec{r} \int_C \left|E_2\right|^2 d\vec{r}\right)^{1/2}} \qquad (2.4)$$

where the upper integral is calculated in the nonlinear material from one unit cell, while the lower integrals are taken over the entire unit cell. Therefore, for estimating the coupling of the modes that satisfy the phase matching condition, we calculate the complex vector field $\vec{E}$ which is physically of the form: $\exp(ikr)$ times a periodic function. We used only these periodic functions, which would correspond to pure Bloch modes, as illustrated in Fig. 2.10, where the real and imaginary part of Bloch modes corresponding to the FW (Fig. 2.10(a) and Fig. 2.10(b)) and SH wave (Fig. 2.10(c) and Fig. 2.10(d)) are represented.

When we calculated the upper integral from the expression of the coupling coefficient from eq. (2.4), over the nonlinear material from one unit cell, the integration did not included the region corresponding to the air hole as illustrated in Fig. 2.11, where the multiplication of the two modes over the entire unit cell (in Fig. 2.11(a)) and the multiplication of the modes only in the nonlinear material (in Fig. 2.11(b)) are represented. This value was obtained multiplying with a function that has the value zero inside the central region corresponding to the air hole, and the value equal with unity in the rest of the unit cell.

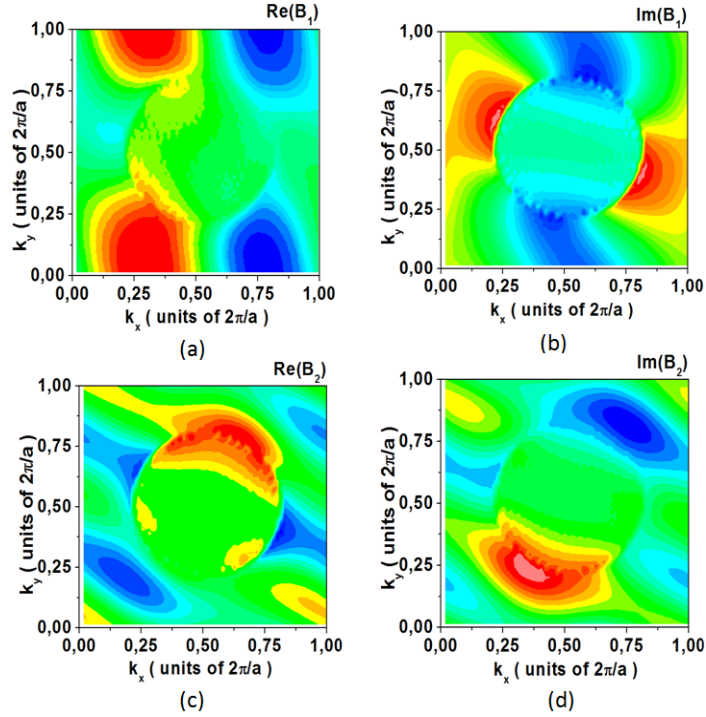

**Fig. 2.10** Real and imaginary part of the Bloch modes for the electric field for the wave vectors corresponding to the phase matching for FW (a) and (b), and for the SH wave (c) and (d), respectively

The coupling coefficient calculated for the wave vector $k_{FW} = 0.387$ ($k_{SH} = 0.774$) corresponding to phase matching for the configuration when the radius $0.25a$ (as illustrated in Fig. 2.9), has the value $K = 0.794$. In the range of hole radius considered from $0.2a$ to $0.3a$ the coupling coefficient varies from $K = 0.88$ to $K = 0.67$. The overlap therefore is sufficiently large to expect an efficient SH generation, i.e. the parametric coupling can be of the same order of magnitude as for the plane waves in homogeneous materials.

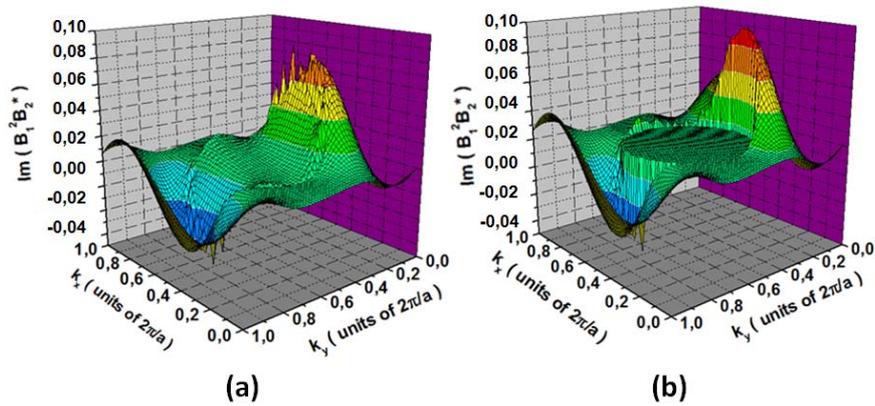

**Fig. 2.11** Illustration of the imaginary part of the product $B_1^2 B_2^*$ over the entire unit cell (a) and only over nonlinear material (b), after multiplying with a function that has the value zero inside the central region corresponding to the air hole, and the value 1 over the rest of the unit cell.

## 2.4 Second harmonic generation

Once the fulfillment of the three necessary conditions has been achieved (non-diffractive propagation for both FW and SH wave, phase matching, and sufficiently large overlap of the two modes), it remains to prove the SH generation directly by the numerical integration of the corresponding equations.

We consider a 2D PC consisting of a periodic lattice of air holes etched in a high refractive index nonlinear material with quadratic nonlinear coefficient. The parameters of the square lattice of air cylinders are: lattice constant $a = 362$ nm and radius of air holes $r = 0.25a$. The FW wavelength is around $\lambda_{FW} = 1.55$ μm, the refractive index of the material is $n = 2.67$ at both FW and SH wavelengths and the effective nonlinear coefficient is $d_{eff} = 97$ pm/V, that corresponds to GaAs in [111] direction of propagation, according to [Sho97].

For the calculation of the SH generation process in the designed structure a nonlinear FDTD method is used. Fig. 2.12 summarizes the results of the simulations for a PC with a length of 30 μm. We consider a narrow Gaussian FW source with a width $\Delta x = 1.5$ μm (corresponding Rayleigh distances are $z_0^{FW} = 2.66$ μm for FW and $z_0^{SH} = 5.32$ μm for the SH wave, much shorter than the crystal length), when the beam is covering the flat segment from the dispersion curve and, for comparison, also the situation when the FW source is wider, with a width $\Delta x = 3$ μm. For the narrow beams, when the phase matching condition is fulfilled (at $\omega_{FW} = 0.235$) the intensity of SH grows along the PC structure with a $z$ dependence close to the $z^2$ law (black star dots line in Fig. 2.12). The fact that the growth law obtained numerically for the narrow beams differ slightly from the $z^2$ dependence (it is $z^\alpha$ with $\alpha \approx 1.8$) indicates that not all interacting components are in phase matching, but some part is phase matched. The growth of the generated SH from a beam of double width (represented with red squares line in Fig. 2.12) shows nearly the $z^2$ law, indicating the precise phase matching. The SH intensity growth of the narrow beam in non-phasematched case is close to $z^1$ law (with cyan rhombi line in Fig. 2.12) as expected for the incoherent SH processes.

It can be noticed that at phase matching, in both cases (for narrow and broader beams), the conversion efficiency is higher than that of narrow beams in homogeneous material (green triangle line in Fig. 2.12). The narrow beams of $\Delta x = 1.5\,\mu m$ spread very fast in propagation and the SH generation process is very weak. We also calculate the SH generated from plane waves in homogeneous material in phase matched conditions, represented with blue circles line in Fig. 2.12. The comparison with the plane wave case shows that the SH generated in PC of smaller beam width (with $\Delta x = 1.5\,\mu m$) was approximately equal to that of the plane wave and the SH efficiency of the broader beam was even larger than that of the plane wave. Due to the calculated overlap integrals of the corresponding Bloch modes in the PC case (around 70%) and also due to the limited width of the plateau for the SH wave, the efficiency in the PC case should be less than that of homogeneous waves. However, due to the decrease of the group velocities in PC (by a factor of approximately 5) given by the decrease of the slope of the dispersion curve, the efficiency respectively increases.

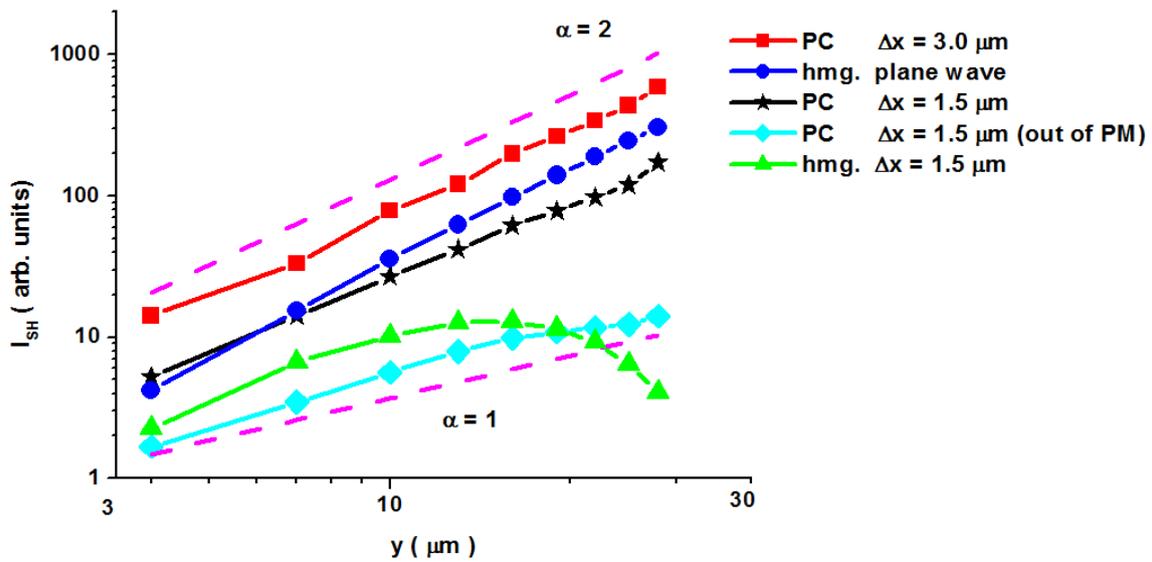

**Fig. 2.12** Intensity of the generated SH along the PC in log-log scale for the input FW beam has a width of $\Delta x = 1.5$ (black line with stars) and $\Delta x = 3$ (red line with squares) under phase matching conditions; the input FW beam is of width $\Delta x = 1.5$ and it is slightly out of phase matching (cyan line with rhombi); the input beam has a width of $\Delta x = 1.5$ in an equivalent homogeneous medium in phase matching (green line with triangles); the input FW is a plane wave at phase matching (blue line with circles)

The transverse distribution of the radiation at the rear face of the PC structure is shown in Fig. 2.13(a) for the case when $\Delta x = 1.5\,\mu m$ and in Fig. 2.13(b) for the case when $\Delta x = 3\,\mu m$. The plots show that the width of SH beam is determined by two factors: the width of the plateau of the corresponding dispersion curve and the width of the FW beam. For a smaller width of FW beam ($\Delta x = 1.5\,\mu m$) the first limiting factor comes into play since a very narrow beam has a broad angular spectrum, as expressed also by eq. (1.15). The effective part of the beam contributing to the SH generation is that which fits within the plateau width. Because the widths of the plateaus for both frequencies were approximately equal (see Fig. 2.9(b)), the FW and SH beams are of nearly equal width, too. In the case of a broader FW beam the plateau of the SH dispersion curve was sufficiently broad to include the whole angular spectrum of the beam and the SH beam width was less than that of the FW beam (roughly with a factor of $\sqrt{2}$), since the SH is proportional to the square of the FW field. In this case the whole FW beam contributes to the SH generation process.

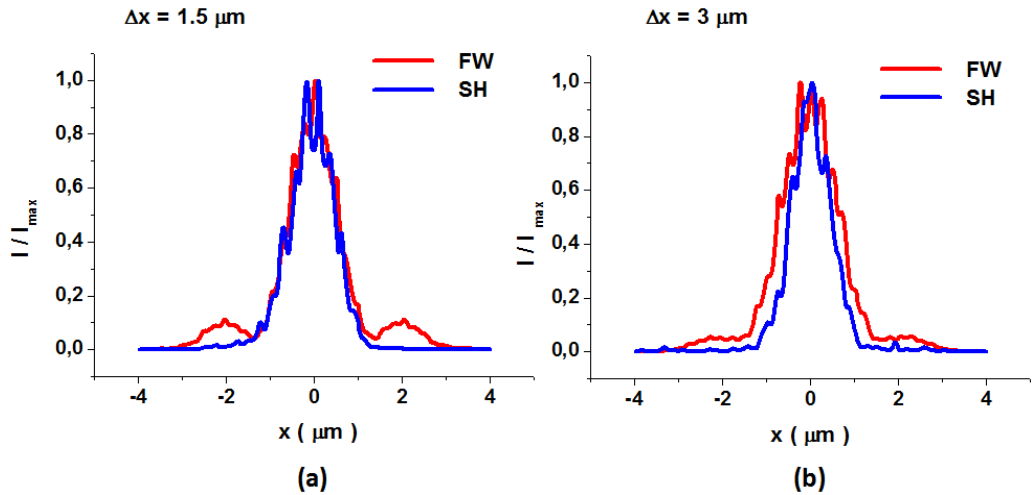

**Fig. 2.13** Transverse distribution of the intensity of FW and SH beams at the exit of the 2D PC for $\Delta x = 1.5\,\mu m$ (a) and $\Delta x = 3\,\mu m$ (b)

We proposed and demonstrated the idea that narrow beams can be used for SH generation in quadratic nonlinear PCs tuned to non-diffractive regimes for both interacting waves (fundamental and second harmonic). We have shown a possibility of such SH process in an idealized case of a 2D PC made of dispersionless material. Moreover, these results are extendable to PCs of different geometries, e.g. of rhombic symmetry.

## 2.5 Conclusions

The main results presented in this chapter can be resumed as follows:
- The possibility to generate SH signal with narrow beams (with width of the order of the wavelength) in 2D PCs made of quadratic nonlinear materials.

- In order to obtain an efficient SH generation process, three main conditions must be met: simultaneous self-collimation for both FW and SH, phase matching of the two waves and a sufficiently strong overlap of the two modes.

- After a systematical study, using dispersion surface calculations and linear FDTD simulations for the propagation of the two waves as main instruments, a configuration that satisfies the conditions mentioned above was found.

- The width of the plateaus from the isofrequency contours limits the beam width that can be effectively used for the SH generation.

- The SH generation process in idealized structures (in particular we neglected the dispersion of material and the finite size of the crystal in the vertical direction) was considered. The application of this idea in real materials (including the dispersion of the material) and in realistic configurations (planar configuration) are presented in Chapter 3 and Chapter 4.

- The basic idea of this study can be also applied to parametric generation processes pumped by extremely narrow beams, as well as other parametric frequency conversion schemes.

Chapter 3

# 3. Vertically Confined Second Harmonic Generation in Subdiffractive Two-Dimensional Photonic Crystals

In the previous chapter it was proved that narrow beams (with widths of few wavelengths) can be used for second harmonic generation in two-dimensional photonic crystals made from quadratic nonlinear material. However, the material considered was an ideal dispersionless material. In this chapter and the next one the case of 2D PCs made in a real dispersive material, AlGaAs, and in realistic configuration, planar waveguide, will be considered.

Material dispersion results in a change of the dispersive properties of the PC shifting the phase matching and self-collimation regimes at different frequencies and different wave vectors (as described with more details in Section 3.1 of this chapter). The planar configuration, used to control the losses in the vertical direction, also brings in an additional dispersion for the effective refractive indices of the guided modes, as it is shown in Appendix B.

As shown in the previous chapter (in Section 2.1) the variation of the radius of air holes leads to a shift of the frequencies and wave vectors corresponding to phase matching and non-diffractive propagation. In this chapter a study is reported on how the variation of air holes radius can compensate the changes introduced by the dispersion and which other parameters of the structure (the geometry of the lattice, the thickness of the planar structure and also combining different guided modes for the two waves) can be varied with the same purpose.

## 3.1 Second harmonic generation in bulk two-dimensional photonic crystals made in AlGaAs

The use of the idea presented in the previous chapter in experiments and applications requires realistic materials. Most real materials show normal dispersion, which means that the refractive index of the material increases for larger frequencies.

For illustration, we follow the changes induced in the dispersion curves calculated for a bulk 2D PC (considered infinite in the direction of holes) when we introduce the dispersion of the material, as represented in Fig. 3.1. Fig. 3.1(a) shows the dispersion curves calculated in a 2D PC made by a square lattice of air holes with radius $r = 0.3a$ (where $a$ is the lattice constant) in an ideal dispersionless material with $n_{FW} = n_{SH} = 2.67$ (where $n_{FW}$ is the refractive index at the FW frequency and $n_{SH}$ is the refractive index at the SH frequency). In Fig. 3.1(b) there are represented the dispersion curves for a 2D PC with the same geometry made in a material that has $n_{FW} = 2.67$ and $n_{SH} = 2.857$ showing a dispersion of 7%, a common value for many dispersive materials, as AlGaAs. In Fig. 3.1(b) it can be noted that the dispersion curve for the SH is shifted down (due to the higher refractive index) with respect to its position from Fig. 3.1(a) and, in this way, the crossing point of the two curves, corresponding to the phase matching condition, is also shifted to different frequencies and different wave vectors.

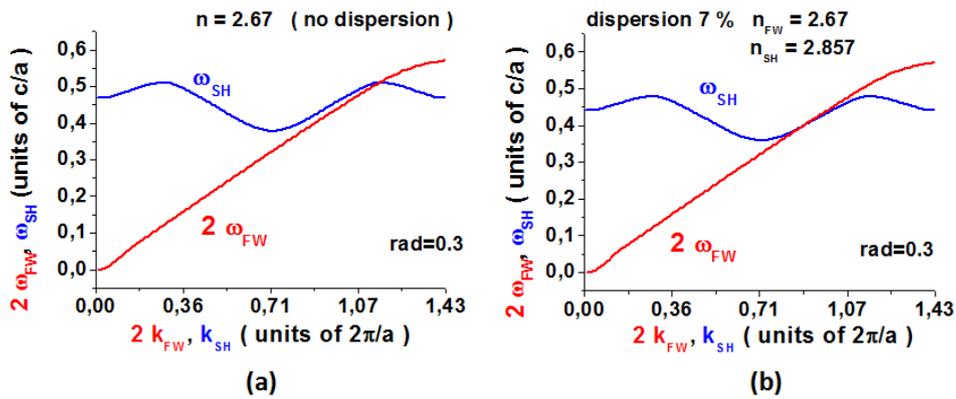

**Fig. 3.1** Dispersion curves calculated in a 2D PC made in dispersionless material (a) and in a 2D PC made in a material showing a dispersion of 7% (b)

When a concrete material for designing a realistic structure is selected, one important factor is the refractive index of the material. A higher refractive index provides a higher index contrast, allowing in this way to emphasize the effects coming from the modulation of the refractive index. The material that considered in our calculations is $Al_{0.3}Ga_{0.7}As$ which, apart from a relative high refractive index, also shows a high quadratic nonlinear coefficient in particular directions of propagation. Another important reason to choose this material is represented by the mature state of the art of the available technologies for sample fabrication from III-V semiconductors such as AlGaAs.

Another important factor for the efficiency of the nonlinear interaction is the polarization of the interacting waves. The first step is to write the polarizability tensor for this material, which has the following expression:

$$\begin{pmatrix} P_x^{(2)} \\ P_y^{(2)} \\ P_z^{(2)} \end{pmatrix} = 2\varepsilon_0 \begin{bmatrix} 0 & 0 & 0 & 0 & d_{14} & 0 \\ 0 & 0 & 0 & -d_{14} & 0 & 0 \\ d_{14} & -d_{14} & 0 & 0 & 0 & 0 \end{bmatrix} \begin{pmatrix} E_x^2 \\ E_y^2 \\ E_z^2 \\ 2E_y E_z \\ 2E_x E_z \\ 2E_x E_y \end{pmatrix} \quad (3.1)$$

where $x$ is the [110] crystallographic direction and $z$ is the [001] direction. For $Al_{0.3}Ga_{0.7}As$ the nonlinear coefficient has the value $d_{14} = 119$ pm/V.

For a TE polarized FW ($E_x$ and $E_y$ components are non-zero and $E_z$ component is zero), the nonlinear polarization becomes:

$$P_z^{(2)} = 2\varepsilon_0 d_{14}(E_x^2 - E_y^2) \quad (3.2)$$

This means that for a TE polarized FW the generated SH has TM polarization (only the $z$ component of the electric field is different from zero).

The refractive index of $Al_{0.3}Ga_{0.7}As$ at the fundamental wavelength ($\lambda_{FW} = 1.55$ μm) is $n_{FW} = 3.224$ and at the SH wavelength ($\lambda_{SH} = 0.775$ μm) it has the value $n_{SH} = 3.452$. Fig. 3.2 shows the dispersion curves for a 2D PC made from a square lattice of air holes with radius $r = 0.3a$ etched in $Al_{0.3}Ga_{0.7}As$. The dispersion curve for FW is calculated for the TE polarization and the dispersion curve for the SH wave is calculated for TM polarization, as follows from eq. (3.2).

As shown in Fig. 3.2, the crossing of the two dispersion curves (indicating the fulfillment of phase matching condition) can be obtained also for a real dispersive material. The isofrequency contours for the phase matched frequencies represented in Fig. 3.2(b) and Fig. 3.2(c) show that for this configuration the diffraction is slightly reduced for the FW, while for the SH wave the diffraction is even stronger with respect to homogeneous material case, illustrated with black dashed curves. The coupling coefficient calculated using the eq. (2.4) gives the value $K = 0.56$, comparable with that obtained in homogeneous case (where the coupling coefficient is equal to unity).

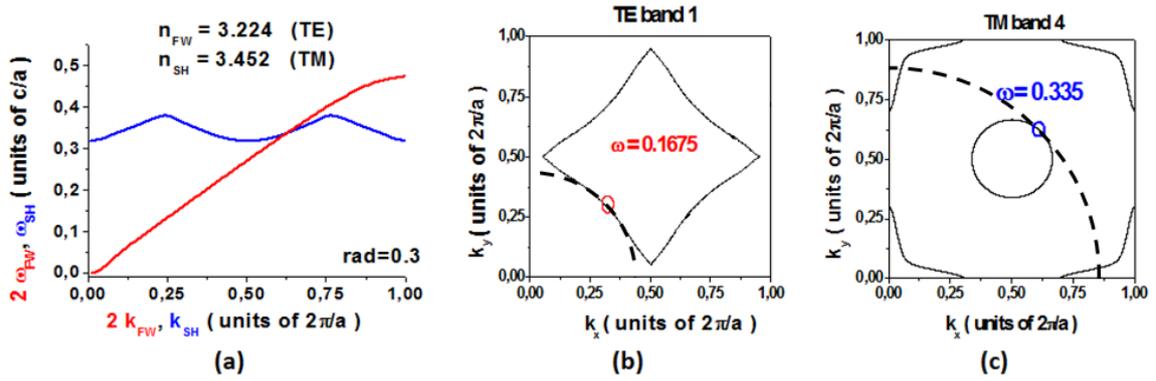

**Fig. 3.2** Dispersion curves for a 2D PC made in $Al_{0.3}Ga_{0.7}As$ (a) and the isofrequency contours for phase-matched frequencies for the FW (b) and SH (c). With black dashed lines there are represented the isofrequency contours for the homogeneous case.

Of course, one could try to tune the structure by varying parameters as holes radius and lattice geometry, in order to obtain also nondiffractive propagation regimes for both FW and SH wave simultaneously with the phase matching. However, this structure, which is considered to be infinite in the third dimension, is a theoretical abstraction. As already mentioned, for a configuration suitable for experiments and applications one should consider also the finite thickness of the structure. For this reason we use a planar geometry which is one of the most frequently used configurations in the study of PCs as it offers the possibility to control the losses in the third dimension. Moreover, this geometry would give more flexibility for the design of the structures, as the slab thickness provides another parameter that could be varied in order to tune the dispersive properties of the crystal.

## 3.2 Second harmonic generation in asymmetric planar two-dimensional photonic crystals

The presence of planar configuration introduces effective refractive indices of the propagation modes different from those in bulk material, as shown in Appendix B. It follows that the effective refractive index of the guided mode is decreased compared to the refractive index of the guiding layer and the decrease is stronger for lower frequencies. Therefore, the dispersion introduced by the planar waveguide is of the sign of the "normal" dispersion, in the sense that it adds (works in the same direction) to the "normal" dispersion of the material.

We start our study for an asymmetric planar 2D PC with a configuration that could be implemented experimentally. The asymmetric planar waveguide is made of a layer of $Al_{0.3}Ga_{0.7}As$ surrounded by a lower layer made by AlOx and an upper layer of air, as illustrated in Fig. 3.3(a). The thickness of the $Al_{0.3}Ga_{0.7}As$ layer is $t_1 = 0.25$ μm and the thickness of the AlOx layer is $t_2 = 2$ μm. The refractive index of AlOx is $n_{AlOx} = 1.6$ (we don't consider here the dispersion of the AlOx layer, taking the same refractive index for both FW and SH wavelengths).

In order to illustrate the effects of the planar configuration over the dispersive properties of the PC and calculate the band diagram of a 2D PC etched in the waveguide represented in Fig. 3.3(a), the effective refractive indices at the FW and SH wavelengths are calculated first using a 2D multilayer mode solver (that is freely available on web[1]). The effective refractive index at the fundamental wavelength $\lambda_{FW} = 1.55$ μm for the TE 0 guided mode represented in Fig. 3.3 (b) has the value $n_{eff}^{FW} = 2.67$ and the effective refractive index at the SH wavelength $\lambda_{SH} = 0.775$ for the TM 0 mode has the value $n_{eff}^{FW} = 3.12$.

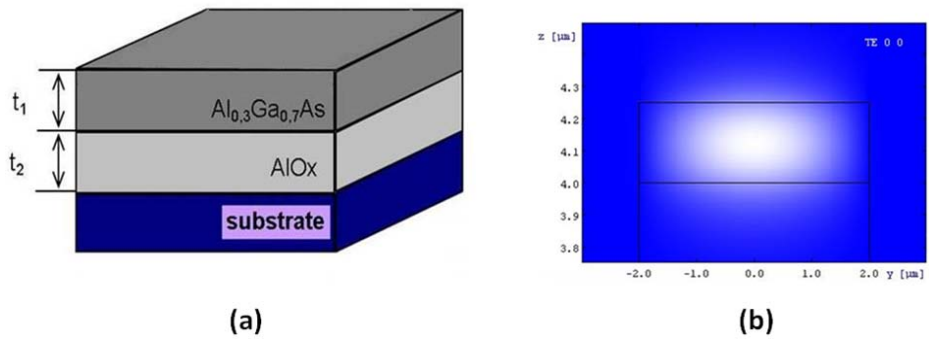

**Fig. 3.3** (a) Representation of the asymmetric waveguide made by a layer of $Al_{0.3}Ga_{0.7}As$ surrounded by a lower layer made by AlOx and an upper layer made by air; (b) representation of the intensity for the TE 00 mode for the FW.

Once the waveguide is selected, a particular 2D lattice should be included in the AlGaAs layer. A 2D PC made by a square lattice of air holes of radius $r = 0.25a$, where $a$ is the lattice constant, is etched in this planar waveguide. In Fig. 3.4, the bands 1, 4 and 6 for this structure are represented, where the red and blue dashed lines correspond to the frequencies for which non-diffractive propagation for the FW (in the first band) and the double of this frequency, which would correspond to the SH, are obtained. It can be noticed the fourth band shifted downward compared to the ideal case of the material without dispersion described in Section 2.1 of the previous chapter.

---

[1] http://wwwhome.math.utwente.nl/~hammer/eims.html

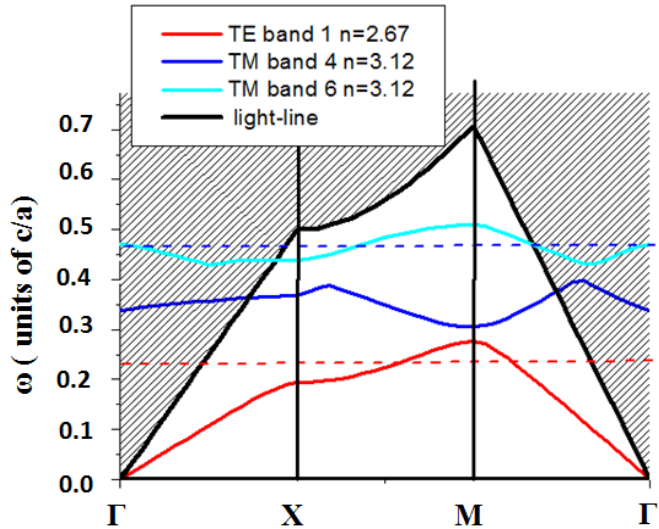

**Fig. 3.4** Band structure calculated for the 2D PC etched in the planar waveguide represented in Fig. 3.3(a)

In this case it can be noticed that the SH frequency appears in the band 6. Hence, in order to obtain simultaneously the phase matching condition and non-diffractive propagation regimes for both FW and SH, the dispersion curves and the isofrequency contours for the corresponding frequencies in band 1 and band 6 should be calculated as described in the previous chapter. In Fig. 3.5 the sub-diffractive areas (frequencies for which the diffraction is reduced to less than 10% with respect to that in homogeneous material) and the phase matching frequencies for different values of the radius of air holes are plotted. It can be noticed that for a radius around $r = 0.35a$ simultaneous phase matching and sub-diffractive propagation for both FW and SH wave are obtained.

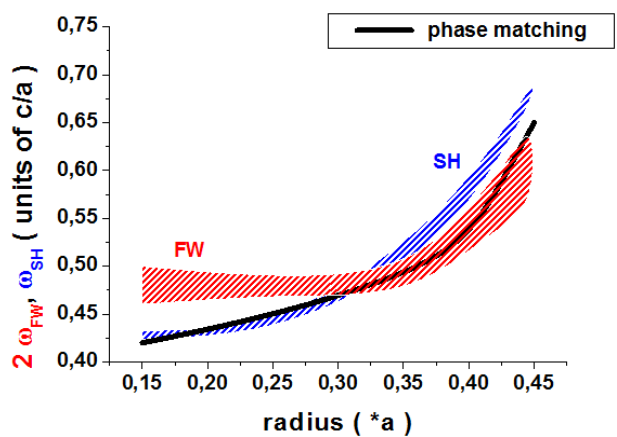

**Fig. 3.5** Representation of the surfaces of small diffraction (10% of transverse beam spreading in homogeneous media) in a plane (radius, frequency) for FW (red) and SH (blue) and the phase-matching curve (black)

Fig. 3.6(a) shows the dispersion curves for the FW and SH wave calculated for the configuration with the radius $r = 0.35a$ where their crossing point gives the phase

matching frequencies and wave vectors. When the isofrequency contours for the phase matched frequencies are plotted it can be noticed that for FW it is obtain indeed a self-collimation regime, as shown in Fig. 3.6 (b) where the red small circle represents the phase matched wave vector. However, in the case of the SH wave, although the phase matching and non-diffractive propagation regime are obtained for the same frequency, as shown in Fig. 3.6 (c), the phase matched wave vector for the SH (represented with a small blue circle) is different than the wave vector for which light can propagate without diffraction at this frequency in this band (where it appears the flat segment).

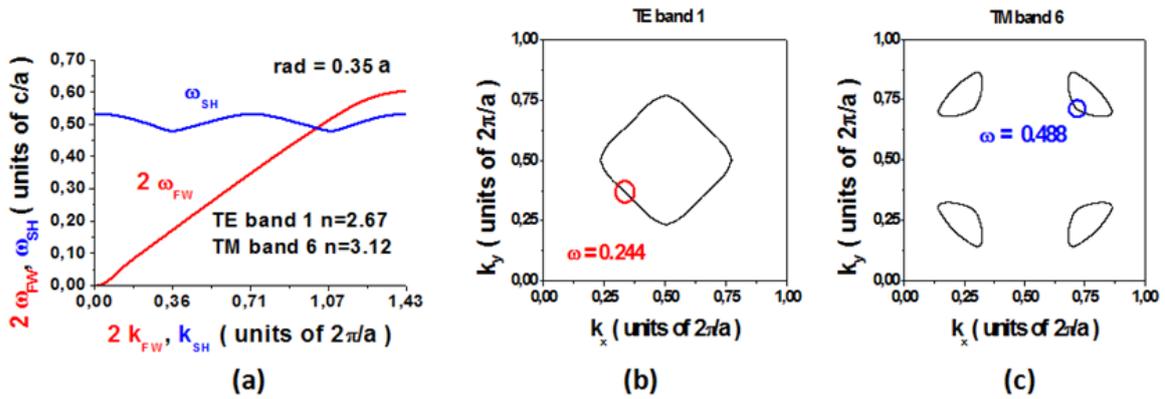

**Fig. 3.6** (a) The dispersion curves for FW in band 1 and SH in band 6 for a value of radius $r = 0.35a$; (b) the isofreqeuncy line for the phase-matched FW frequency and (c) the isofrequency line for the phase-matched SH frequency. With red and blue small circles are represented the phase matched wave vectors for the FW and SH wave, respectively

Nevertheless, one could try to shift the crossing point of the two dispersion curves (corresponding to the phase matching condition) to a region where a nondiffractive propagation regime is obtained for both waves. As already showed, one way to obtain this change in the dispersion curve is to vary the radius of the air holes. In Fig. 3.7 there are represented the dispersion curves for the case when the radius of air holes is $r = 0.45a$ together with the isofrequency contours for the phase-matched frequencies for the FW (in Fig. 3.7 (a)) and for the SH wave (in Fig. 3.7 (b)). It can be noticed that for this case there is a flattening of the isofrequency contours for both waves at the phase matched frequencies and at the phase matched wave vectors. However, the diameter of the air holes is very big in this case compared to the lattice constant, fact that makes the fabrication of such sample to be difficult.

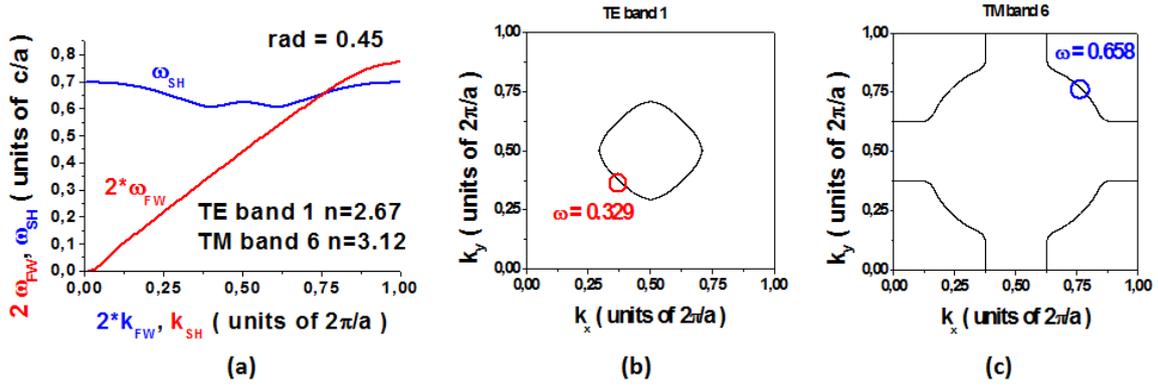

**Fig. 3.7** (a) The dispersion curves for FW in band 1 and SH in band 6 for a value of radius $r = 0.45a$; (b) the isofreqeuncy line for the phase-matched FW frequency and (c) the isofrequency line for the phase-matched SH frequency. With red and blue small circles are represented the phase matched wave vectors for the FW and SH wave, respectively

Thus, the study of the asymmetric waveguide does not lead to a satisfactory implementation of the idea promoted in the previous chapter. Another promising configuration in order to obtain good conditions to improve the efficiency of the nonlinear parametric process is the symmetric waveguide, as described in the next section.

## 3.3 Second harmonic generation in symmetric planar two-dimensional photonic crystals

The proposed structure is a symmetric waveguide made by a layer of $Al_{0.3}Ga_{0.7}As$ surrounded on both sides by air (an $Al_{0.3}Ga_{0.7}As$ membrane), schematically shown in Fig. 3.8(a) together with the variation of the effective refractive index for both FW, for the TE 0 mode, and SH wave, for the TM 0 mode, as a function of the thickness of the $Al_{0.3}Ga_{0.7}As$ layer (Fig. 3.8(b)). It can be seen that when the thickness of the membrane is increased, the difference between the effective refractive indices decreases. However, we are limited on this direction by the fact that we want to keep a monomode waveguide for the FW in order to avoid the interferences between different modes at the FW wavelength. Therefore, the maximum thickness of the waveguide is considered the value for which it is still monomode at the FW wavelength (for a larger value it will appear the next guided mode for the FW), which in our case is $d_{max} = 0.5\,\mu m$.

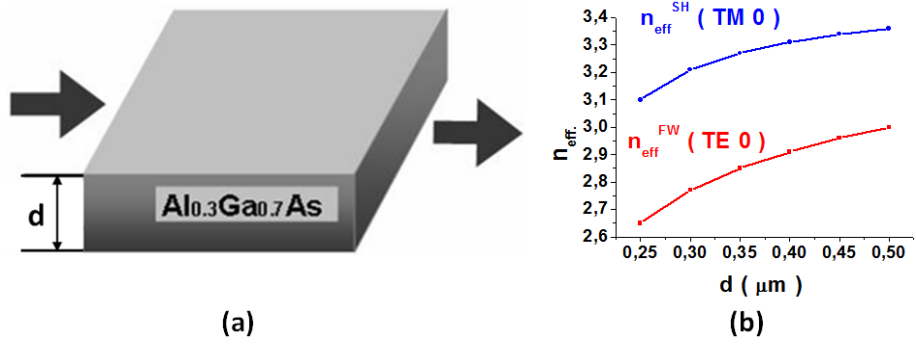

**Fig. 3.8** (a) Representation of the symmetric waveguide made by a layer of $Al_{0.3}Ga_{0.7}As$ surrounded on both sides by air; (b) the variation of the effective refractive indices with the thickness of the $Al_{0.3}Ga_{0.7}As$ layer

In the previous chapter, for the case of the ideal material without dispersion a structure that offers simultaneously phase matching and self-collimation propagation regimes for both FW and SH wave was obtained. Inclusion of material dispersion results in a more complicated situation. The introduction of the planar configuration complicates the situation even more, as it introduces additional dispersion acting in the same direction as the dispersion of the material. Hence, in order to apply the idea described in the previous chapter to realistic materials and configurations there are two options: one of them is to try to minimize the overall dispersion of the structure (material + waveguide) and then adjust the parameters of the structure in order to compensate the effects introduced by dispersion, and the other is to use the dispersion introduced by the waveguide (as it depends on the thickness of the structure) as another parameter for tuning the configuration in order to obtain the desired dispersive properties.

We will proceed by choosing first the thickness of the waveguide in such a way that the difference between the two effective refractive indices is minimal and the waveguide is still mono-mode. This gives a waveguide thickness of $d = 0.5$ μm. The effective refractive indices for this configuration are $n_{eff}^{FW} = 3.006$ and $n_{eff}^{SH} = 3.366$. We consider a 2D PC made by a square lattice of air holes of radius $r = 0.25a$ (where $a$ is the lattice constant) etched in this waveguide, schematically represented in Fig. 3.9(a). The dispersion curves for the FW, which has the frequency in the band 1 for TE polarization, and for the SH wave, which has the frequency in band 4 for TM polarization, are plotted in Fig. 3.9(b). It can be seen that the two curves intersect giving the phase-matched frequencies and wave vectors. However, the intersection of the dispersion curves occurs in the lower part of the curve for the SH indicating, as shown in Fig. 3.9(c), that the isofrequency contour for the phase-matched frequency will be given by a small circle. This means not only a strong diffraction for the SH wave, but also that there will be only a small range of wave vectors for which light can propagate in this band. Therefore, the crossing point of the two curves should be shifted at frequencies for which broader segments in the isofrequency contours can be obtained.

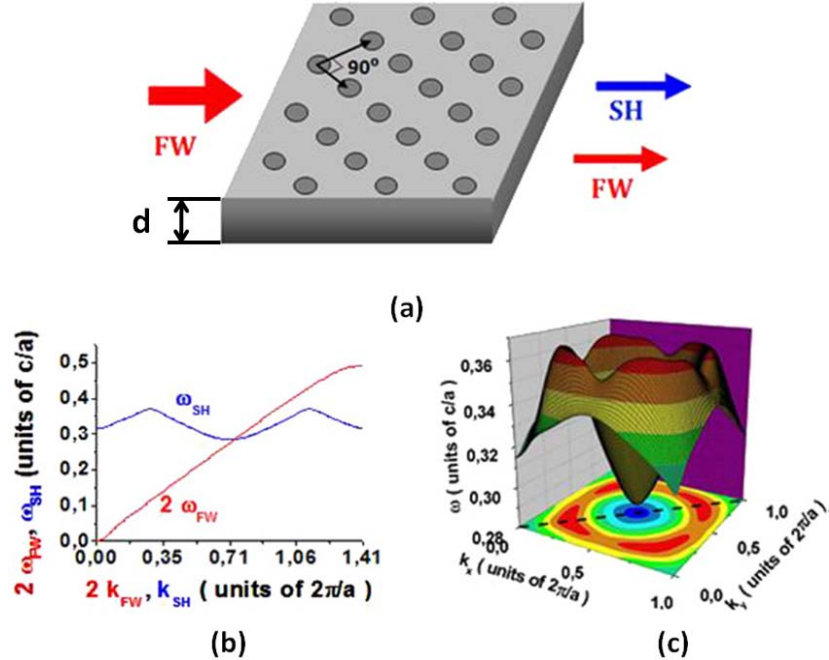

**Fig. 3.9** (a) Schematic representation of the planar 2D PC; (b) The dispersion curves for the FW and SH wavesquare lattice case considering band 1 in TE polarization and band 4 in TM polarization; (c) Representation of the fourth band for TM polarization

This shift of the dispersion curves could be obtained by changing the radius of the air holes, but using a very small or very large radius for the air holes will make the fabrication of this structure very difficult. Therefore, for a further engineering of the dispersion relation another parameter is used, the geometry of the 2D lattice.

For starting, we consider a rhombic lattice with an angle of $60^0$ between the lattice vectors, which is schematically represented together with the directions of light propagation in Fig. 3.10(a) (the direction of the propagation is along the long diagonal of the rhombus). From the dispersion curves plotted in Fig. 3.10(b) the frequencies and wave vectors corresponding to phase-matching are obtained. When the corresponding bands and their position with respect to the light cone are plotted in Fig. 3.10(c), we represent also the phase-matched frequencies and wave vectors with small red and blue circles for the FW and SH, respectively. It can be seen that the FW frequency appears below the light line, which means that the FW will be well guided and there will be no out-of-plane losses for this wave. However, the SH frequency appears above the light line at the wave vector for which the phase matching is satisfied and this leads to out-of-plane losses for the SH wave. The iso-frequency contours for the phase-matched FW and SH frequencies are represented in Fig. 3.10(d) and Fig. 3.10(e), respectively. It can be noticed that, while the diffraction for the FW is similar to that in homogeneous material, for the SH wave the diffraction is even stronger in this configuration (it is approximately 10 times stronger compared to that in homogeneous material).

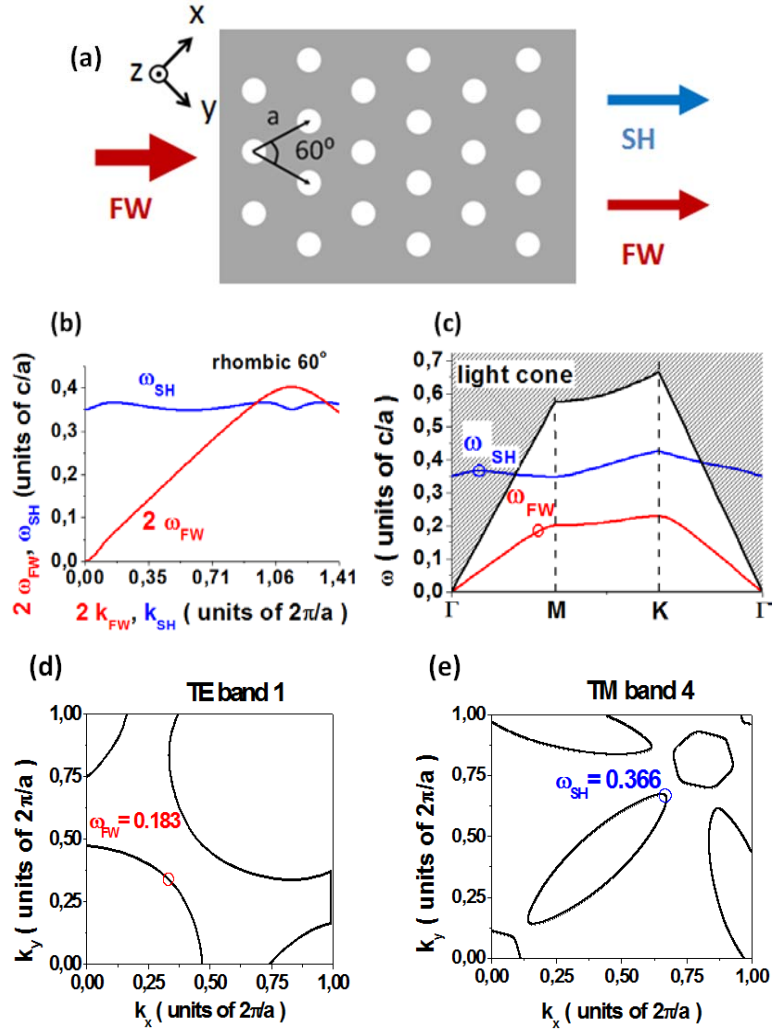

**Fig. 3.10** Schematic representation of the rhombic lattice with an angle of 60° between the lattice vectors (a), the dispersion curves for the FW and SH wave in PC with such a lattice (b), the corresponding bands and their positions with respect to the light cone (c) and the iso-frequency line for the FW (d) and the SH (e).

As it can be noted from Fig. 3.10(e), the strong diffraction for the SH wave can be avoided if the perpendicular direction of propagation, corresponding to the short diagonal of the rhombus is considered, as schematically shown in Fig. 3.11(a). The dispersion curves calculated for this configuration plotted in Fig. 3.11(b) give at their crossing point the phase matched frequencies and wave vectors. When plotting the corresponding bands and their position with respect to the light cone in Fig. 3.11(c) it can be noticed that both frequencies appear below the light cone, indicating that for this configuration the out-of-plane losses vanish for both FW and SH wave. This fact means not only that the FW will be collimated and will continue generating SH along the propagation through the crystal, but also that the SH wave won`t radiate out-of-plane and it will accumulate during the propagation creating conditions for a high efficiency of the process. The iso-frequency contours represented in Fig. 3.11(d) and Fig. 3.11(e) for the phase matched FW and SH frequencies, respectively, show that the diffraction for both FW and SH wave remain similar to that in homogeneous material.

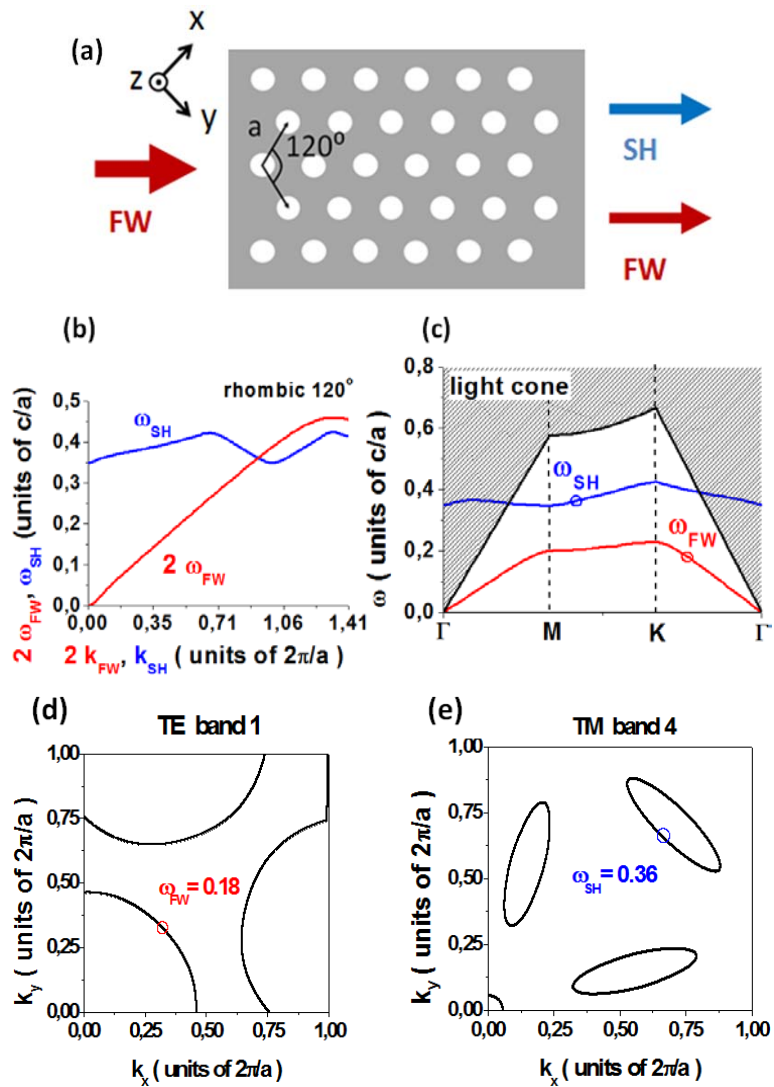

**Fig. 3.11** Schematic representation of the rhombic lattice and of the direction of propagation (a), the dispersion curves (b) for the FW and SH wave in PC with such a lattice, the corresponding bands and their positions with respect to the light cone (c) and the iso-frequency line for the FW (d) and the SH (e).

The parametric coupling coefficient for the two modes calculated with formula (2.4) for this configuration has the value $K = 0.05$, a value that is small with respect to the values obtained in the case of ideal dispersionless material, but under the conditions of no out-of-plane losses for both FW and SH wave and a decreasing of the group velocity for the FW wave with a factor of approximately 5 provided by this configuration, the overall properties of the structure warrant an experimental realization of this idea.

Nevertheless, the situation could be further improved by achieving nondiffractive propagation regimes for FW and SH wave maintaining conditions of no out-of-plane losses. In the next section it is presented the study made in order to obtain such situation.

## 3.4 Vertically confined self-collimated beams second harmonic generation in planar two-dimensional photonic crystals

We consider again an $Al_{0.3}Ga_{0.7}As$ membrane surrounded on both sides by air and this time we use the dispersion of the waveguide as another parameter for tuning the structure in order to obtain the desired dispersive properties. The dispersion of the waveguide can be adjusted not only by changing the thickness of the slab, but also by choosing different propagating modes for the two waves.

Until now, the fundamental guided mode for both waves were considered, the TE 0 mode for FW and the TM 0 mode for the SH wave. Now, we continue considering the TE 0 mode for the FW wave (represented in Fig. 3.12(a)), since we want to keep a mono-mode waveguide for the FW, but for the SH we will consider the TM 2 mode, represented in Fig. 3.12(c). This particular guided mode for the SH was chosen because it has a lower refractive index compared to the fundamental mode and permits a shift of the first photonic band at higher frequencies, making possible in this way to obtain the SH frequency in the first band and, consequently, below the light cone, avoiding the out-of-plane losses at this frequency, too. Another reason for choosing this guided mode is that it ensures a relative good coupling with the TE 0 mode of the FW mode (of approx. 41%, as detailed later in this section). In Fig. 3.12(b) and Fig. 3.12(d) the distribution of the electromagnetic field in the vertical direction for the TE 0 mode and TM 2 mode, respectively, are represented. If the fundamental wavelength is $\lambda_{FW} = 1.55$ μm (and the SH wavelength $\lambda_{SH} = 0.775$ μm) for a thickness of the waveguide of 350 nm the effective refractive indices for the corresponding modes are $n_{FW}^{eff} = 2.85$ and $n_{SH}^{eff} = 1.425$, for the FW and SH, respectively.

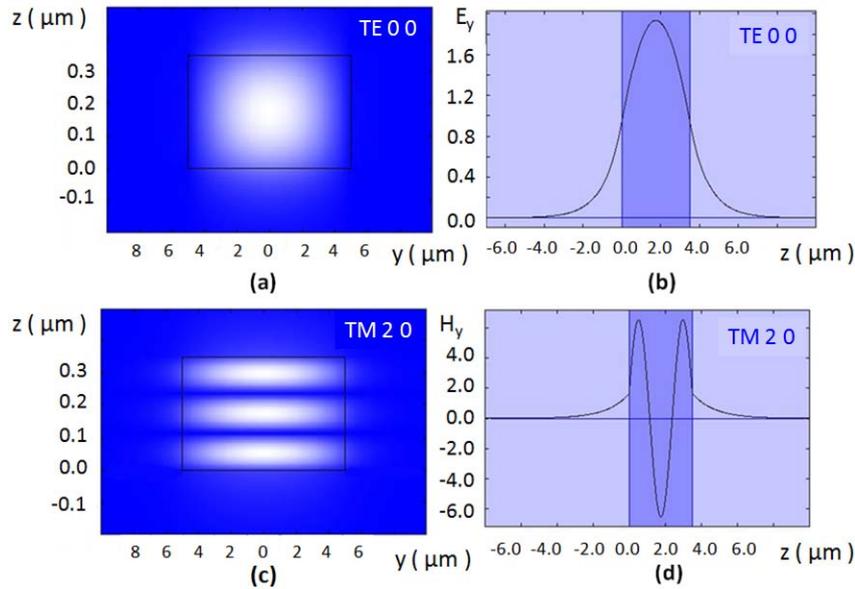

**Fig. 3.12** Representation of the TE 0 guided mode (a) and TM 2 guided mode (c) and the distribution of the field in the third direction for the TE 0 mode (b) and the TM 2 mode (d)

It can be noticed that in this case the effective refractive index for the SH wave is smaller than the effective refractive index for the FW. This means that, when a PC made by a lattice of air holes is etched in this waveguide, the first band for the SH mode (TM 2 mode) will be at higher frequencies with respect to the first band corresponding to the FW mode (TE 0 mode), as illustrated in Fig. 3.13.

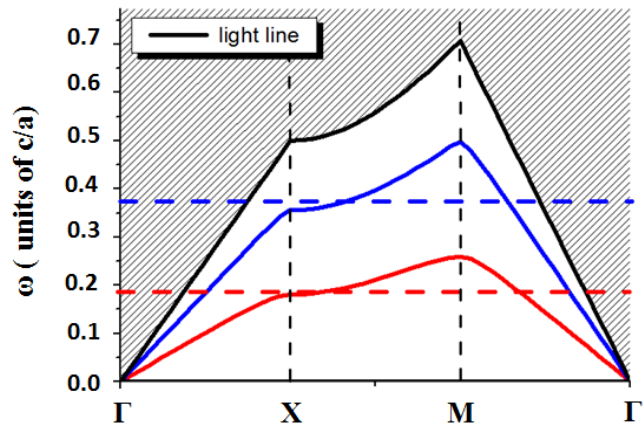

**Fig. 3.13** Representation of the first photonic band for the FW mode (with red line) and SH mode (with blue line) when the FW refractive index is higher than the SH one and their position with respect to the light cone; with dashed red and blue lines there are represented the FW and SH frequencies, respectively

Moreover, as the SH effective refractive index is half of the FW effective refractive index, it means that one can obtain also the SH frequency in the first band, for the particular guided mode. Therefore, as both FW and SH frequencies appear in the first band, both will be below the light cone and there will be no out-of-plane losses, for neither of the two waves. This fact means, as mentioned before, not only that the FW will be collimated and will continue generating SH along the propagation through the crystal, but also that the SH wave won`t radiate out-of-plane and it will accumulate during the propagation creating conditions for a high efficiency of the process.

We consider a square lattice of air holes with the radius $r = 0.25a$ etched in the waveguide described above. From the dispersion curves represented in Fig. 3.14(a) the phase matched frequencies and wave vectors (at the crossing point of the two curves) are obtained and when plotting the corresponding bands and their position with respect to the light cone shown in Fig. 3.14(b) one can see that, indeed, both frequencies appear below the light line indicating that the out-of-plane losses will vanish for both waves. The iso-frequency contours for the phase matched FW and SH frequencies plotted in Fig. 3.14(c) and Fig. 3.14(d) show that the diffraction is decreased with respect to the homogeneous case but it is still far from the self collimation regime.

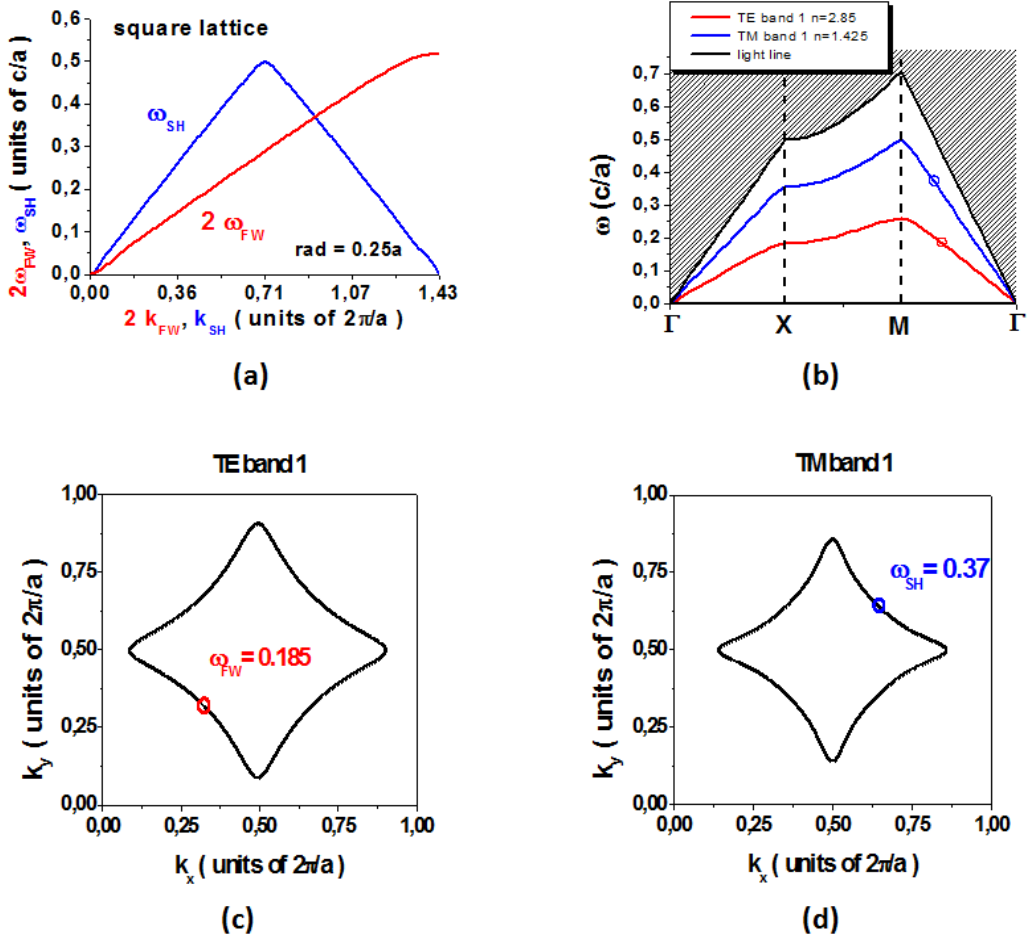

**Fig. 3.14** Representation of the dispersion curves (a), the corresponding bands and their position with respect to the light cone, where with red and blue circles are represented the FW and SH frequencies, respectively (b) and the isofreqeuncy contours for the FW and SH wave in (c) and (d), respectively

In order to obtain flatter segments in the iso-frequency contours, we should change the parameters of the structure. As already shown, one way to engineer the shape of the iso-frequency lines is to change the geometry of the lattice. We consider a rhombic lattice instead of the square lattice with the angle between the lattice vectors as parameter for adjusting the optical properties of the structure. After the dispersion curves and iso-frequency contours are calculated for different values of the lattice angle, a configuration that offers simultaneously non-diffractive propagation conditions for both FW and SH wave at the phase matched frequencies is obtained. The dispersion curves for the rhombic lattice with an angle of $76^0$ are represented in Fig. 3.15(a) and the iso-frequency contours for the FW and SH are plotted in Fig. 3.15(b) and Fig. 3.15(c) respectively. From the iso-lines it can be seen the appearance of flat segments for both waves at the phase-matched frequencies and wave vectors (represented with red arrow for the FW and blue arrow for the SH wave), indicating the self-collimation regimes. Moreover, the negative slope of the SH dispersion curve at the phase matching point predicts that the SH wave will propagate backward, in opposite direction with respect to the FW.

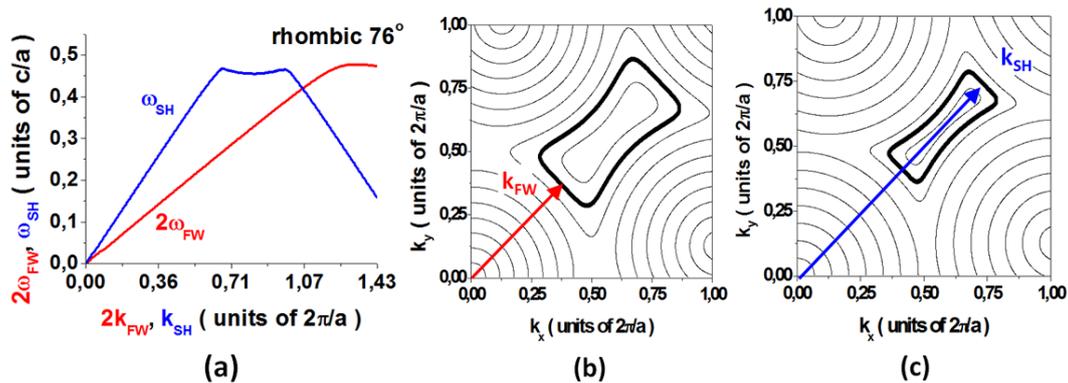

**Fig. 3.15** Dispersion curves (a) and iso-frequency contours for the phase matched FW (b) and SH (c) frequencies; with red and blue arrows are represented the phase matched wave vectors for the FW and SH wave in (b) and (c), respectively

The dispersion bands represented in Fig. 3.15 are calculated in 2D using the effective refractive indices of the waveguide, determined as explained in Appendix B of this chapter. Additionally, the 3D photonic bands for the planar 2D PC were calculated in order to confirm the self-collimation regimes at the phase matched frequencies. The 3D calculations were made in collaboration with Timothy Karle from Laboratory for Photonics and Nanostructures (CNRS-UPR20) Marcoussis, France, using a modified version of the GME code (that is freely available on web[2]). The isofrequency contours in the central plane of the slab are represented in Fig. 3.16(a) and Fig. 3.16(b) for the FW and SH wave, respectively. For the FW contours it can be noted the similarity of the contours with the 2D calculations, as the only difference is given by the slightly shifted orientation of the wave vector axis used in Fig. 3.16(a) and (b) with respect to the orientation used in Fig. 3.15(b) and Fig. 3.15(c). Nevertheless, the isofrequency contours calculated for SH are substantially changed with respect to the 2D calculations. The main reason is that while the waveguide is monomode for the FW wavelength, at the SH wavelength there are more guided modes allowed and when performing 3D calculations the isofrequency contour reflects all these wave vectors for which light can propagate through the structure at the given frequency. Even under these conditions, one can observe that the phase matched frequency and wave vector correspond to a nondiffractive regime also for the SH wave.

---

[2] http://fisicavolta.unipv.it/nanophotonics/Gme.htm

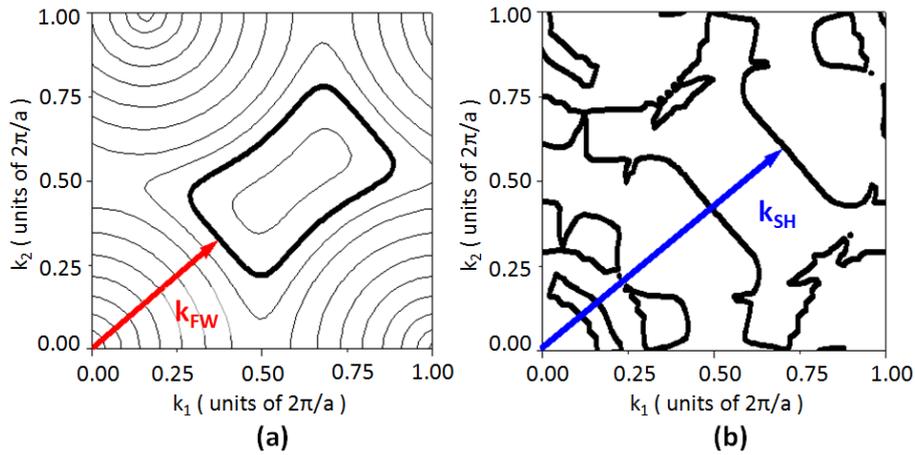

**Fig. 3.16** Isofrequency contours calculated in 3D for the FW (a) and SH wave (b); with red and blue arrows are represented the phase matched FW and SH wave vectors in (a) and (b) respectively

In order to illustrate the nondiffractive propagation for the two waves, linear 2D FDTD simulations for light propagation in this structure were performed. The effective refractive indices calculated above are used, $n_{FW}^{eff} = 2.85$ and $n_{SH}^{eff} = 1.425$ for the FW and the SH respectively. A crystal with a length of 60 μm and a width of 20 μm is considered, with an etched rhombic lattice of air holes with the radius $r = 0.25a$, and $a = 322.4$ nm. The angle between the lattice vectors is $76^0$ and the direction of propagation is along the short diagonal of the rhombus. A Gaussian source of 1.5μm width is placed inside the crystal close to the left side, as illustrated in Fig. 3.17(a) and Fig. 3.17(b). First, the source is set to radiate at the FW wavelength (1.55 μm) and the distribution of the intensity obtained is represented in Fig. 3.17(a). Next, the source is set to radiate at the SH wavelength (0.775 μm) and the distribution of intensity is represented in Fig. 3.17(b). It can be seen that after 60 μm both beams remain well collimated (the Rayleigh range corresponding to a Gaussian beam of 1.5 μm width at the FW and SH wavelengths are approximately 2.8 μm).

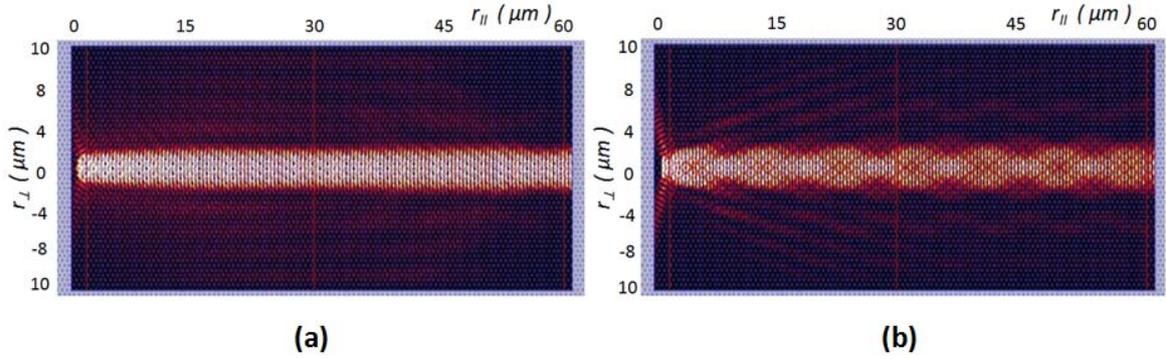

**Fig. 3.17** (a) Representation of the intensity when the source has the FW wavelength; (b) representation of the intensity when the source has the SH wavelength

Moreover, when the length of the crystal is increased to 200 μm and the FDTD simulations are made for the propagation of the SH wave, shown in Fig. 3.18(a), the beam is propagating without diffractive broadening, as illustrated also by the transversal distribution of the intensity after 198 μm, represented in Fig. 3.18(b).

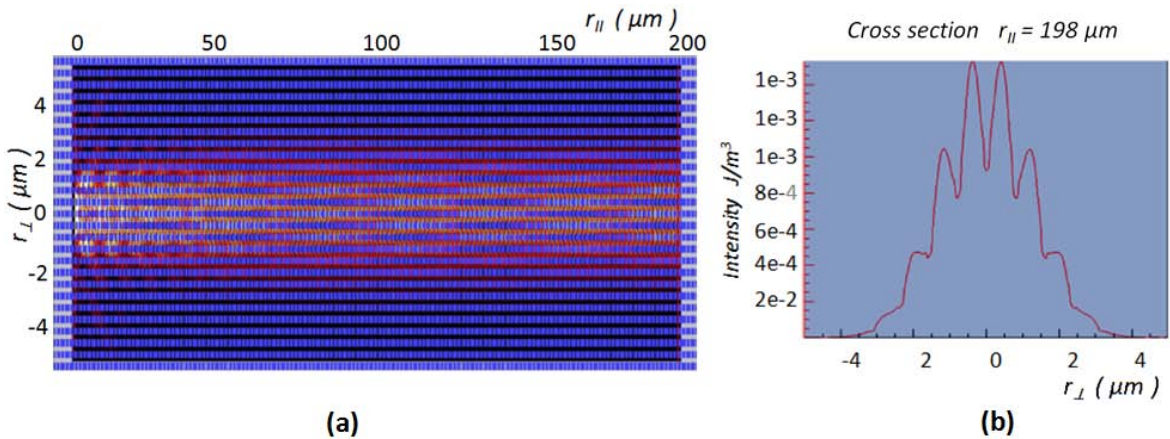

**Fig. 3.18** (a) Representation of the intensity when the source has the SH wavelength; (b) Transversal distribution of the intensity after 198 μm

Once shown that the structure satisfies simultaneously the phase matching and the nondiffractive propagation conditions for both FW and SH wave, the next step is to estimate if there will be out-of-plane losses. We could see that for the square lattice configuration when we consider the TE 0 guided mode for the FW and the TM 2 mode for the SH the out-of-plane losses vanish for both waves, as illustrated in Fig. 3.14(b). However, now we are using a rhombic lattice with an angle of $76^0$ and in order to estimate if there will be or not out-of-plane losses we need to plot again the corresponding bands and their position with respect to the light cone. Therefore, the first Brillouin zone and then the irreducible Brillouin zone should be calculated for the rhombic lattice with an angle of $76^0$. Fig. 3.19 shows the iso-frequency contours from the first band with TE polarization for a generic 2D PC made by a lattice of air holes having such symmetry. With blue line it is represented the first Brillouin zone and with green line the irreducible Brillouin zone. Therefore, in order to calculate the bands when the lattice has this symmetry, the dispersion $\omega(k)$ for the points along the contour represented with the green line was calculated.

In Fig. 3.20(a) the irreducible Brillouin zone for this lattice is represented with the notation used for each vertex and in Fig. 3.20(b) the photonic bands for the two modes and their position with respect to the light cone are plotted. The phase matched frequencies and wave vectors for the FW and SH wave are represented with red and blue small circles, respectively. One can see that both frequencies are below the light cone indicating that there will be no out-of-plane losses. Therefore, the vertical confinement of both beams also for the rhombic lattice with the angle of $76^0$ is obtained.

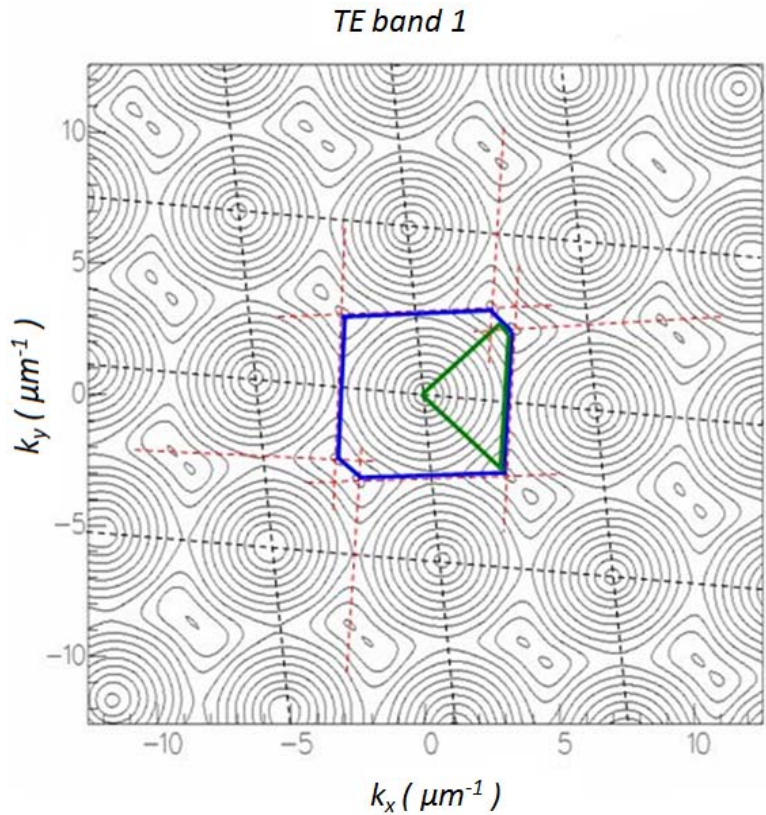

**Fig. 3.19** Iso-frequency lines from the first band with TE polarization for a generic 2D PC made by a rhombic lattice of air holes with an angle of $76^0$ between the lattice vectors. With blue line it is represented the first Brillouin zone and with green line it is represented the irreducible Brillouin zone. The lattice constant is a=1 μm and the radius of the air holes is $r = 0.25a$

As described in Section 2.3, another condition for an efficient SHG process is to have a sufficiently strong overlap between the two interacting modes. Therefore, in order to evaluate the efficiency of nonlinear coupling the properly normalized cross-correlation between the two fields is calculated:

$$K = \frac{\left|\int_M \left(E_1^2 E_2^*\right) dr\right|}{\left(\int_C \left|E_1^4\right| dr \int_C \left|E_2\right|^2 dr\right)^{1/2}} \tag{3.3}$$

where $E_1$ and $E_2$ represent the FW and SH fields, respectively. The upper integral is calculated in the nonlinear material in one unit cell, while the lower integrals are taken in the entire unit cell. First, the coupling coefficient in 2D is calculated, representing the 2D overlap of the Bloch modes and we obtained the value $K_B \approx 0.37$. Due to different vertical distribution of the two fields, as they belong to different guided modes, the overlap of the TE$_0$ mode and TM$_2$ mode is $K_W \approx 0.41$. The full 3D overlap results in $K \approx 0.15$.

Therefore, the proposed structure ensures phase matching, self collimation regimes for both waves, a relative large nonlinear coupling of modes and vanishing of out-of-plane losses at both FW and SH frequencies.

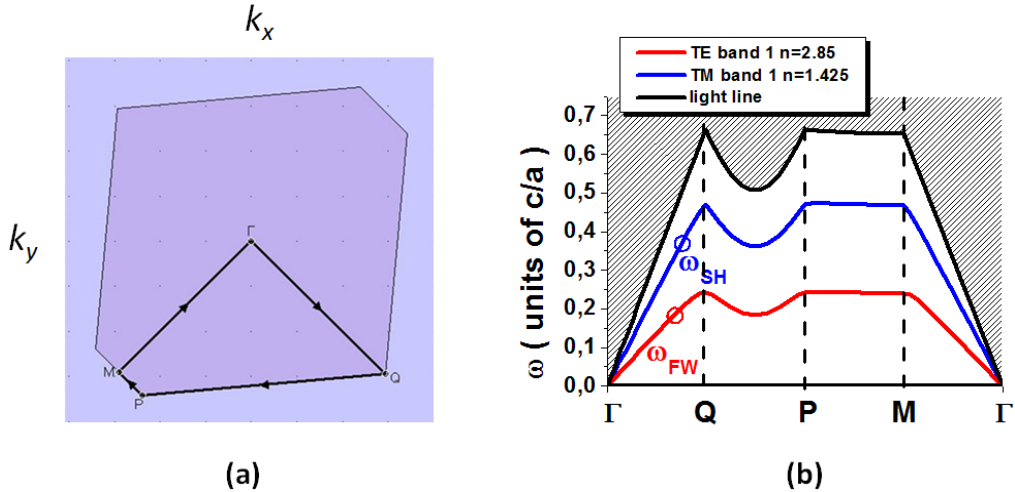

**Fig. 3.20** Representation of the irreducible Brillouin zone for the rhombic lattice with the angle of $76^0$ (a) and the photonic bands corresponding to FW and SH wave and their position with respect to the light cone; with small red and blue circles are represented the phase matched frequencies and wave vectors for the FW and SH wave, respectively.

Once we found a particular structure with proper conditions for SH generation with narrow beams, next it follows to confirm these results by means of numerical simulations. The nonlinear calculations were made in collaboration with Fabrice Raineri and Timothy Karle from Laboratory for Photonics and Nanostructures (CNRS-UPR20) Marcoussis, France, using a nonlinear 2D FDTD code described in [Rai02]. The nonlinear FDTD method that was used works in the non-depleted pump approximation and neglects intra-pulse chromatic dispersion. It uses two parallel linear FDTD codes, one of them at the FW wavelength and the other at the SH wavelength. The quadratic nonlinearity is only taken into account for the SH, which is not coupled back at the FW. Chromatic dispersion is considered simply by taking the actual refractive indices at the FW and SH wavelengths. This artificial separation of FW and SH propagation allows easy identification of the FW and SH field distributions and other relevant physical parameters.

The structure used in the simulations is 14 μm wide and 54 μm long. A 2 μm beamwidth short pulse FW source centered at 1550 nm, with bandwidth of 150 nm corresponding to pulse duration of approximately 23 fs was used. The field distributions for the FW and SH wave are plotted in Fig. 3.21(a) and Fig. 3.21(b) respectively at different moments during the propagation. As the FW propagates, the short pulse broadens weakly due to his nonzero spectral width and obtains a characteristic spatiotemporal shape for the nondiffractively propagating pulses [Sta06]. The SH wave is generated at the position of the FW pulse and, as the time increases, the SH beam spreads along the structure. In the last frame of Fig. 3.21 the FW beam has already left the structure at the right hand boundary, while the SH beam is still inside the structure and propagates toward the left border, evidencing the opposite direction of propagation of SH with respect to FW propagation direction, as predicted also from Fig. 3.15. It can be noticed that also the SH beam is highly collimated showing no significant diffraction along the distance used in simulation.

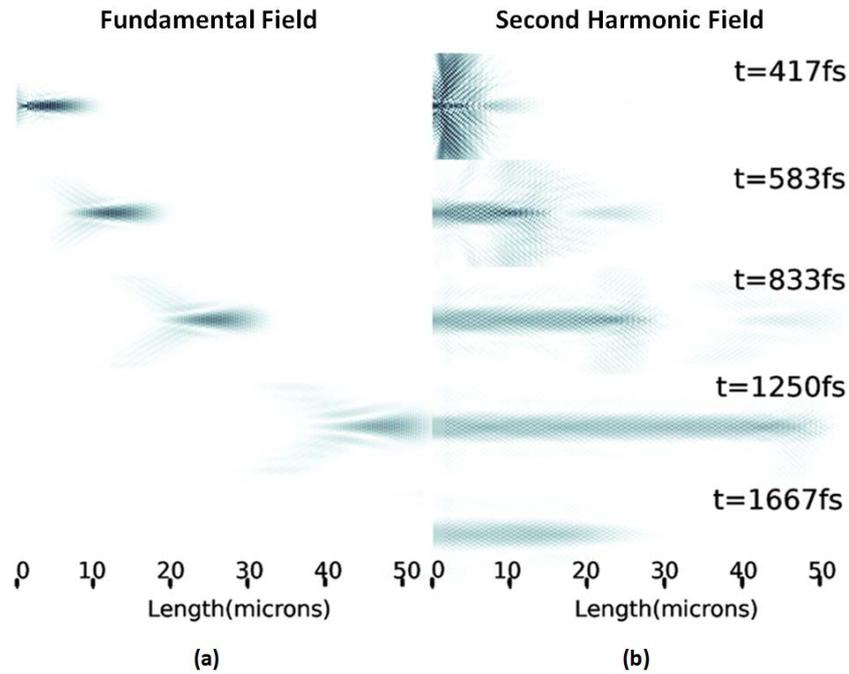

**Fig. 3.21** Spatial distribution of the FW (a) and the SH (b) fields at different moments during the propagation

The spectrum of the backward generated SH, represented in Fig. 3.22 , shows a resonance at the phase matched SH wavelength, very close to the expected value from the linear calculations.

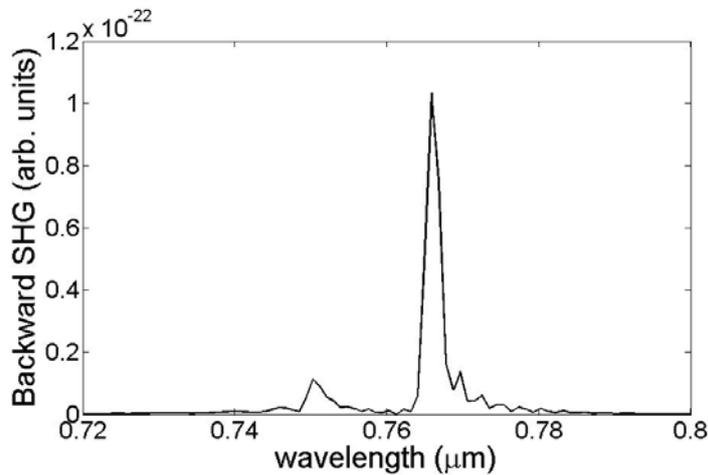

**Fig. 3.22** Spectrum of the backward generated SH

Next, we made the simulation in the continuous wave regime. The field distribution for the FW and SH wave are shown in Fig. 3.23 after the stationary regime was reached. The two plots confirm the strong collimation of the two waves, as the Rayleigh length calculated in homogeneous material for the FW and SH are approximately 5 μm for a width of the beams of 2 μm.

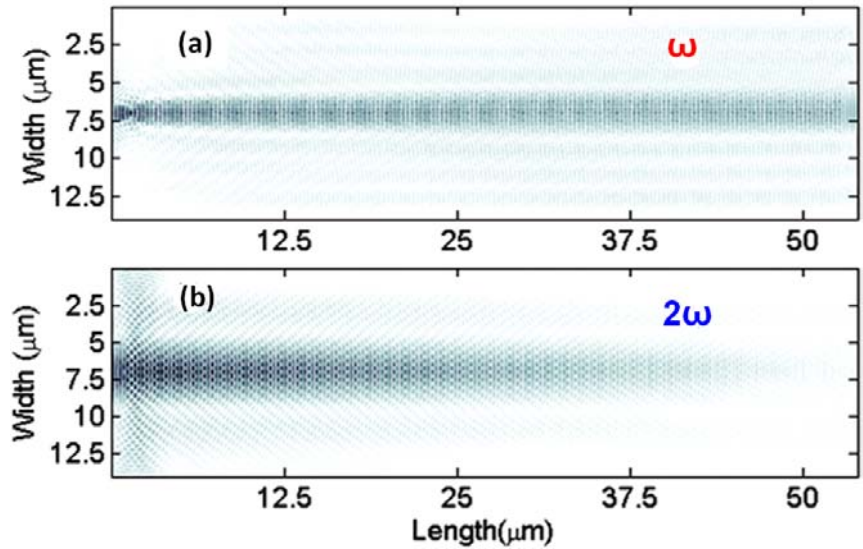

**Fig. 3.23** Spatial distribution of the FW field (a) and SH field (b) after the stationary regime was reached

The collimation of the FW beam can be observed also in the field amplitude cross section represented in Fig. 3.24 (a), where it can be noted that the amplitude of the FW field is constant as it propagates through the structure, except the first 10 μm, where occurs the coupling to other modes for part of the radiation. The cross section for the SH field, represented in Fig. 3.24 (b) shows an almost linear growth with the length in the backward direction evidencing the phase matching of the two waves.

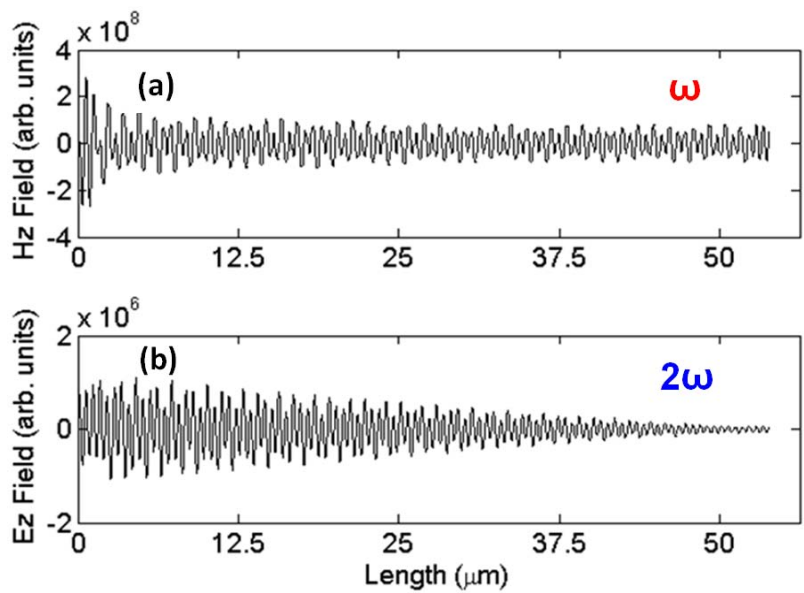

**Fig. 3.24** Field amplitude cross section for the FW field (a) and the SH wave field (b)

With an input power of 1 GW/cm$^2$, a conversion efficiency of 6.3x10$^{-4}$ for the 2D calculations is obtained. To calculate the conversion efficiency of the real structure, one must consider the overlap of the TE 0 and TM 2 modes, which is $K_W = 0.41$, as calculated before. Since the conversion depends on the square of the overlap coefficient, this results in a conversion of approximately $1.06 \times 10^{-4}$ for the real structure.

For comparison, the conversion efficiency in an ideal homogeneous material with the same nonlinear coefficient at the phase matching is calculated. The following formula for the nonlinear length (after which the conversion should be 1) is used:

$$l = \left( \frac{2n_1^2 n_2 \varepsilon_0 c}{I} \right)^{1/2} \frac{c}{2\omega_1 d_{eff}} \qquad (3.4)$$

where $I$ is the FW wave intensity, $n_1$ and $n_2$ are the FW and SH refractive indices, respectively, $\omega_1$ is the frequency of the FW and $d_{eff}$ is the nonlinear coefficient.

The nonlinear length obtained has the value $l=6.11 \times 10^{-4}$ m, which means that after 54 μm (the distance used in the FDTD simulations) the conversion should be $7.8 \times 10^{-3}$. Taking into account the overlap of the two guided modes the conversion obtained in homogeneous material for the interaction of the TE 0 and TM 2 modes becomes $1.3 \times 10^{-3}$. Therefore, the conversion efficiency in PC is lower than in homogeneous materials by a factor of approximately 12, caused mainly by the overlap of the Bloch modes for the two waves in PC (where it is $K_B = 0.37$, as calculated before) compared to the homogeneous case.

In conclusion, we have suggested and proved the design of a 2D PC structure with real material parameters capable of generating second harmonic using very narrow fundamental beams, with diameters comparable to the wavelength of light. The characteristics of the beams and the conversion efficiency are obtained by a full nonlinear study. The SH generation occurs in the backward propagation direction and both fundamental and SH beams propagate under the light cone (without radiative losses) and in a non-diffractive regime.

Concerning the experimental part of the work, the fabrication of the sample presented above and the experiments are carried out in collaboration with Laboratoire de Photonique et de Nanostructures (CNRS UPR20) Marcousis, France. At the moment of presenting the thesis the first samples are in process of fabrication and characterization with the participation of Timothy J. Karle, Fabrice Raineri, Isabelle Sagnes, Remy Braive and Rama Raj from the French group laboratory. The PC, shown in Fig. 3.25, is made of a 350 nm $Al_{0.3}Ga_{0.7}As$ layer grown on an $Al_{0.9}Ga_{0.1}As$ sacrificial layer. The pattern, represented by a rhombic lattice of air holes with radius $r = 0.25a$ and lattice constant $a = 322.4$ nm, was made using E-beam lithography.

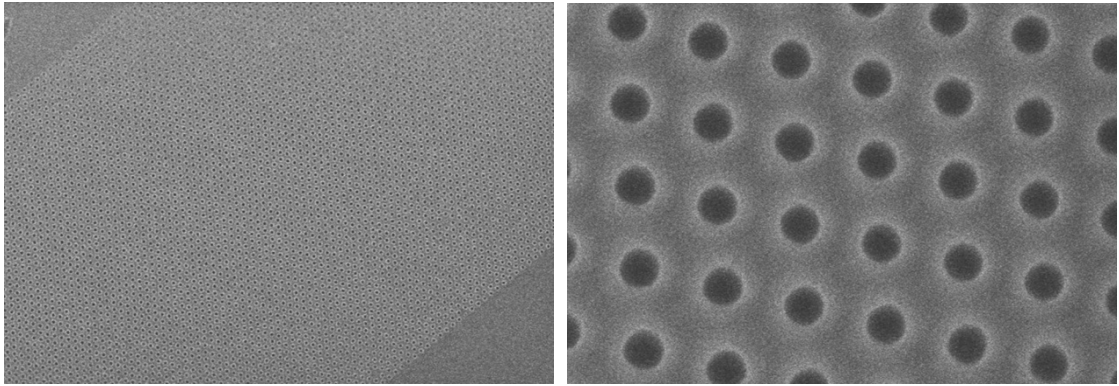

**Fig. 3.25** SEM images of the fabricated photonic crystal

## 3.5 Conclusions

From the results presented in this chapter the following conclusions can be stressed:
- In the case of the bulk 2D PC, the introduction of the material dispersion shifts the position of the dispersion curves and the phase matching can be obtained by changing the air holes radius and lattice geometry. Anyway, the bulk configuration is studied with just a theoretical purpose, since the realistic conditions impose a planar configuration.
- When the planar 2D PCs are used, the effective refractive indices for the guided modes change the dispersion relation and for the asymmetric configurations the phase matching can be obtained for the square lattice considering the FW in the TE 1 band (for the TE 0 guided mode) and the SH wave in the TM 6 band (for the TM 0 guided mode). However, simultaneous self-collimation regimes for FW and SH wave were not found.
- For the case of symmetric planar 2D PC, we obtained phase matching considering the FW in the TE 1 band (for the TE 0 guided mode) and the SH wave in the TM 4 band (for the TM 0 guided mode). Changing the lattice geometry, for the rhombic lattice with an angle of $120^0$ between the lattice vectors we maintain the phase matching condition with the frequencies and wavevectors of both waves below the light cone, indicating the vanishing of the out-of-plane losses. Anyway, the diffraction is similar to the homogeneous case for both FW and SH wave.

- In the case of symmetric planar 2D PCs considering the TE 0 guided mode for the FW and the TM 2 guided mode for the SH wave, we obtained the phase matching condition when both FW and SH frequencies are in their respective first photonic band, indicating the vanishing of the out-of-plane losses for both waves. For the case of rhombic lattice with an angle of $76^0$ we obtain, in addition, simultaneous self-collimation regimes for both waves. The nonlinear FDTD calculations confirm the nondiffractive propagation regimes and the backward propagation direction of the generated SH wave and permit the evaluation of the conversion efficiency.

Chapter 4

# 4. Broad Spectral Range and Broad Angular Range Phase Matched Second Harmonic Generation in Planar Two-Dimensional Photonic Crystals

The phase matching is a crucial condition for efficient nonlinear wave mixing, such as second harmonic generation. The PM is absent in isotropic dispersive homogeneous materials with normal dispersion due to the frequency dependent refractive index: the modulus of the wave vector of SH frequency $k_{SH}$ is always larger than the double of fundamental wave vector $2|k_{FW}|$. As already shown in the previous chapter, the use of wave guiding geometries (e.g. the planar waveguide) usually does not improve the PM because the effective refractive index $n_{eff}$ of a guided mode also increases with frequency. This results in an increase of the phase mismatch, since the dispersion of the waveguide adds to the one of the material.

One possibility to obtain PM relies on the use of photonic crystals, where the periodic modulation of the refractive index makes that the dispersion curve can be strongly distorted in both frequency- and space- domain. Therefore, adjusting the parameters of the PC structure one could extend the PM in the frequency-domain, leading to broad spectral range PM, and in the space-domain, leading to a broad angular range PM. The work presented in this chapter is dedicated to find such configuration that ensures both broad angular range and broad spectral range PM.

## 4.1 Broad spectral range and broad angular range phase matching in photonic crystals

The idea is to match the top of the first band for the FW frequency $\omega_1$ with the top of the second propagation band for the SH, which is close to $2\omega_1$. The magnitude of the distortion $\omega_2 - 2\omega_1$ of the dispersion curves in PC is of order of the width of the band-gap $\Delta\omega_{1,2}$, and for high contrast materials can reach 10-20%. Usual materials show a difference in the refractive indices at the FW and SH frequencies of ~5-10%. A planar waveguide imposes an additional difference of ~5-10%. This means that planar PCs could be ideal structures for obtaining PM as the difference in the refractive indices between FW and SH has a value of ~10-20% that can be well matched by the distortion of the dispersion curves.

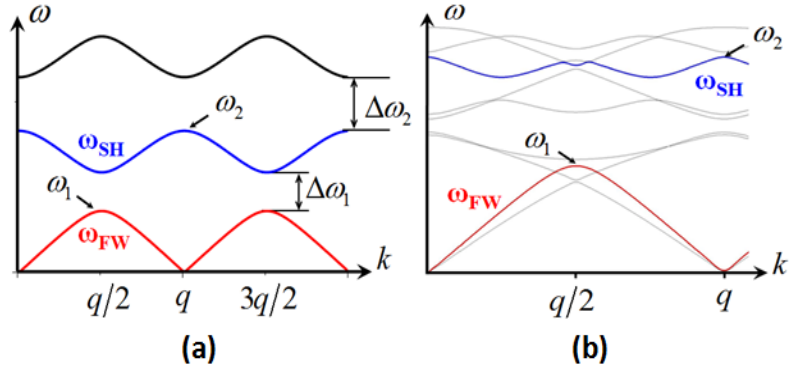

**Fig. 4.1** (a) Illustration of PM in 1D PCs when the FW frequency is at the top of the first photonic band and the SH frequency is at the top of the second band; (b) The PM in 2D PC when the FW frequency is in the first band for TM polarization is matched with the SH frequency is in the sixth band for TE polarization

An important advantage in this situation is that the crossing of the two dispersion curves is obtained in regions where they have similar slopes. This means that the PM (or sufficiently small phase mismatch) can be obtained for a larger range of frequencies. Achieving a broad spectral range PM is very convenient since any real light beam has a spectral width (it doesn't contain only one frequency, the phase matched one) and therefore more spectral components of the beam will participate efficiently in the nonlinear interaction in this case.

In materials with normal dispersion only the PM of the collinear plane wave components is possible, as shown in Fig. 4.2(a). However, for flattened dispersion curves there can be phase matched plane wave components propagating at different angles leading to more dense interactions, as illustrated in Fig. 4.2(b). Obtaining PM in a broad angular spectrum allows the use of very narrow beams, which permits an increasing of the intensity by focusing the light beam. Therefore, the parametric coupling between beams should be more efficient for the flattened dispersion case because the SH conversion of narrow beams asymptotically increases as first power of interaction length for conventionally diffracting materials (Fig. 4.2(a)) and as the second power for the flattened diffraction curves or surfaces (Fig. 4.2(b)).

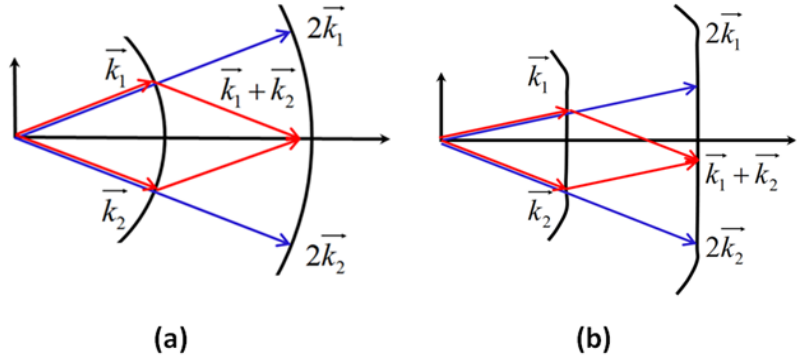

**Fig. 4.2** The PM for conventional diffraction (in homogeneous materials) can be achieved only for the collinear plane wave components (a), and for all components for flattened diffraction surfaces in PCs (b).

Therefore, here we basically combine two ideas: 1) the dispersion of the material and of the planar waveguide can be compensated and PM can be in principally reached in PCs for plane waves propagating along a particular direction due to distortion of *dispersion* curves belonging to different bands; 2) when PM is reached in one particular direction it can be extended in angular space, due to the distortion (the flattening, in this case) of the *diffraction* curves. In addition, the PM can be obtained also in a broad spectral range given by the similar slopes of the dispersion curves of the two waves. As a result an efficient geometry for the parametric wave mixing, in particular for the SH generation of narrow beams, is proposed.

## 4.2 Phase matched second harmonic generation in planar two-dimensional photonic crystals

A planar waveguide made by an $Al_{0.3}Ga_{0.7}As$ membrane is considered. The TE polarized FW propagates in AlGaAs in the [100] crystallographic direction and generates a TM polarized field at the SH frequency due to the particular expression of the polarizability tensor for this material, as described in the previous chapter. The refractive index of $Al_{0.3}Ga_{0.7}As$ at the FW wavelength ($\lambda_{FW} = 1.55$ µm) is $n_{FW} = 3.224$ and at the SH wavelength ($\lambda_{SH} = 0.775$ µm) it is $n_{SH} = 3.452$. The thickness of the waveguide is 0.5 µm and the effective refractive indices are $n_{FW}^{eff} = 3.006$ (for the TE 0 mode) and $n_{SH}^{eff} = 3.366$ (for the TM 0 mode). On the waveguide it is etched a square lattice of air holes with radius $r = 0.25a$, where $a$ is the lattice constant. In Fig. 4.3(a) it is represented schematically this planar 2D PC. The dispersion curves in this configuration for the FW in the first photonic band for TE polarization and for the SH wave in the sixth band for TM polarization are calculated. The dispersion curves are plotted in Fig. 4.3(b) and one can see that indeed the two curves intersect giving the phase matched frequencies and wave vectors, but their intersection is not in a region where they have similar slopes, situation that is expected in order to obtain a broader spectral range phase matching.

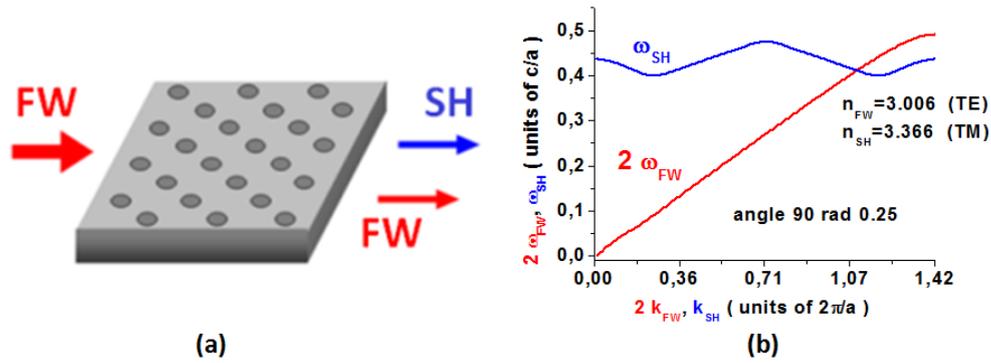

**Fig. 4.3** Schematic representation of the planar 2D PC made by a lattice of air holes made in an AlGaAs membrane (a) and the dispersion curves calculated for the square lattice case, when the radius of the air holes is $r = 0.25a$ (b)

Therefore, a systematic study varying different parameters of the structure as the lattice constant and radius is made in order to bring the crossing point of the two dispersion curves in a region where they have similar slopes. The main steps of this study are summarized in Fig. 4.4 where the dispersion curves calculated for different configurations are represented. Different geometries are used: rhombic lattice with the angle between the lattice vectors having the values $120^0$ (in Fig. 4.4(a)), $80^0$ (in Fig. 4.4(c) and Fig. 4.4(d)) and $70^0$ (in Fig. 4.4(b)), and different radii for the air holes: $r = 0.25a$ (in Fig. 4.4(a), Fig. 4.4(b) and Fig. 4.4(c)) and $r = 0.35a$ (in Fig. 4.4(d)). Finally, a configuration for which we could match the regions with similar slopes of the two dispersion curves and which predicts a broad angle PM, as illustrated in Fig. 4.4(d), is obtained.

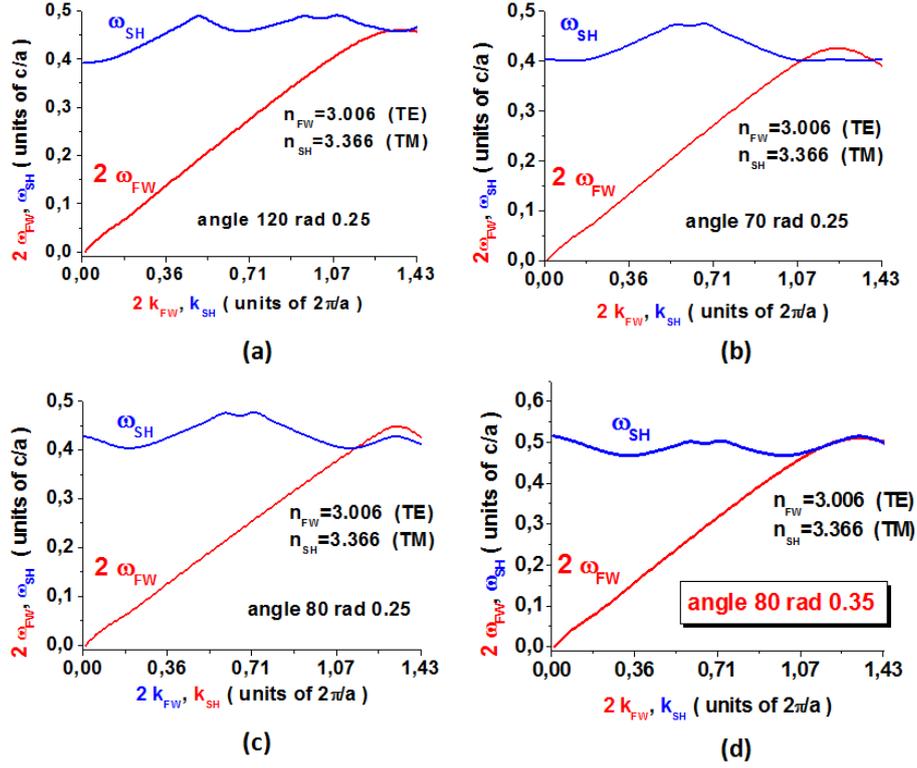

**Fig. 4.4** Dispersion curves calculated for the FW in the first band, represented with red line, and for the SH wave in the sixth band, represented with blue line, for different configurations where we vary the geometry of the rhombic lattice and the radius of the air holes: the rhombic lattice with an angle of $120^0$ between the lattice vectors and a radius of $r=0.25a$ (a), the rhombic lattice with an angle of $70^0$ with the radius of $r=0.25a$ (b), the rhombic lattice with an angle of $80^0$ with a radius of $r=0.25a$ (c) and the rhombic lattice with an angle of $80^0$ and a radius of $r=0.35a$ (d)

For the configuration with the rhombic lattice with an angle of $80^0$ and a radius of $r=0.35a$, the dispersion of the material and the dispersion introduced by the waveguide add in order to match the top of the FW band with the top of the SH band, in a similar way to the phase matching obtained in 1D PC. In Fig. 4.5 it is schematically shown the contribution of the material and waveguide dispersion, separately. Fig. 4.5(a) shows the dispersion curves for the FW and SH wave in an ideal dispersionless material in bulk configuration, without material or waveguide dispersion, where one can see the position of the two curves, which are shifted one with respect to another. Inclusion of material dispersion in bulk configuration (without waveguide dispersion) results in a change of dispersion curves as illustrated in Fig. 4.5(b). One can note the displacement downward of the SH curve, in such a way that the two curves get closer one another. In Fig. 4.5(c) the two curves are calculated for non-dispersive material, but in the planar configuration, thus it is included only the dispersion introduced by the waveguide. It can be noticed again that the two curves get closer one another. And finally, in Fig. 4.5(d) there are plotted the dispersion curves for the real dispersive material ($Al_{0.3}Ga_{0.7}As$) in the planar configuration, which means that there are included both the material and waveguide dispersion. It can be noticed again that changing the refractive index results in a shift of the bands, upward for lower values of the refractive index, while the shape of the dispersion curve depends mainly on the periodicity (lattice geometry) and size of the air holes, since the two curves remain essentially unchanged, when it is varied only the refractive index.

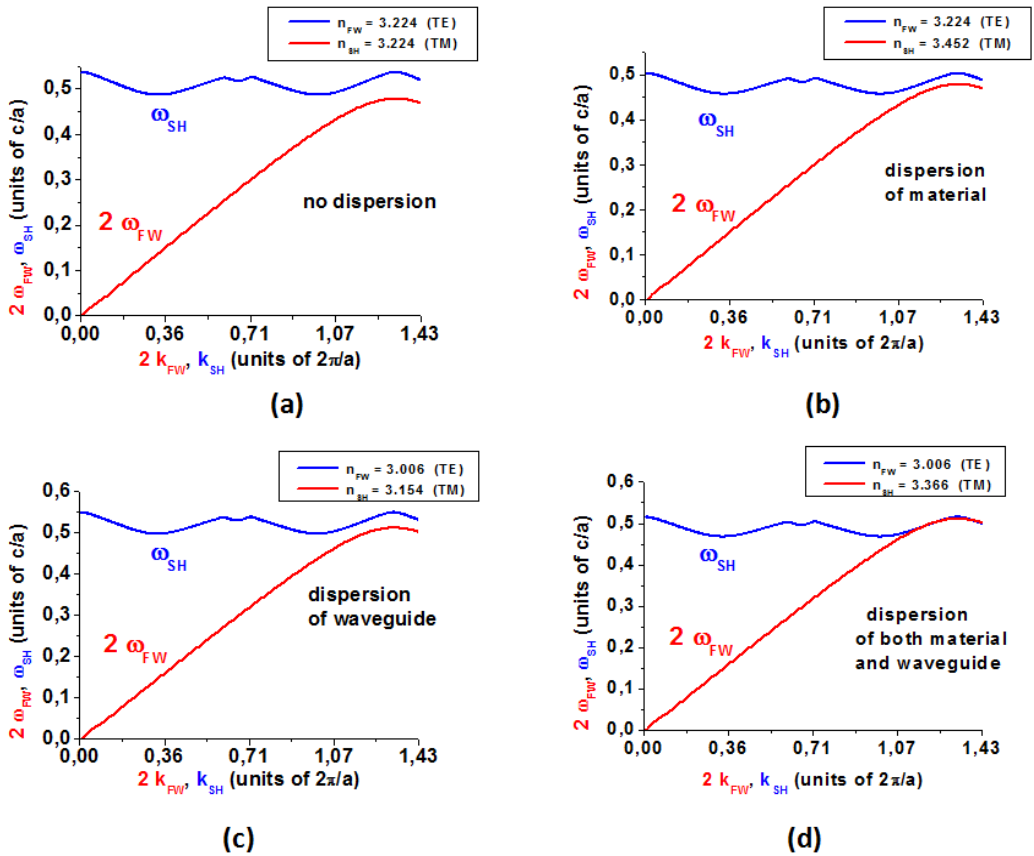

**Fig. 4.5** The dispersion curves calculated for the FW and SH wave when there is no dispersion included (a), when it is included only the dispersion of the material (b), when it is included only the dispersion of the waveguide (c) and when there are included both the dispersion of the material and the dispersion of the waveguide

Once a broad spectral range PM is obtained, next step is to calculate the iso-frequency contours for the phase matched frequencies because the presence of flat segments at the phase matched wave vectors would indicate that the PM is obtained in a broad angular range, too. In Fig. 4.6(a) and Fig. 4.6(b) there are represented the iso-frequency contours for the FW and SH wave, respectively. The thick lines indicate the phase matched frequencies, while the red and blue arrows indicate the phase matched wave vectors. It can be seen that for both waves flat regions in the iso-frequency contours are obtained, which show that the phase matching is obtained also in a broad angular range, the size of the plateau giving the minimum width of the beam that can be phase matched in this configuration.

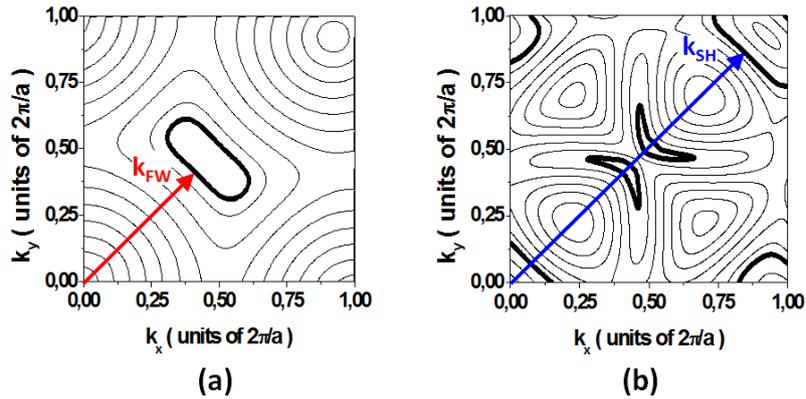

**Fig. 4.6** The isofrequency contours for FW in the first band for TE polarization (a) and for the SH wave in the sixth band for TM polarization (b); thick lines are represent the phase matched frequencies and red and blue arrows show the phase matched wave vectors for the FW and SH wave, respectively

## 4.3 Out-of-plane scattering in two-dimensional photonic crystals

Another aspect that needs to be accounted is related to the finite thickness of the structure. In a planar waveguide, one has to calculate if the waves will be well guided inside the planar structure or there will be also out-of-plane scattering. The existence of the out-of-plane scattered waves at the FW frequency has a negative effect on the efficiency of the nonlinear process, since the part of the energy corresponding to the waves that will escape from the waveguide will participate no more to the SH generation process. Therefore, it is important to obtain a good guiding of the FW, while an out-of-plane scattering at the SH frequency requires that the detection should be made not only at the end of the waveguide, but also in the upper and lower planes with respect to the planar structure. In some situations this could be even useful as it permits the out-of-plane detection of the SH wave.

The presence of the out-of-plane scattering can be estimated from the position of the photonic bands for the two waves with respect to the light cone. Therefore, the irreducible Brillouin zone for the rhombic lattice with an angle of $80^0$ is calculated and represented in Fig. 4.7(a), where the same notation for the vertices as for the case of rhombic lattice with an angle of $76^0$ presented in the previous chapter is used. Calculation of the bands corresponding to the FW (the first band for the TE polarization) and SH wave (the sixth band for TM polarization) represented in Fig.

4.7(b) shows that the FW phase matched frequency and wave-vector (represented with a small red circle) appear below the light cone. This means that there will be no out-of-plane scattering at the FW frequency, so the FW will continue to generate SH along the propagation through the crystal. However, the SH phase matched frequency and wave vector represented with small blue circle in Fig. 4.7(b) appear inside the light cone, indicating that a part of the generated SH will be scattered out-of-plane requiring that the detection of the SH should be made also in the upper and lower plane with respect to the planar crystal. This fact will give rise to new and interesting geometries for the detection of SH in order to maximize the efficiency of the nonlinear device.

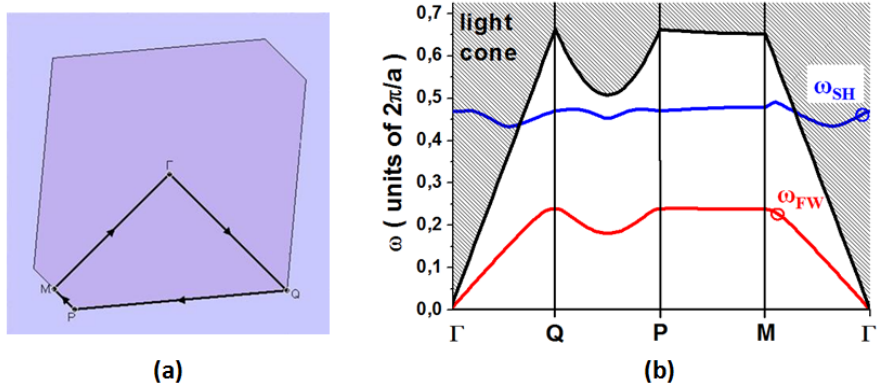

**Fig. 4.7** Representation of the irreducible Brillouin zone for the rhombic lattice with an angle of $80^0$(a) and the photonic bands corresponding to the FW and SH wave and their position with respect to the light cone (b) where with small red and blue circle are represented the phase matched frequencies for the FW and SH wave, respectively

Therefore, a PC structure for which the crossing of the dispersion curves at similar slopes is obtained and the presence of flat segments in the iso-frequency contours at the phase matched frequencies and wave vectors predict that the phase matching condition is satisfied in both broad spectral range and a broad angular range. Next, nonlinear FDTD calculations are performed in order to illustrate the advantages that the described structure presents for the nonlinear parametric process.

## 4.4 FDTD method for simulation of light propagation in quadratic nonlinear materials

For calculating the SH generation we made a 2D nonlinear FDTD code consisting of two linear 2D FDTD schemes which propagate in parallel the FW and SH wave [Rai02]. The coupling between the two schemes is made through the nonlinear polarization term of second order. In order to reduce the time and resources needed for calculations, two simplifications are made. First, the quadratic nonlinear polarization term is considered only for the SH wave. Therefore, the depletion of the FW and interactions such as the nonlinear dephasing induced by the succession of two second order nonlinear interactions ("cascading") are neglected. This reduction is reasonable while the conversion efficiency is smaller than 10%, fact which is verified in our simulations due to the relative short distances of interaction. The other simplification is that the dispersion around the FW and SH wavelengths is not considered and, in this way, the time needed for calculations is strongly reduced.

From the Maxwell's curl equations (1.31) and (1.32) and the constitutive relation (1.33) we obtain for the SH wave:

$$\frac{\partial \vec{H}_{SH}}{\partial t} = -\frac{1}{\mu_0} \nabla \times \vec{E}_{SH} \qquad (4.1)$$

$$\frac{\partial \vec{E}_{SH}}{\partial t} = -\frac{1}{\varepsilon} \nabla \times \vec{H}_{SH} - \frac{1}{\varepsilon} \frac{\partial \vec{P}^{(2)}}{\partial t} \qquad (4.2)$$

where $P^{(2)} = \varepsilon_0 \chi^{(2)} E_{FW}^2$ is the second order nonlinear polarization.

In this case, the nonlinear polarization appears as a source term for the SH wave. To calculate the SH fields we follow the scheme represented in Fig. 4.8, where the electromagnetic field for the FW is calculated at each time step through the structure using the usual FDTD method. The results permit to calculate the nonlinear source term $P^{(2)}$ and therefore calculate the SH field at each moment inside the computational area.

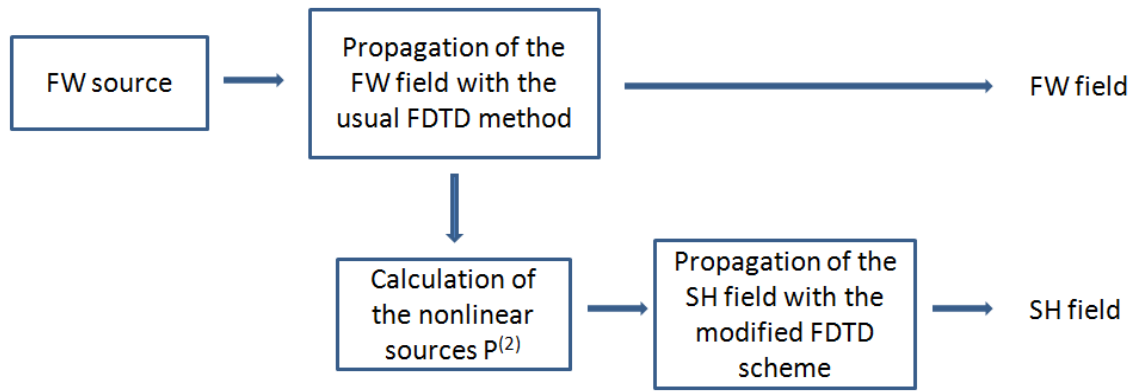

**Fig. 4.8** Scheme of the nonlinear decoupled FDTD method

A more detailed description of the nonlinear FDTD method that we used is presented in Appendix C.

## 4.5 Nonlinear FDTD simulations for second harmonic generation in planar two-dimensional photonic crystals

Next, the structure presented in Section 4.2 of this chapter is considered: the planar 2D PC made by a rhombic lattice of air holes etched in a $0.5$ thick AlGaAs membrane surrounded by air, represented in Fig. 4.3(a). The TE polarized FW propagates in AlGaAs in the [100] crystallographic direction and generates a TM polarized wave at the SH frequency. The effective refractive indices of the waveguide at the FW wavelength (chosen to be $\lambda_{FW} = 1.55$ μm) is $n_{FW}^{eff} = 3.006$ (for the TE 0 guided mode) and at the SH wavelength ($\lambda_{FW} = 0.775$ μm) is $n_{FW}^{eff} = 3.366$ (for the TM 0 mode). The angle between the lattice vectors is $80^0$ and the radius of the holes is $0.35a$ where $a = 346$ nm is the lattice constant.

Using the FDTD code presented in the previous section the SH generation in the structure described is calculated for different wavelengths of the FW wave and different angles between the lattice vectors. For each value of the angle between the lattice vectors the generation of SH for different values of the FW wavelength is calculated and a plot similar to that represented in Fig. 4.9(a) for the case of the lattice with an angle of $81.2^0$, is obtained. The presence of the resonance shows the PM condition and the wavelength (frequency) for which it is obtained in this particular configuration. Next, the maximum value for the SH intensity from this graph is taken together with the maximum value obtained for the SH intensity when different values for the angle between the lattice vectors were considered and they were plotted in Fig. 4.9(b).

The dispersion curves calculated when the angle between the lattice vectors is varied around the value for which the broad spectral range PM is obtained, represented in Fig. 4.10, show that when the angle is smaller than $83^0$ the crossing point of the two dispersion curves is getting closer to the region where the curves have similar slopes. The phase matching is obtained in this region, indicating a broad spectral range phase matching for angles around $81^0$, as shown in Fig. 4.10(b) and Fig. 4.10(c). The broad spectral range phase matching for this configuration results in an increasing of the generated SH intensity, as illustrated in Fig. 4.9(b).

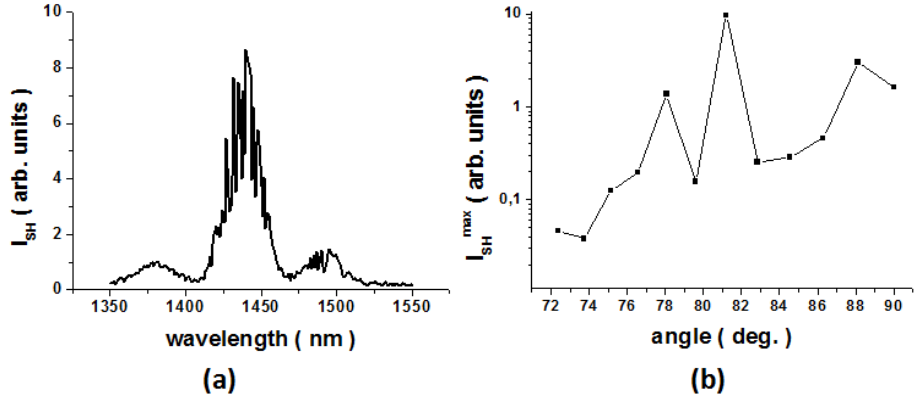

**Fig. 4.9** (a) Intensity of the generate SH for the rhombic lattice of angle 81.2⁰ when we vary the wavelength of the FW and (b) representation of the maximum value for the generated SH for different values of the angle between the lattice vectors

The width of the beams used for these simulations were sufficiently big compared to the length of propagation in order to avoid the effects due to the diffractive broadening of the beams. The relative small difference between the value of the angle at which the maximum SH intensity shown in Fig. 4.9(a) is obtained and the value for which broad angular range phase matching should appear resulting from the linear calculations (band solving) illustrated in Fig. 4.10 is given by the resolution used in calculations and by the variation of the refractive index due to the change in the carrier density, described by the $\alpha-$factor [Hen82] or by the more general Kramers-Kronig relations.

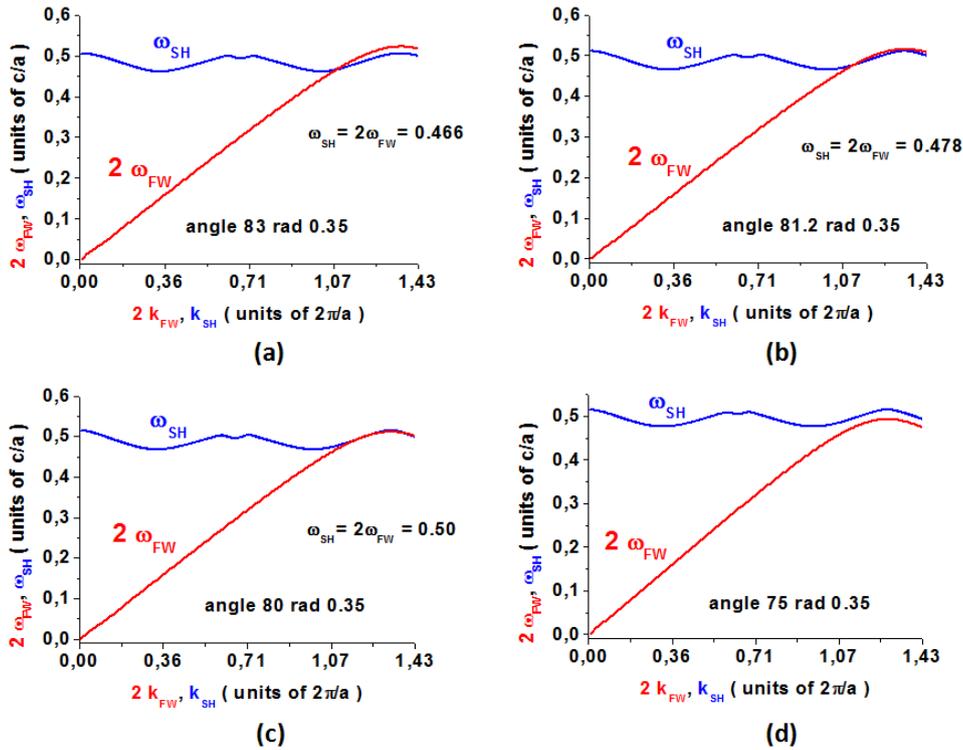

**Fig. 4.10** Representation of the dispersion curves when the angle between the lattice vectors has the value 83⁰ (a), 81.2⁰ (b), 80⁰ (c) and 75⁰(d).

In Fig. 4.11(a) and Fig. 4.11(b) there are represented the field distributions for the FW and SH wave, respectively, in the configuration for which the maximum conversion is obtained, with the angle $81.2^0$. The width of the FW source is 1.5 μm and it can be seen that both waves propagate without diffraction along the distance considered. The cross-section for the FW and SH fields along the propagation direction at $r_\perp = 0$, represented in Fig. 4.11(c) and Fig. 4.11(d) show beatings at both frequencies, indicating the presence of more wave-vector components for each of them.

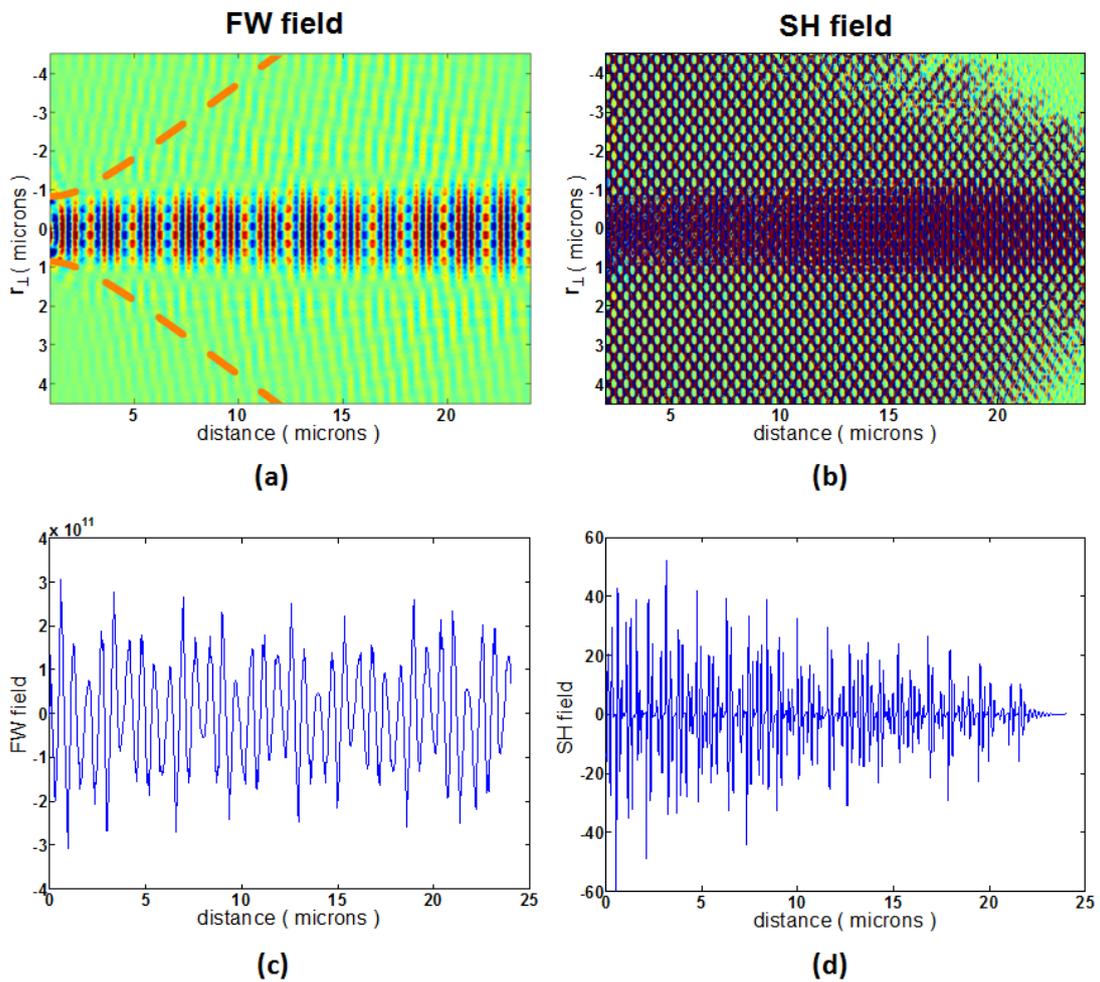

**Fig. 4.11** Distribution of the fields for FW (a) and SH wave (b) and the cross-sections along the propagation direction at $x = 0$ for both FW (c) and SH (d)

In order to obtain the spectral distribution for the two waves, the Fourier transform is applied to the complex field calculated from the real field data plotted in Fig. 4.11. In Fig. 4.12(a) and Fig. 4.12(b) the spectra calculated for the FW and SH wave, respectively, are represented for the case when the FW source is placed at the beginning of the structure. The presence of different wave vector components for both waves, which give the standing-wave like distribution from Fig. 4.11(c) and Fig. 4.11(d), can be noticed.

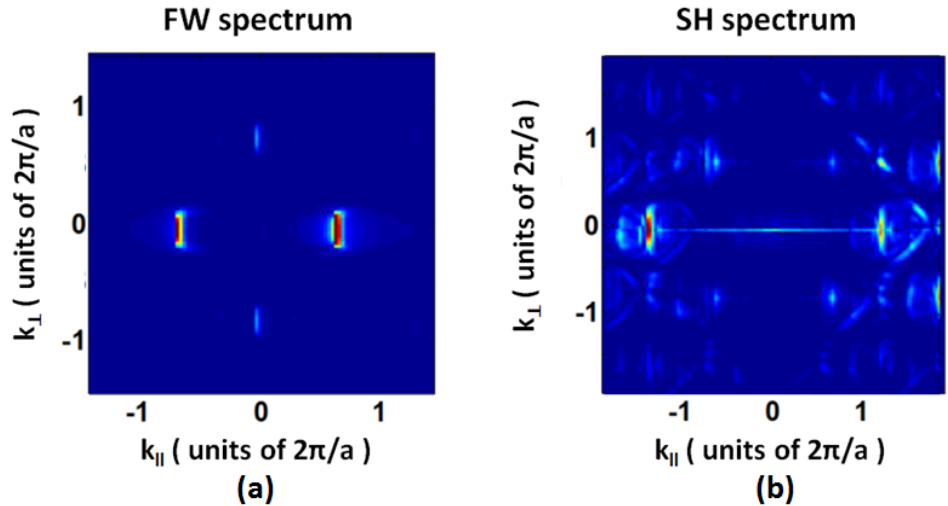

**Fig. 4.12** The spectra calculated for FW (a) and SH wave (b) obtained applying Fourier transform on the complex field data

Next, in order to illustrate better the wave vectors for which light can propagate inside this structure, at both FW and SH frequencies, the FW source is placed inside the structure calculating the spectra represented in Fig. 4.13(a) and Fig. 4.13(c) for the FW and SH wave, respectively. In Fig. 4.13(b) and Fig. 4.13(d) the isofrequency contours calculated for the same configuration are represented, and it can be seen a good correspondence between the two sets of plots. The direction of propagation is along the $k_{II}$ axis in Fig. 4.13(a) and Fig. 4.13(c), and along the diagonal from the point (0,0) to the point (1,1) in Fig. 4.13(b) and Fig. 4.13(d). Noting also that the scale is different in the two sets of figures, it can be observed that the width of the isofreqeuncy plateaus and their positions are similar. The width of the beams used in these calculations was relatively small (1.5 μm) in order to obtain wave vector components in a wider angular range, and be able to illustrate a more rich spectral distribution of components. Therefore, the components corresponding to strongly curved segments in the isofrequency contours from Fig. 4.13(d) appear with less intensity in the spectra from Fig. 4.13(c).

As shown in Fig. 4.4(d), the dispersion curves of the FW and SH wave for this configuration cross at similar slopes and this fact predicts that there will be a relative broad spectral range phase matching. The results of the FDTD calculations represented in Fig. 4.9(a) confirm the wide spectral range phase matching and the resonance has a FWHM (full width at half maximum) of approximately 20 nm. In Fig. 4.14(a) the SH intensity when the FW frequency is varied around the phase matched frequency is represented. The SH intensity obtained in a homogeneous medium with the same nonlinear coefficient and using plane wave source is calculated as well. Then, the SH intensity obtained in PC is expressed in units of the SH intensity obtained in the homogeneous material. It can be noticed that at the phase matching the SH intensity obtained in the PC is almost 100 times stronger than the SH intensity obtained in homogeneous case, difference given mainly by the decreasing of the group velocity in the PC.

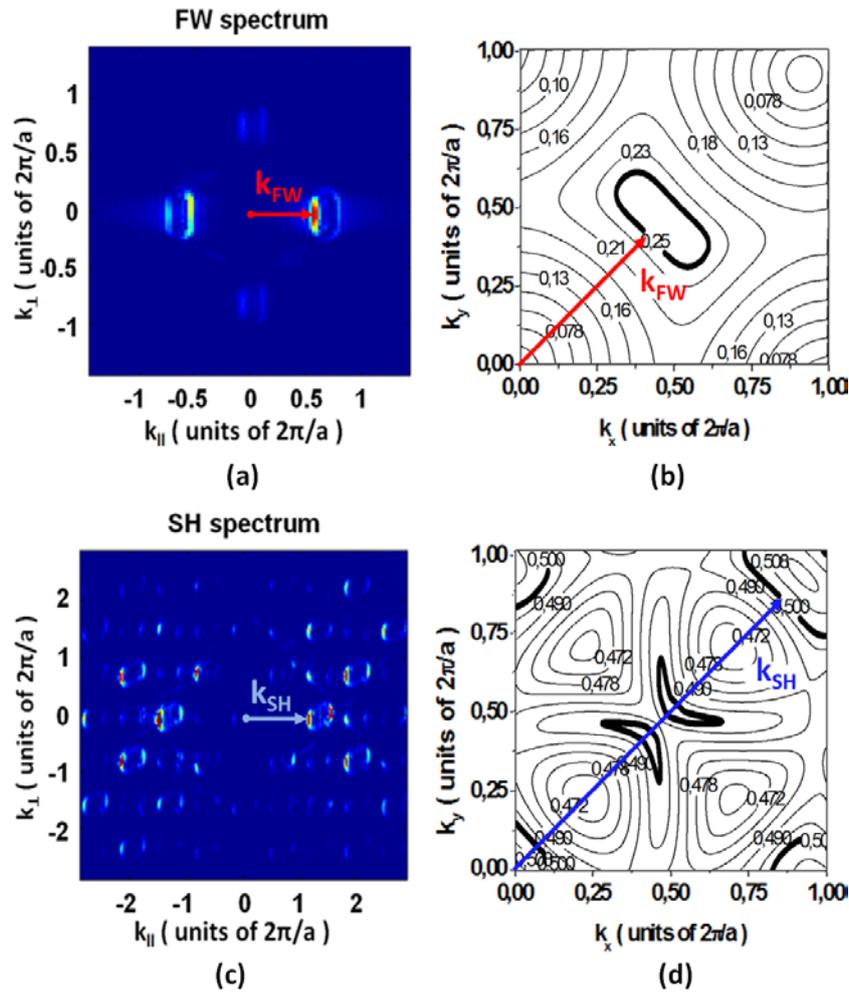

**Fig. 4.13** The spectra calculated for FW (a) and SH wave (c) when the FW source is placed in the middle of the structure, and the iso-frequency contours calculated for the same structure for the FW (b) and SH wave (d); with red and blue arrows are represented the phase matching wavevectors for FW and SH wave, respectively

The generated SH intensity is calculated also for different widths of the FW source. As expected, it can be noticed that the conversion efficiency depends on the width of the initial FW beam. The SH generation for extremely narrow beams becomes less efficient, as the width of the spatial spectra becomes broader than the plateau of the dispersion curve. The width of the plateau for SH wave in Fig. 4.13(c) and Fig. 4.13(d) corresponds to the beam width of approximately 2 μm. This means that the conversion efficiency sharply decreases for beams narrower than 2 μm, as illustrated also in Fig. 4.14(b). For large values of the width, the SH intensity does not change significantly as the entire width of the spatial spectra falls on the central part of the plateau.

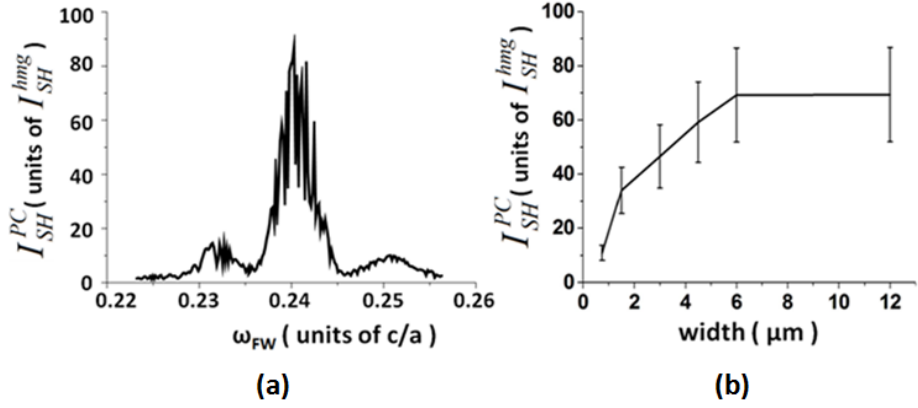

**Fig. 4.14** The intensity of the SH depending on the frequency of FW (a), and on the width of the FW beam at resonant frequency (b). The intensity is normalized to that of SH wave generated by the plane wave in homogeneous media of the same length and the same nonlinearity at PM.

## 4.6    Conclusions

The main conclusions of the results presented in this chapter can be summarized as follows:

- The distortion of dispersion surfaces in 2D PC allow a broad spectral range phase matching given by the similar slopes of the two dispersion curves at the phase matched wavevectors. In addition, the simultaneous self-collimation regimes allow a broad angular range phase matching that permits the use of narrow beams for the nonlinear interactions.

- We designed a planar 2D PC where both broad angular range and broad spectral range phase matching are expected due to the characteristics of the dispersion surfaces at the phase matched wavevectors.

- The position of the corresponding photonic bands with respect to the light cone show the presence of the out-of-plane scattering for the SH wave. Anyway, the out-of-plane scattering vanish for the FW, which means that the FW will continue to generate SH along the propagation through the crystal.

- The nonlinear FDTD calculations confirm both broad angular range and broad spectral range phase matching. Consequently, a significant increase of the conversion efficiency with respect to the plane wave case is obtained.
- The configuration can be easily tuned to achieve phase matching in other nonlinear materials (different refractive indices) and in other planar waveguides (different thickness, and therefore different effective refractive indices). The tuning of the main parameters, as the radius of the holes and the lattice geometry, allows the compensation of dispersion in a relative large range, from 0% to around 20%, therefore the idea can be applied to nearly all known materials.

## 5. General Conclusions and Future Work

The goal of the work presented in this thesis is to achieve efficient parametric interaction of narrow beams (in particular second harmonic generation) using quadratic nonlinear photonic crystal tuned to self-collimation (nondiffractive) regimes. The first part concerns the fundamentals of this idea showing that simultaneously self-collimation regimes for both fundamental and second harmonic waves can be obtained in an idealized 2D PC (neglecting the dispersion of the material and the finite size of the crystal) and lead to an amplification of the nonlinear interaction. The second part is more applied, dedicated to analyze the realization of this idea in realistic materials and configurations. The tasks have been completed and two realizable configuration have been suggested, one of them presently being in process of fabrication and characterization.

The main objectives accomplished are:

1) predicting and numerically proving that in 2D PCs made of ideal dispersionless materials the nondiffractive propagation regimes for both FW and SH wave, the phase matching and an efficient parametric coupling of the two modes can be obtained simultaneously. The nonlinear FDTD calculations made in the 2D PC confirm the self-collimation regimes of the two waves and the enhancement of conversion efficiency with respect to that obtained in homogeneous material using plane wave source. The enhancement is given also by the decreasing of the group velocity obtained in PCs;

2) a systematical study using different lattice geometries (square and rhombic), different radii of the air holes and combining different waveguiding modes with the purpose to obtain the phase matching, self-collimation regimes simultaneously and vanishing of the out-of-plane losses for both FW and SH wave in planar 2D PCs. Using the parameters of a real dispersive material ($Al_{0.3}Ga_{0.7}As$), a planar 2D PC for which the out-of-plane losses vanish for both FW and SH wave in conditions of diffraction similar to that in homogeneous material was designed;

3) to develop a 2D nonlinear FDTD program capable of making simulations of light propagation in quadratic nonlinear dispersive materials, feature that is usually absent in the commercial programs;

4) to show that 2D PCs made of materials with normal dispersion allow broad angular range and broad spectral range phase matching simultaneously, leading to significant increasing of the efficiency of the nonlinear process. A planar 2D PC made in AlGaAs that ensures both broad angular range and broad spectral range phase matching was proposed and the nonlinear FDTD simulations confirm the predicted propagation regimes for the FW and SH wave and the increasing of conversion efficiency with respect to the case of plane waves in homogeneous materials;

5) the design and numerical characterization of an AlGaAs membrane where both FW and SH wave propagate without diffraction and without out-of-plane losses; the conversion efficiency and the characteristics of the backward generated second harmonic are obtained by nonlinear FDTD calculations.

The most direct and natural continuation of this work is the fabrication of the photonic crystal structures presented trying to verify experimentally the theoretical results. Another straightforward continuation is to consider the more general case of optical parametric amplification where in conditions of nondiffractive propagation regimes for all waves involved the efficiency of the process could be enhanced.

Next, the results could be extended to 3D structures, like woodpiles or opals made from, or filled by, quadratic nonlinear materials.

Finally, the idea developed in this thesis could be applied in fields beyond optics, acoustics in particular. As acoustic materials are dispersion-free, the first objective presented above can be directly applied for building efficient second harmonic generation and parametric amplification schemes of sound beams in sonic crystals.

# Appendix A

## Finite-difference time-domain method

The two curl equations of Maxwell's equations (1.1)-(1.4) written in a Cartesian coordinate system give:

$$-\frac{\partial B_x}{\partial t} = \frac{\partial E_z}{\partial y} - \frac{\partial E_y}{\partial z} \qquad (A.\,1)$$

$$-\frac{\partial B_y}{\partial t} = \frac{\partial E_x}{\partial z} - \frac{\partial E_z}{\partial x} \qquad (A.\,2)$$

$$\frac{\partial B_z}{\partial t} = \frac{\partial E_x}{\partial y} - \frac{\partial E_y}{\partial x} \qquad (A.\,3)$$

$$\frac{\partial D_x}{\partial t} = \frac{\partial H_z}{\partial y} - \frac{\partial H_y}{\partial z} - J_x \qquad (A.\,4)$$

$$\frac{\partial D_y}{\partial t} = \frac{\partial H_x}{\partial z} - \frac{\partial H_z}{\partial x} - J_y \qquad (A.\,5)$$

$$\frac{\partial D_z}{\partial t} = \frac{\partial H_y}{\partial x} - \frac{\partial H_x}{\partial y} - J_z \qquad (A.\,6)$$

where $\vec{A} = (A_x, A_y, A_z)$. We denote a grid point of the space as:

$$(i, j, k) = (i\Delta x, j\Delta y, k\Delta z) \qquad (A.\,7)$$

and for any function of space and time we put:

$$f(i\Delta x, j\Delta y, k\Delta z, n\Delta t) = f_{i,j,k}^n \qquad (A.\,8)$$

The Taylor's series expansion of first order for a function $f(x, y, t)$ about the point $x$ to the point $x \pm \Delta x$ at a fixed $y$, $z$ and $t$ gives:

$$f(x+\Delta x, y, z, t) = f(x, y, z, t) + \frac{\partial f(x, y, z, t)}{\partial x}\Delta x + o\left([\Delta x]^2\right) \quad \text{(A. 9)}$$

$$f(x-\Delta x, y, z, t) = f(x, y, z, t) - \frac{\partial f(x, y, z, t)}{\partial x}\Delta x + o\left([\Delta x]^2\right) \quad \text{(A. 10)}$$

Adding these two expressions we obtain the central difference approximation:

$$\frac{\partial f(x, y, z, t)}{\partial x} = \frac{f(x+\Delta x, y, z, t) - f(x-\Delta x, y, z, t)}{2\Delta x} + o\left([\Delta x]^2\right) \quad \text{(A. 11)}$$

In a similar way we obtain:

$$\frac{\partial f(x, y, z, t)}{\partial y} = \frac{f(x, y+\Delta y, z, t) - f(x, y-\Delta y, z, t)}{2\Delta y} + o\left([\Delta y]^2\right) \quad \text{(A. 12)}$$

$$\frac{\partial f(x, y, z, t)}{\partial z} = \frac{f(x, y, z+\Delta z, t) - f(x, y, z-\Delta z, t)}{2\Delta z} + o\left([\Delta z]^2\right) \quad \text{(A. 13)}$$

$$\frac{\partial f(x, y, z, t)}{\partial t} = \frac{f(x, y, z, t+\Delta t) - f(x, y, z, t-\Delta t)}{2\Delta t} + o\left([\Delta t]^2\right) \quad \text{(A. 14)}$$

Using the central difference approximation and the notation (A.8) in equation (A.1) we have:

$$\frac{B_{x\ i,j+1/2,k+1/2}^{n+1/2} - B_{x\ i,j+1/2,k+1/2}^{n-1/2}}{\Delta t} = \frac{E_{y\ i,j+1/2,k+1}^{n} - E_{y\ i,j+1/2,k}^{n}}{\Delta z} - \frac{E_{z\ i,j+1/2,k+1}^{n} - E_{z\ i,j,k+1/2}^{n}}{\Delta y}$$

(A. 15)

and in a similar way it can be obtained the finite difference equations for (A.2)-(A.3). For equation (A.4) we have:

$$\frac{D_{x\ i+1/2,j,k}^{n} - D_{x\ i+1/2,j,k}^{n-1}}{\Delta t} = \frac{H_{z\ i+1/2,j+1/2,k}^{n-1/2} - H_{z\ i+1/2,j-1/2,k}^{n-1/2}}{\Delta y} - \frac{H_{y\ i+1/2,j,k+1/2}^{n-1/2} - H_{y\ i+1/2,j,k-1/2}^{n-1/2}}{\Delta z} - J_{x\ i+1/2,j,k}^{n-1/2}$$

(A. 16)

and the equations corresponding to (A.5)-(A.6) can be constructed in a similar way.

The grid size should be such that over one step the electromagnetic field does not change significantly. This means that, to have significant results, the dimension of the grid must be only a fraction of the wavelength. For computational stability, it is necessary to satisfy a relation between the space increment and time increment. Stability ensures that the energy is bounded within the system, and this means that for any value of $n$:

$$\frac{\left|f^{n+1}\right|^2}{\left|f^n\right|^2} \leq 1 \qquad (A.\ 17)$$

where $f$ represents the electric and magnetic fields. A detailed analysis of the stability problem can be found in [Taf05].

In order to avoid the reflections at the boundaries of the computational space, it is needed the use of some absorbing boundary conditions. A commonly used way to truncate the computational region is the perfect matched layer (PML), which is an artificial absorbing layer for wave equations. The main property of PML that distinguishes it from an ordinary absorbing material is that it is designed so that waves incident upon the PML from a non-PML medium do not reflect at the interface and this allows PML to strongly absorb outgoing waves from the interior of a computational region without reflecting them back. More technically, the PML represent an analytical continuation of the wave equation into complex coordinates, replacing the propagating (oscillating) waves by exponentially decaying waves [Taf05].

In the case of 2D structures, the dielectric constant is modulated only in the plane of propagation (the $xy$-plane), and it is constant in the third direction (the $z$-direction). Therefore, the electromagnetic field is constant along this direction, and the derivative with respect to this coordinate is zero $\partial/\partial x = 0$. This will simplify the equations (A.1)-(A.6):

$$\begin{cases} \dfrac{\partial H_x}{\partial t} = -\dfrac{1}{\mu_0}\dfrac{\partial E_z}{\partial y} & \dfrac{\partial E_x}{\partial t} = \dfrac{1}{\varepsilon}\dfrac{\partial H_z}{\partial y} \\ \dfrac{\partial H_y}{\partial t} = \dfrac{1}{\mu_0}\dfrac{\partial E_z}{\partial x} & \dfrac{\partial E_y}{\partial t} = -\dfrac{1}{\varepsilon}\dfrac{\partial H_z}{\partial x} \\ \dfrac{\partial H_z}{\partial t} = \dfrac{1}{\mu_0}\left(\dfrac{\partial E_x}{\partial y} - \dfrac{\partial E_y}{\partial x}\right) & \dfrac{\partial E_x}{\partial t} = \dfrac{1}{\varepsilon}\left(\dfrac{\partial H_y}{\partial x} - \dfrac{\partial H_x}{\partial y}\right) \end{cases} \qquad (A.\ 18)$$

It can be seen that there are two groups of independent components: the *TE* components ($H_z, E_x$ and $E_y$) and the *TM* components ($E_z, H_x$ and $H_y$). Therefore, the propagation of the electromagnetic field can be calculated independently for the two polarizations, *TE* and *TM*.

For the *TE* polarization we have:

$$H_x = H_y = 0, \quad E_z = 0 \tag{A.19}$$

$$\frac{\partial E_x}{\partial t} = \frac{1}{\varepsilon}\frac{\partial H_z}{\partial y} \tag{A.20}$$

$$\frac{\partial E_y}{\partial t} = -\frac{1}{\varepsilon}\frac{\partial H_z}{\partial x} \tag{A.21}$$

$$\frac{\partial H_z}{\partial t} = \frac{1}{\mu_0}\left(\frac{\partial E_x}{\partial y} - \frac{\partial E_y}{\partial x}\right) \tag{A.22}$$

and for *TM* polarization we have:

$$E_x = E_y = 0, \quad H_z = 0 \tag{A.23}$$

$$\frac{\partial H_x}{\partial t} = -\frac{1}{\mu_0}\frac{\partial E_z}{\partial y} \tag{A.24}$$

$$\frac{\partial H_y}{\partial t} = \frac{1}{\mu_0}\frac{\partial E_z}{\partial x} \tag{A.25}$$

$$\frac{\partial E_x}{\partial t} = \frac{1}{\varepsilon}\left(\frac{\partial H_y}{\partial x} - \frac{\partial H_x}{\partial y}\right) \tag{A.26}$$

When writing the finite difference equations, we have for the *TE* waves:

$$E_{x\ i+1/2,j}^{n+1} = E_{x\ i+1/2,j}^{n} + \frac{\Delta t}{\varepsilon \Delta}\left(H_{z\ i+1/2,j+1/2}^{n+1/2} - H_{z\ i+1/2,j-1/2}^{n+1/2}\right) \tag{A.27}$$

$$E_{y\ i,j+1/2}^{n+1} = E_{y\ i,j+1/2}^{n} + \frac{\Delta t}{\varepsilon \Delta}\left(H_{z\ i-1/2,j+1/2}^{n+1/2} - H_{z\ i+1/2,j+1/2}^{n+1/2}\right) \tag{A.28}$$

$$H_{z\ i+1/2,j+1/2}^{n+1/2} = H_{z\ i+1/2,j+1/2}^{n-1/2} + \frac{\Delta t}{\varepsilon \Delta}\left(E_{z\ i,j+1/2}^{n} - E_{z\ i+1,j+1/2}^{n} + E_{x\ i+1/2,j+1}^{n} - E_{x\ i+1/2,j}^{n}\right) \tag{A.29}$$

and for the *TM* waves we have:

$$H_{x\ i,j+1/2}^{n+1/2} = H_{x\ i,j+1/2}^{n-1/2} + \frac{\Delta t}{\varepsilon \Delta}\left(E_{z\ i,j}^{n} - E_{z\ i,j+1}^{n}\right) \quad \text{(A. 30)}$$

$$H_{y\ i+1/2,j}^{n+1/2} = H_{y\ i+1/2,j}^{n-1/2} + \frac{\Delta t}{\varepsilon \Delta}\left(E_{z\ i+1,j}^{n} - E_{z\ i,j}^{n}\right) \quad \text{(A. 31)}$$

$$E_{z\ i,j}^{n+1} = E_{z\ i,j}^{n} + \frac{\Delta t}{\varepsilon \Delta}\left(H_{z\ i+1/2,j}^{n+1/2} - H_{z\ i-1/2,j}^{n+1/2} + H_{x\ i,j-1/2}^{n+1/2} - H_{x\ i,j+1/2}^{n+1/2}\right) \quad \text{(A. 32)}$$

where $\Delta = \Delta x = \Delta y$.

# Appendix B

## Effective refractive indices in planar waveguides

Planar optical waveguides are key devices to construct integrated optical circuits and semiconductor lasers, as they permit to confine light in the vertical direction. A planar waveguide, represented in Fig. B. 1, consists of a core layer surrounded by substrate and cladding layers with lower refractive indices than the core. The presence of the planar cavity introduces effective refractive indices of the propagation modes different from those in the bulk material.

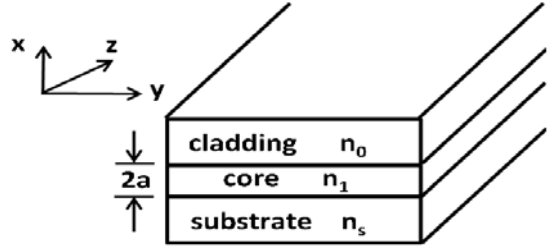

**Fig. B. 1** Schematic representation of an optical planar waveguide

We set the permittivity and permeability as $\varepsilon = \varepsilon_0 n^2$, where $n$ is the refractive index, and $\mu = \mu_0$ in the curl Maxwell's equations:

$$\nabla \times \vec{E} = -\mu_0 \frac{\partial \vec{H}}{\partial t} \tag{B. 1}$$

$$\nabla \times \vec{H} = \varepsilon_0 n^2 \frac{\partial \vec{E}}{\partial t} \tag{B. 2}$$

We are interested in the propagation of plane waves of the form:

$$\vec{E}(x, y, t) = \vec{E}(x, y)e^{j(\omega t - \beta z)} \tag{B. 3}$$

$$\vec{H}(x, y, t) = \vec{H}(x, y)e^{j(\omega t - \beta z)} \tag{B. 4}$$

Substituting eq. (B.3) and (B.4) in eq. (B.1) and (B.2) we obtain the following set of equations for the electromagnetic field components:

$$\begin{cases} \dfrac{\partial E_z}{\partial y} + j\beta E_y = -j\omega\mu_0 H_x \\ -j\beta E_x - \dfrac{\partial E_z}{\partial x} = -j\omega\mu_0 H_y \\ \dfrac{\partial E_y}{\partial x} - \dfrac{\partial E_x}{\partial y} = -j\omega\mu_0 H_z \end{cases} \quad \text{(B. 5)}$$

$$\begin{cases} \dfrac{\partial H_z}{\partial y} + j\beta H_y = j\omega\varepsilon_0 n^2 E_x \\ -j\beta H_x - \dfrac{\partial H_z}{\partial x} = j\omega\varepsilon_0 n^2 E_y \\ \dfrac{\partial H_y}{\partial x} - \dfrac{\partial H_x}{\partial y} = -j\omega\varepsilon_0 n^2 E_z \end{cases} \quad \text{(B. 6)}$$

In a planar waveguide, as represented in Fig. B. 1, the electromagnetic fields $\vec{E}$ and $\vec{H}$ do not vary along the $y$-axis. Therefore, we consider $\partial E/\partial y = 0$ and $\partial H/\partial y = 0$. Introducing these relations in eq. (B.5) and (B.6) we obtain two independent electromagnetic modes that are denoted TE (transverse electric, as the electric field lies in the plane perpendicular to the direction of propagation) and TM (transverse magnetic, as the magnetic field lies in the plane perpendicular to the direction of propagation), respectively.

The TE mode satisfies the following wave equation:

$$\frac{\partial^2 E_y}{\partial x^2} + (k^2 n^2 - \beta^2) E_y = 0 \quad \text{(B. 7)}$$

where

$$H_x = -\frac{\beta}{\omega\mu_0} E_y \quad \text{(B. 8)}$$

$$H_z = \frac{j}{\omega\mu_0} \frac{\partial E_y}{\partial x} \quad \text{(B. 9)}$$

and

$$E_x = E_z = 0 \quad \text{and} \quad H_y = 0 \quad \text{(B. 10)}$$

The TM mode satisfies the following wave equation:

$$\frac{\partial}{\partial x}\left(\frac{1}{n^2}\frac{\partial H_y}{\partial x}\right) + \left(k^2 - \frac{\beta^2}{n^2}\right)H_y = 0 \quad (B.11)$$

where

$$E_x = \frac{\beta}{\omega\varepsilon_0 n^2}H_y \quad (B.12)$$

$$E_z = -\frac{j}{\omega\varepsilon_0 n^2}\frac{\partial H_y}{\partial x} \quad (B.13)$$

and

$$H_x = H_z = 0 \text{ and } E_y = 0 \quad (B.14)$$

Propagations constant and electromagnetic fields for TE and TM modes can be obtained by solving eq. (B.7) and (B.11). We consider the slab waveguide with uniform refractive-index profile in the core, as represented in Fig. B. 2. Considering the fact that the guided electromagnetic fields are confined in the core and exponentially decay in the cladding, the electric field distribution is expressed as:

$$E_y = \begin{cases} A\cos(\kappa a - \varphi)e^{-\sigma(x-a)} & x > a \\ A\cos(\kappa x - \varphi) & -a \leq x \leq a \\ A\cos(\kappa a + \varphi)e^{\xi(x+a)} & x < -a \end{cases} \quad (B.15)$$

where $\kappa, \sigma$ and $\xi$ are wave numbers along the $x$-axis in the core and cladding regions and are given by:

$$\begin{cases} \kappa = \sqrt{k^2 n_1^2 - \beta^2} \\ \sigma = \sqrt{\beta^2 - k^2 n_0^2} \\ \xi = \sqrt{\beta^2 - k^2 n_s^2} \end{cases} \quad (B.16)$$

The boundary conditions that we have to use are that the electric field component $E_y$ and the magnetic field component $H_z$ should be continuous at the boundaries of core-cladding interfaces $x = \pm a$. Neglecting the terms independent of $x$, the boundary condition for $H_z$ is given by the continuity of $\partial E_y/\partial x$ as:

$$\frac{\partial E_y}{\partial x} = \begin{cases} -\sigma A \cos(\kappa a - \varphi) e^{-\sigma(x-a)} & x > a \\ -kA \sin(\kappa a - \varphi) & -a \le x \le a \\ \xi A \cos(\kappa a + \varphi) e^{\xi(x+a)} & x < -a \end{cases} \quad \text{(B. 17)}$$

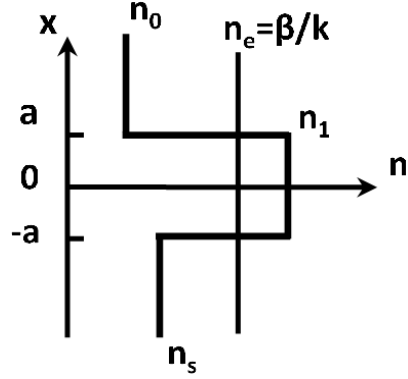

**Fig. B. 2** Refractive index profile in a planar waveguide

From the conditions that $\partial E_y / \partial x$ are continuous at $x = \pm a$ we obtain the following equations:

$$\begin{cases} \kappa A \sin(\kappa a + \varphi) = \xi A \cos(\kappa a + \varphi) \\ \sigma A \cos(\kappa a - \varphi) = \kappa A \sin(\kappa a - \varphi) \end{cases} \quad \text{(B. 18)}$$

Eliminating the constant A we have

$$\tan(u + \varphi) = \frac{w}{u} \quad \text{(B. 19)}$$

$$\tan(u - \varphi) = \frac{w'}{u} \quad \text{(B. 20)}$$

where

$$\begin{cases} u = \kappa a \\ w = \xi a \\ w' = \sigma a \end{cases} \quad \text{(B. 21)}$$

From eq. (B.19) and (B.20) we obtain the eigenvalue equation:

$$u = \frac{m\pi}{2} + \frac{1}{2}\tan^{-1}\left(\frac{w}{u}\right) + \frac{1}{2}\tan^{-1}\left(\frac{w'}{u}\right) \quad \text{with } m = 0,1,2,... \quad (B.22)$$

The normalized transverse wave numbers $u$, $w$ and $w'$ are related by the following equations:

$$u^2 + w^2 = k^2 a^2 \left(n_1^2 - n_s^2\right) \equiv v^2 \quad (B.23)$$

$$w' = \sqrt{\gamma v^2 + w^2} \quad (B.24)$$

$$\gamma = \frac{n_s^2 - n_0^2}{n_1^2 - n_s^2} \quad (B.25)$$

where $v$ is the normalized frequency and $\gamma$ is a measure of the asymmetry of the cladding refractive index.

Next, from equations (B.22) and (B.23) there can be obtained, either by a graphical or a numerical method [Oka06] the propagation constant $\beta$ for different guided modes, when $m = 0,1,2,...$. Therefore, the effective refractive index is given by the relation:

$$n_e = \frac{\beta}{k} \quad (B.26)$$

In order to obtain the wave guiding, as illustrated also in Fig. B. 2, the following condition should be satisfied:

$$n_s \leq n_e \leq n_1 \quad (B.27)$$

where $n_s$ represents the refractive index of the substrate and $n_1$ is the refractive index of the core layer.

When $n_e < n_s$, the electromagnetic field is dissipated as it corresponds to a radiation mode (non-guided mode). Since the condition $\beta = kn_s$ represents the critical condition for which the field becomes non-guided mode, it is called cutoff condition. Also, another parameter, $b$, called normalized propagation constant is relevant and it is defined by:

$$b = \frac{n_e^2 - n_s^2}{n_1^2 - n_s^2} \tag{B. 28}$$

The conditions for the guided modes are expressed by:

$$0 \leq b \leq 1 \tag{B. 29}$$

and the cutoff condition becomes:

$$b = 0 \tag{B. 30}$$

Whereas in slab waveguides the TE and TM modes are defined with respect to the direction of propagation, in 2D PC the TE and TM modes are defined with respect to the axis along the inclusions (cylinders, or air holes), noted as $z$-axis in Fig. B. 3. Consequently, the TE PC mode has only the components $E_x$, $E_y$ and $H_z$, and it is different from the TE slab mode. However, in the plane bisecting the slab through the field amplitude maximum (the peak point) of $H_z$ (or $E_x$, $E_y$), $H_x = H_y = 0$ and $E_z = 0$ [Qiu02]. Thus, in this plane (the maximum plane) TE slab modes are the same as TE PC modes and similar conclusions can also be obtained for TM modes.

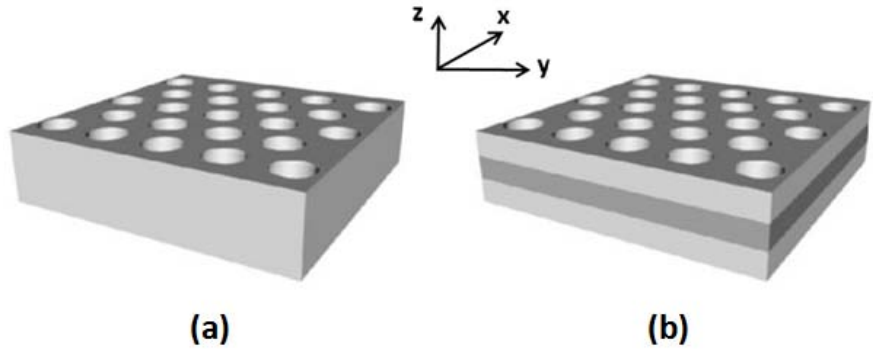

**Fig. B. 3** Schematic representation of a 2D PC(a) and of a planar 2D PC (b)

In planar 2D PC, the TE and TM modes are no longer separable due to the lack of translational symmetry in the $x-y$ plane. However, the modes in planar PC have strong similarity to the modes in unperturbed slab waveguides, as shown in [Joh99]. Similar to the situation above, in the maximum plane TE-like modes and TM-like modes are purely TE and TM modes, respectively. Hence, in this plane, the 3D slab PC could be considered as a pure 2D PC and, in order to correctly reconstruct the phase velocity, the material index in 2D PC must be replaced by the effective refractive index of the guided modes in the unperturbed planar waveguides.

Successful comparison of full-vectorial 3D FDTD results with 2D band structures showing the validity of the effective index method is reported in [Qiu02]. The effective index method has been proved as effective and efficient approach by comparing calculated transmission spectra through 2D PC devices with experimental results [Ben99, Oli01].

# Appendix C

## Nonlinear decoupled FDTD method

From the Maxwell's curl equations (1.31) and (1.32) and the constitutive relation (1.33) we obtain for the SH wave:

$$\frac{\partial \vec{H}_{SH}}{\partial t} = -\frac{1}{\mu_0} \nabla \times \vec{E}_{SH} \qquad (C.1)$$

$$\frac{\partial \vec{E}_{SH}}{\partial t} = -\frac{1}{\varepsilon} \nabla \times \vec{H}_{SH} - \frac{1}{\varepsilon} \frac{\partial \vec{P}^{(2)}}{\partial t} \qquad (C.2)$$

where $P^{(2)} = \varepsilon_0 \chi^{(2)} E_{FW}^2$ is the second order nonlinear polarization.

Therefore, the equations used for calculating the SH field for TE and TM polarizations become:

$$\frac{\partial E_z}{\partial t} = \frac{1}{\varepsilon(x,y)}\left(\frac{\partial H_y}{\partial x} - \frac{\partial H_x}{\partial y}\right) - \frac{1}{\varepsilon(x,y)}\frac{\partial P^{(2)}}{\partial t} \qquad TM \qquad (C.3)$$

$$\begin{cases} \dfrac{\partial E_x}{\partial t} = \dfrac{1}{\varepsilon(x,y)}\dfrac{\partial H_z}{\partial y} - \dfrac{1}{\varepsilon(x,y)}\dfrac{\partial P^{(2)}}{\partial t} \\ \dfrac{\partial E_y}{\partial t} = -\dfrac{1}{\varepsilon(x,y)}\dfrac{\partial H_z}{\partial x} - \dfrac{1}{\varepsilon(x,y)}\dfrac{\partial P^{(2)}}{\partial t} \end{cases} \qquad TE \qquad (C.4)$$

After discretization, using the same notation as in Appendix A, these equations become for TM polarization:

$$E_{z\ i,j}^{n+1} = E_{z\ i,j}^{n} + \frac{\Delta t}{\varepsilon_{i,j}\Delta}\left(H_{z\ i+1/2,j}^{n+1/2} - H_{z\ i-1/2,j}^{n+1/2} + H_{x\ i,j-1/2}^{n+1/2} - H_{x\ i,j+1/2}^{n+1/2}\right) - \frac{1}{\varepsilon_{i,j}}\left(P_{z\ i,j}^{(2),n+1} - P_{z\ i,j}^{(2),n}\right)$$

(C.5)

and for TE polarization:

$$\begin{cases} E_x{}^{n+1}_{i+1/2,j} = E_x{}^n_{i+1/2,j} + \frac{\Delta t}{\varepsilon_{i,j}\Delta}\left(H_z{}^{n+1/2}_{i+1/2,j+1/2} - H_z{}^{n+1/2}_{i+1/2,j-1/2}\right) - \frac{1}{\varepsilon_{i,j}}\left(P_z^{(2),n+1/2}{}_{i+1/2,j} - P_z^{(2),n-1/2}{}_{i+1/2,j}\right) \\ E_y{}^{n+1}_{i,j+1/2} = E_y{}^n_{i,j+1/2} + \frac{\Delta t}{\varepsilon_{i,j}\Delta}\left(H_z{}^{n+1/2}_{i-1/2,j+1/2} - H_z{}^{n+1/2}_{i+1/2,j+1/2}\right) - \frac{1}{\varepsilon_{i,j}}\left(P_z^{(2),n+1/2}{}_{i,j+1/2} - P_z^{(2),n-1/2}{}_{i,j+1/2}\right) \end{cases}$$

(C. 6)

The computational domain is truncated using perfectly matched layer (PML) absorbing boundaries. The formulation used in this code is based on the original split-field Berenger PML [Ber94], where certain components of the electromagnetic field are split into subcomponents. This splitting introduces an additional degree of freedom in specifying the material parameters which permits that waves of arbitrary frequency and angle of propagation rapidly decay and yet maintain the velocity and field impedance of the lossless dielectric case.

We consider the TE polarization case, when we have the field components $E_x, E_y$ and $H_z$. Here, the PML formulation specifies four rather than usual three coupled field equations because $H_z$ is split into two subcomponents $H_{zx}$ and $H_{zy}$:

$$\varepsilon_d \frac{\partial E_x}{\partial t} + \sigma_y E_x = \frac{\partial(H_{zx} + H_{zy})}{\partial y} \tag{C. 7}$$

$$\varepsilon_d \frac{\partial E_y}{\partial t} + \sigma_x E_y = -\frac{\partial(H_{zx} + H_{zy})}{\partial x} \tag{C. 8}$$

$$\mu_d \frac{\partial H_{zx}}{\partial t} + \sigma_x^* H_{zx} = -\frac{\partial E_y}{\partial x} \tag{C. 9}$$

$$\mu_d \frac{\partial H_{zy}}{\partial t} + \sigma_y^* H_{zy} = \frac{\partial E_x}{\partial y} \tag{C. 10}$$

Electric loss $\sigma$ and magnetic loss $\sigma^*$ are assigned to the electric and magnetic field components as indicated to cause exponential decay of propagating fields in the PML region. Moreover, if we take:

$$\frac{\sigma_x}{\varepsilon_d} = \frac{\sigma_x^*}{\mu_d} \quad \text{and} \quad \frac{\sigma_y}{\varepsilon_d} = \frac{\sigma_y^*}{\mu_d} \tag{C. 11}$$

where $\varepsilon_d$ and $\mu_d$ are the PML permittivity and permeability, this decay can take place in a frequency-independent manner without affecting the wave impedance.

We consider a 2D dielectric waveguide (with $\varepsilon_d$ and $\mu_d$) where the TE polarized wave in propagating along the $+x$ direction. At $x_{max}$ the waveguide is loaded with PML that has $\sigma_x$ and $\sigma_x^*$ matched to eq. (C.11) with $\sigma_y = \sigma_y^* = 0$ to permit reflectionless transmission across the dielectric PML interface. Nevertheless, the layer is not without reflection and an apparent reflection can be defined, which is a function of two parameters, the PML thickness $\delta$ and the conductivity profile $\sigma(\rho)$, where $\rho$ is the distance from the interface. This conductivity is either $\sigma_x$ in the layer normal to $x$ direction or $\sigma_y$ in a layer normal to $y$ direction.

We consider that the loss in the PML region increases quadratically with the distance $\rho$. This means that the loss rises from zero at $\rho = 0$ (the interface between the waveguide and the PML) to a maximum value $\sigma_{max}$ at $\rho = \delta$:

$$\sigma(\rho) = \sigma_{max} \left(\frac{\rho}{\delta}\right)^2 \qquad (C.\,12)$$

Then, $\sigma_{max}$ can be chosen to adjust the apparent reflection coefficient:

$$R = e^{-\frac{2\sigma_{max}\delta}{3\varepsilon_d c}} \qquad (C.\,13)$$

to some desired low level, say $10^{-4}$.

# List of Publications

**Articles:**

"Lossless backward second-harmonic generation of extremely narrow subdiffractive beams in two-dimensional photonic crystals"
C. Nistor, C. Cojocaru, T.J.Karle, F. Raineri, J. Trull, R. Raj, and K. Staliunas
Phys. Rev. A 82, 033805 (2010)

"Broad angle phase matching in subdiffractive photonic crystals"
Cristian Nistor, Crina Cojocaru, Jose Trull, and Kestutis Staliunas
Optics Communications 283, 3533-3535 (2010)

"Phase matched second harmonic generation in planar two-dimensional photonic crystals"
Cristian Nistor, Crina Cojocaru, Yurii Loiko, Jose Trull, Kestutis Staliunas
J. Opt. A: Pure Appl. Opt. 11, 114016 (2009)

"Second-harmonic generation of narrow beams in subdiffractive photonic crystals"
C. Nistor, C. Cojocaru, Yu. Loiko, J. Trull, R. Herero and K. Staliunas
Physical Review A 78, 053818 (2008)

**Conference proceedings:**

"Broadband phase-matched second-harmonic generation for narrow beams in planar two-dimensional photonic crystal"
C. Nistor, C. Cojocaru, J. Trull, and K. Staliunas
Proccedings of SPIE 7713, 771315 (2010);

"Second Harmonic Generation in Planar Two-Dimensional Photonic Crystals without Out-of-Plane Losses"
C. Nistor, C. Cojocaru, Y. Loiko, J. Trull, and K. Staliunas

ICTON 2009 11[th] International Conference on Transparent Optical Networks Vol. 1-2 pages: 936-939 (2009)

"Actively induced reflection via a quadratic optical interaction in a semiconductor photonic crystal: Application to ultra fast all-optical modulation and switching"
C. Cojocaru, J. Trull, C. Nistor, R. Herrero, and K. Staliunas
ICTON 2006: 8[th] International Conference on Transparent Optical Networks Vol. 2, pages: 79-82 (2006)

**Other conference participations:**

"Vertically Confined Backward Second Harmonic Generation for Narrow Beams in Planar Two-Dimensional Photonic Crystals"
C. Nistor, T.J. Karle, F. Raineri, C. Cojocaru, J. Trull, R. Raj, and K. Staliunas
Poster at PECS-IX 2010, 9[th] International Conference on Photonic and Electromagnetic Crystal Structures, Granada, Spain, 26-30 September, 2010

"Vertically confined phase matched beams second harmonic generation in planar two-dimensional photonic crystals"
C. Nistor, C. Cojocaru, J. Trull, and K. Staliunas
Oral presentation at LPHYS' 09: 18[th] International Laser Physics Workshop, Barcelona July 13-17, 2009

"Second harmonic generation of narrow beams in subdiffractive photonic crystals"
Cristian Nistor, Yurii Loiko, Crina Cojocaru, Jose Trull, R. Herrero and Kestutis Staliunas
Oral presentation at CLEO Europe - EQEC 2009 June 14 – 19, 2009 Munich, Germany

"Narrow beams phase-matched second harmonic generation in two-dimensional photonic crystals"
C. Nistor, Y. Loiko, C. Cojocaru, J. Trull, R. Herrero and K. Staliunas
Poster at "A Future in Light" International Conference on Photonics March 26-27, 2009 Metz, France

"Phase matched second harmonic generation in planar two-dimensional photonic crystals"
C. Nistor, C. Cojocaru, Y. Loiko, J. Trull, and K. Staliunas
Poster at 1st International Workshop on Theoretical and Computational Nano-Photonics, Bad Honnef, Germany, Dec. 03-05, 2008

"Second harmonic generation of narrow beams in subdiffractive photonic crystals"
C Nistor, K. Staliunas, C.Cojocaru, J.Trull, R.Herrero
Poster at Conferencia Española de Nanofotónica, April 2-4, 2008 Tarragona, Spain

"Ultra-fast reflectivity and transmission tuning via a quadratic nonlinearity in a semiconductor photonic crystal"
C. Cojocaru, J. Trull, C. Nistor, R. Herrero, K. Staliunas
Oral presentation at: EOS Topical Meeting on Nanophotonics, Metamaterials and Optical Microcavities, October 16-18, 2006 Paris, France

"Ultra-fast tuning of a photonic crystal via a quadratic nonlinearity"
C. Cojocaru, J. Trull, C. Nistor, K. Staliunas, R. Vilaseca
Oral presentation at XXII Trobades Cientifiques de la Mediterránea, October 9-11, 2006 Menorca, Spain